\documentclass[11pt,a4paper]{article}
\usepackage[T1]{fontenc}
\usepackage[utf8]{inputenc}
\usepackage{lmodern}
\usepackage[margin=27mm]{geometry}
\usepackage{amsmath,amssymb,amsthm,mathtools}
\usepackage{microtype,booktabs,array,tabularx,longtable,enumitem}
\usepackage{xcolor,tikz}
\usepackage{comment,xspace}
\usetikzlibrary{arrows.meta,positioning,calc,patterns,decorations.pathreplacing}
\usepackage[colorlinks=true,linkcolor=blue!50!black,citecolor=green!35!black,urlcolor=blue!60!black]{hyperref}
\usepackage[nameinlink,noabbrev]{cleveref}
\hypersetup{pdftitle={Nearly Settling the Complexity of Connected epsilon-Envy-Free Cake Cutting},pdfauthor={Simon Mackenzie and Mashbat Suzuki}}
\newtheorem{theorem}{Theorem}[section]
\newtheorem{lemma}[theorem]{Lemma}
\newtheorem{proposition}[theorem]{Proposition}
\newtheorem{corollary}[theorem]{Corollary}

\theoremstyle{definition}
\newtheorem{definition}[theorem]{Definition}

\theoremstyle{remark}

\newcommand{\eps}{\varepsilon}
\newcommand{\R}{\mathbb R}

\newcommand{\ConnEF}{\textup{\textsc{ConnEF}}}

\DeclareMathOperator{\poly}{poly}
\DeclareMathOperator{\clip}{clip}

\DeclareMathOperator{\sgn}{sgn}
\setlist{itemsep=3pt,topsep=5pt}
\title{\bfseries Nearly Settling the Complexity of\\ Connected $\boldsymbol{\varepsilon}$-Envy-Free Cake Cutting}
\author{Simon Mackenzie\\UNSW Sydney\\\texttt{simon.william.mackenzie@gmail.com}
\and Mashbat Suzuki\\UNSW Sydney\\\texttt{mashbats@gmail.com}}
\date{}
\begin{document}
\maketitle
\begin{abstract}
Envy-free cake cutting has been one of the central problems in fair division. In cake cutting, agents' preferences over the cake are accessed through Robertson--Webb (RW) queries. For three or more agents, no algorithm that makes finitely many RW queries can always find a contiguous, exactly envy-free allocation. For contiguous $\varepsilon$-envy-free allocations, however, Hollender and Rubinstein~(JACM, 2025) showed that $\poly(\log(1/\eps))$ RW queries suffice for four agents, leaving open whether such a bound is possible for five or more agents. 

In this paper, we establish a striking change in query complexity: $\poly(\log(1/\eps))$ RW queries suffice for up to five agents, whereas $(1/\varepsilon)^{\Omega(1)}$ queries are necessary for seven or more agents, even when valuations are additive. Thus, we resolve the five-agent case and, more broadly, determine whether polylogarithmic query complexity is possible for every number of agents except six.

We also explore the computational complexity of finding contiguous $\varepsilon$-envy-free allocations. When monotone Lipschitz valuations are given by polynomial-time evaluation procedures (e.g., rational piecewise-linear circuits and polynomial-time Turing machines), we show that an $\varepsilon$-envy-free division can be computed in polynomial time for five agents. In contrast, even for additive valuations, finding an $\varepsilon$-envy-free division is UEOPL-hard for every fixed number of agents $n\ge7$ and PPAD-complete for every fixed $n\ge13$.

\end{abstract}
\newpage 

\noindent\begin{minipage}{\linewidth}
\section*{Use of AI tools}
We used OpenAI Codex and Anthropic Claude, including Claude Code, throughout this project. These systems developed the mathematical proofs and wrote most of the manuscript. We designed and organised the research process and actively guided the investigation. As the work developed, we responded to intermediate results with further mathematical intuitions and suggestions, redirected unpromising approaches and chose which ideas to pursue.

We ran many instances of these systems in parallel, assigning them specific research and review tasks. Some coordinated the work and assigned further tasks; others explored competing approaches, combined results or examined proposed arguments. Results and criticisms were shared between the Codex and Claude efforts. The systems developed the algorithms, lower-bound constructions and reductions, proved supporting lemmas, searched for counterexamples, repaired proof gaps and refined the bounds. They also searched the literature, checked attribution and model assumptions, and wrote and revised the exposition, diagrams and LaTeX manuscript. Additional model responses contributed to the investigation of the unresolved six-agent case but did not lead to successful resolution of that case.

The systems wrote and ran experimental code, exact-arithmetic checkers, allocation enumerators and separate implementations of proposed algorithms, and compiled and checked the manuscript. Much of the proof checking was also performed by AI systems assigned separate review tasks. The systems also wrote the Lean formalisation described in Section~\ref{sec:formal}. 
\end{minipage}

\section{Introduction}
\label{sec:introduction}

Cake cutting is a central problem in fair division, modelling how a divisible resource should be allocated among agents with different preferences~\cite{RW}. The resource, represented by the interval $[0,1]$, may be time on a shared facility or a tract of land. A fundamental fairness requirement is \emph{envy-freeness}: every agent should value her own allocation at least as much as any other agent's allocation.

A natural requirement in many applications is that each agent's allocation be contiguous. For example, an agent using a shared facility may need one uninterrupted period of access. We therefore study \emph{contiguous}, also called \emph{connected}, cake cutting: the entire cake is partitioned into intervals, one for each agent.

Classical results show that a contiguous, exactly envy-free allocation always exists for nonatomic additive valuations~\cite{Stromquist,Woodall,Su}. The existence proofs use topological tools such as Brouwer's fixed-point theorems and Sperner's lemma. Thus, contiguity and envy-freeness can be achieved simultaneously for any number of agents. The algorithmic question is how to find such an allocation when the agents' valuations are accessible only through queries.

The standard framework for studying this question is the \emph{Robertson--Webb} (RW) query model~\cite{RW}. An algorithm may ask an agent for the value of an interval through an \emph{evaluation query}, or for an endpoint at which an interval reaches a prescribed value through a \emph{cut query}. Despite the existence guarantee, Stromquist~\cite{Stromquist2008} proved that, for three or more agents, no algorithm can always find a contiguous, exactly envy-free allocation while terminating after finitely many RW queries on every instance. This impossibility holds even when the number of queries is allowed to depend on the instance.

This motivates allowing a small amount of envy. Let $v_i$ denote agent $i$'s nonatomic additive valuation, normalised so that $v_i([0,1])=1$. For a tolerance $\eps>0$, a contiguous allocation is \emph{$\eps$-envy-free} if each agent values every other agent's piece at most $\eps$ more than her own. Writing $A_i$ for the piece assigned to agent $i$, this condition is
\begin{align*}
    v_i(A_i) \ge v_i(A_j)-\eps
    \qquad\text{for all agents }i,j.
\end{align*}

Approximation makes finite-query algorithms possible, shifting the focus to how the required number of queries grows as $\eps$ decreases. For any fixed number of agents, a contiguous $\eps$-envy-free allocation can be found by brute-force search using $\poly(1/\eps)$ RW queries~\cite{BN}. In this setting, we call an algorithm \emph{efficient} if it uses only $\poly(\log(1/\eps))$ RW queries, an exponential improvement over the brute-force bound.

Deng, Qi, and Saberi~\cite{DQS} gave an
$O(\log^2(L/\eps))$ value-query algorithm for three agents with
monotone valuations that are positive on every interval of positive
length and have a known Lipschitz bound $L$.
For additive valuations in the Robertson--Webb model,
Br\^anzei and Nisan~\cite[Theorem~4.2]{BN} obtained an
$O(\log(1/\eps))$ bound by simulating the moving-knife procedure of
Barbanel and Brams~\cite{BB}. For four agents, Hollender and Rubinstein~\cite{HR} gave a deterministic protocol using $O(\log^2(1/\eps))$ queries. For five or more agents, however, it remained open whether a deterministic protocol could achieve $\poly(\log(1/\eps))$ query complexity. Hollender and Rubinstein asked:
\begin{quote}
\emph{What is the complexity of the problem for five agents with monotone valuations? Can the problem be solved efficiently, or can we show a lower bound? Can we at least show a lower bound for some larger number of players?}
\end{quote}

In this work, we answer these questions by showing that $\poly(\log(1/\eps))$ RW queries suffice for up to five agents, whereas $(1/\eps)^{\Omega(1)}$ queries are necessary for seven or more. This leaves only the six-agent case unresolved. Our upper bound for five agents also applies to monotone valuations with a known Lipschitz bound, while our lower bound holds even for additive valuations.

We also establish a corresponding separation in computational complexity when valuations are given by polynomial-time evaluation procedures. For five agents with monotone Lipschitz valuations, a contiguous $\eps$-envy-free allocation can be computed in polynomial time. In contrast, even for additive valuations, finding such an allocation is Unique End of Potential Line (UEOPL)-hard for every fixed $n\ge 7$ and PPAD-complete for every fixed $n\ge 13$.

\subsection{Our Results}\label{sec:intro-results}

We nearly settle the query and computational complexity of contiguous $\eps$-envy-free cake cutting for a fixed number of agents.

\paragraph{Query complexity.} We show that there is a stark change in query complexity between five and seven agents. 

\medskip

\noindent\textit{Five agents.} 
For five agents with arbitrary nonatomic additive valuations, we give a
deterministic algorithm using $O(\log^2(1/\eps))$ RW queries
(Theorem~\ref{five:main}). 

Our approach also extends to monotone valuations
with a known Lipschitz bound $L$, yielding an algorithm that uses only
$O(\log^3(L/\eps))$ evaluation queries
(Theorem~\ref{extension:monotone-five}). This answers the question posed by
Hollender and Rubinstein~\cite[Section~1]{HR}.

\medskip

\noindent\textit{Seven or more agents.} We show that in settings with seven or more agents, there are 
enough interacting envy constraints to encode a long hidden path, and finding an approximately envy-free allocation requires locating that path’s endpoint. However, each Robertson–Webb query reveals only a small amount of local information about the path, forcing deterministic protocols to make $\Omega(\eps^{-1/3})$ RW queries in the worst case (Theorem~\ref{seven:main}). 

This lower bound holds even when every valuation has a piecewise-constant density bounded above and away from zero by constants depending only on $n$. Hence the lower bound also applies to monotone valuations with a known Lipschitz bound. 
The same construction gives an $\Omega_n(\eps^{-1/3})$ lower bound on
the expected number of queries of randomized protocols that succeed
with probability at least $2/3$ on every input
(Proposition~\ref{seven:randomized}).
\Cref{tab:oracle} summarises the query bounds for additive valuations. 

\begin{table}[ht]
\centering
\caption{Robertson--Webb query complexity of finding a
contiguous $\eps$-envy-free allocation for additive valuations.
The number of agents $n$ is fixed, and hidden constants may depend on $n$.
Our results are displayed in bold.}
\label{tab:oracle}
\medskip
\small
\renewcommand{\arraystretch}{1.2}
\setlength{\tabcolsep}{7pt}
\begin{tabular}{@{}cccl@{}}
\toprule
Agents $n$ & Upper bound & Lower bound & Complexity regime \\
\midrule
$2$
    & $2$ queries, exact
    & ---
    & Constant \\
\addlinespace
$3$
    & $O(\log(1/\eps))$
    & $\Omega(\log(1/\eps))$
    & Logarithmic, tight \\
\addlinespace
$4$
    & $\boldsymbol{O(\log(1/\eps))}$
    & $\Omega(\log(1/\eps))$
    & Logarithmic, tight \\
    & {\footnotesize (Theorem~\ref{four:main})} & & \\
\addlinespace
$5$
    & $\boldsymbol{O(\log^2(1/\eps))}$
    & $\Omega(\log(1/\eps))$
    & Polylogarithmic \\
    & {\footnotesize (Theorem~\ref{five:main})} & & \\
\addlinespace
$6$
    & $O(1/\eps)$
    & $\Omega(\log(1/\eps))$
    & \emph{Open} \\
\addlinespace
$n\ge7$
    & $O(1/\eps)$
    & $\boldsymbol{\Omega(\eps^{-1/3})}$
    & Polynomial \\
    & & {\footnotesize (Theorem~\ref{seven:main})} & \\
\bottomrule
\end{tabular}

\medskip
\begin{minipage}{\linewidth}
\footnotesize
\textit{Previous results.}
All unbolded asymptotic bounds are from~\cite{BN}.
For four agents, the previous upper bound was
$O(\log^2(1/\eps))$~\cite{HR}.
The exact two-agent allocation is obtained by cut and choose.
\end{minipage}
\end{table}

Br\^anzei and Nisan's three-agent bounds and logarithmic lower bound~\cite{BN},
together with our upper bounds for four and five agents and our lower bound
from seven agents, give the following classification. Only the six-agent case remains open.

\begin{theorem}[Query complexity, informal]
For every fixed number of agents $n\ge3$ with nonatomic additive valuations,
the deterministic Robertson--Webb query complexity of finding a contiguous
$\eps$-envy-free allocation satisfies the following:
\begin{itemize}
    \item If $3\le n\le5$, the query complexity is
    $\bigl(\log(1/\eps)\bigr)^{\Theta(1)}$.
    \item If $n\ge7$, the query complexity is
    $(1/\eps)^{\Theta(1)}$.
\end{itemize}
For five agents, a polylogarithmic query upper bound also holds for monotone
valuations with a known Lipschitz bound.
\end{theorem}

\paragraph{Computational complexity.} Our query lower bound also gives a running-time lower bound for deterministic
algorithms that access valuations only through RW queries. It does not,
however, directly establish computational hardness when valuations are given
by polynomial-time evaluation procedures: an algorithm can inspect their
descriptions and exploit information unavailable through oracle access. We
therefore study this setting separately, following Hollender and
Rubinstein~\cite[Section~3]{HR}. Running time is measured in the total
input length, including a Lipschitz bound and the binary tolerance $\eps$.

\medskip

\noindent\textit{Five agents.} 
We show that the key ideas in the query-efficient algorithm can be used to compute a contiguous $\eps$-envy-free allocation in polynomial time for five agents with monotone Lipschitz valuations given by polynomial-time evaluation procedures (Theorem~\ref{circuit:five}).

\medskip

\noindent\textit{Seven or more agents.} We show that in settings with seven or more agents, 
the problem of computing a contiguous $\eps$-envy-free allocation is UEOPL-hard for every fixed $n\ge7$ (Theorem~\ref{ueopl:hardness}). 
For every fixed $n\ge13$, we strengthen
this result to PPAD-completeness (Theorem~\ref{circuit:ppad13}).

Together with the previously known computational complexity bounds for $n=3$ and $n=4$, our results
yield the following classification, leaving the six-agent case unresolved.

\begin{theorem}[Computational complexity, informal]
Fix the number of agents $n\ge3$. Valuations are monotone and
Lipschitz, represented by polynomial-time evaluation procedures. A Lipschitz
bound and the rational tolerance $\eps>0$ are given in binary.
\begin{itemize}
    \item If $3\le n\le5$, a contiguous $\eps$-envy-free allocation can
    be computed in polynomial time.
    \item If $n\ge7$, finding such an allocation is UEOPL-hard, even
    for additive valuations.
    \item If $n\ge13$, finding such an allocation is PPAD-complete, even
    for additive valuations.
\end{itemize}
\end{theorem}

Although UEOPL-hardness is weaker than PPAD-hardness, it already provides
substantial evidence against polynomial-time solvability. The class UEOPL
contains several longstanding computational problems, including the P-matrix
linear complementarity problem and simple stochastic games~\cite{FGMS}.
A polynomial-time algorithm for any UEOPL-hard problem would therefore yield
polynomial-time algorithms for all of these problems. Thus, an efficient
algorithm for contiguous envy-free cake cutting with seven agents would
resolve major open questions well beyond fair division.

The path-following structure of UEOPL helps explain why uniqueness alone need
not make a problem easy. On instances satisfying the uniqueness promise, the
canonical problem asks for the endpoint of a unique path represented by
circuits, along which a potential strictly increases~\cite{FGMS}. Both the
next step and the potential can be computed efficiently, but the path may
contain exponentially many vertices. Uniqueness removes ambiguity about which
path to follow, and the increasing potential guarantees termination; neither
guarantees rapid progress. Moreover, evaluating the potential at a given vertex
does not tell us how to find a vertex with a prescribed potential, so it does
not by itself enable binary search.

Further evidence comes from cryptography. Bitansky, Paneth, and Rosen
establish hardness of Sink-of-Verifiable-Line under suitable subexponential
security assumptions on indistinguishability obfuscation and injective one-way
functions~\cite{BPR}. Hub\'a\v{c}ek and Yogev reduce Sink-of-Verifiable-Line
to End-of-Metered-Line~\cite{HY}. As Fearnley, Gordon, Mehta and
Savani observe~\cite{FGMS}, the reduction produces a unique line on these
promised instances. The conditional hardness therefore carries over to
UEOPL and hence to our cake-cutting problem. Under these assumptions, access to the valuation
descriptions does not suffice to guarantee an efficient algorithm.

\medskip

\noindent\textbf{Organisation.}
Section~\ref{sec:models} fixes the valuation, query and computational models.
Section~\ref{five:section} develops the common five-agent construction and
its query and polynomial-time implementations. Section~\ref{seven:section}
proves the query lower bound for seven or more agents.
Section~\ref{circuit:section} establishes UEOPL hardness from seven agents
and PPAD completeness from thirteen.
The appendices give the four-agent algorithm, the extension to monotone
valuations, and the relations between query models and valuation classes.

\subsection{Technical overview}\label{sec:intro-tech}

\paragraph{Five agents: one search, two implementations.}
\label{sec:intro-algorithms}
The query and computational upper bounds use the same family of
partitions and the same planar search. The main geometric step is to
make two pieces assignable before searching for owners of the other
three. Write the five consecutive pieces as $L,P,U,M,R$. We call
$P,U$ \emph{private} because their owners are selected at each
partition; $L,M,R$ are \emph{public} because their ownership remains
to be resolved. These names describe their roles in the algorithm,
not restrictions on who may receive them.

For each agent, compare a private piece with her favourite public
piece and mark the cut where their values agree. Equalising suitable
order statistics of these marks produces a two-parameter family of
partitions. At every partition in this family, two distinct agents can
receive the private pieces as overall favourites, and each remaining
agent has a public piece as an overall favourite. Envy-freeness now
amounts to assigning distinct public pieces to the three remaining
agents.

Their excess demand for the three public pieces is a vector whose
coordinates sum to zero, so it lies in a plane. The boundary conditions
force a nonzero winding number. Repeatedly bisecting the parameter
square preserves this certificate until a small cell is found; Hall's
matching theorem then gives an assignment with small envy. The
important algorithmic property is that on each nondegenerate horizontal row, each
agent's values for $L,M,R$ respectively increase, decrease and stay
constant. Only constantly many preference changes need to be located
to compute the row's contribution to the winding number.

The implementations differ in how they evaluate this common search.
With Robertson--Webb access, we construct a nearby auxiliary profile
from lazily queried quantiles. Its relevant roots and preference changes
can be found exactly with logarithmically many queries, even when the
original cumulative functions have flat intervals. With polynomial-time
evaluation procedures and a Lipschitz bound, we instead interpolate
rounded values on a dyadic grid and add a small positive tilt. The
resulting auxiliary valuations support the same search, with polynomially
bounded bit lengths throughout. This implementation also covers
nonadditive monotone valuations. Section~\ref{five:section} develops the
shared argument and its computational consequence;
Appendix~\ref{extension:monotone-section} gives the monotone interpolation
and its error analysis.

\paragraph{Seven agents: hiding a path in envy comparisons.}
\label{sec:intro-lower}
For a query lower bound, it is not enough to encode a difficult
configuration: \emph{every} approximately envy-free allocation must
reveal it, and even an exact cut query must reveal little information.
We start with a hidden path whose last vertex requires many local
probes to find. Two polygonal curves encode the path so that every
near-meeting point reveals this vertex. Seven additive valuations then
encode the two coordinates of the curves in the envy inequalities.
Summing these inequalities controls the cuts independently of the
assignment of pieces to agents. Consequently every valid allocation
decodes to a near-meeting point.

The encoding also has a local inverse: an exact evaluation or cut query
can be answered with only constantly many probes to the hidden path.
This is essential, since an exact real answer could otherwise disclose
far more than one bit. Relating the length of the hidden path to the
required envy tolerance yields the polynomial inverse-precision lower
bound. Additional agents can be incorporated without losing either
the decoding property or the local query simulation.

\paragraph{Computational hardness: succinct constructions.}
\label{sec:intro-hardness}
For computational hardness, locality of oracle answers is replaced
by a succinct, polynomial-time evaluable description. Our reductions
produce rational piecewise-linear cumulative circuits, a restricted
class of the permitted descriptions. For seven agents we start
from the Jordan Curve problem, which Hollender, Maystre and Risse proved
UEOPL-hard~\cite[Theorem~4.1]{HMR}. We encode the two grid walks in
seven additive valuations. From any sufficiently accurate envy-free
allocation we can recover a grid vertex visited by both walks.

For PPAD hardness, a separate thirteen-agent construction encodes a
two-dimensional Brouwer problem. Of three table evaluations, at least two
interpolate directions from the grid square containing the encoded point.
Two feedback equations translate their combined directions into envy comparisons. Both reductions must work
for every ownership order. The standard approximate PPAD-membership result completes
the classification, giving PPAD completeness from thirteen agents. Section~\ref{circuit:section} presents these
computational reductions together.

\subsection{Formal verification}\label{sec:formal}
The accompanying Lean~4 development~\cite{Lean4}, using
Mathlib~\cite{Mathlib}, verifies the query theorems
\ref{five:main}, \ref{extension:monotone-five}, \ref{seven:main}
and~\ref{four:main}. It also verifies the query-model transfers in
Propositions~\ref{prop:rw-transfer}, \ref{prop:value-transfer},
\ref{prop:five-restricted} and~\ref{prop:monotone-to-additive}
and Corollary~\ref{cor:cut-only}, including the additive-source
strengthening of Proposition~\ref{prop:monotone-to-additive}.
The randomized extension in Proposition~\ref{seven:randomized} is not
included in this formalisation.

For computational complexity, the development verifies the
piecewise-linear circuit constructions and decoding arguments underlying
Theorems~\ref{ueopl:hardness} and~\ref{circuit:ppad13}, together with
PPAD membership for that circuit class. The extension to the
polynomial-time evaluation model of Definition~\ref{circuit:problem},
including the bit-complexity argument in Theorem~\ref{circuit:five}
and the broader membership statement in
Proposition~\ref{circuit:membership}, is not included in the formal
verification. Thus the computational statements in the broader model
are established by the proofs in this paper, not by the current Lean
development.

The formal computational statements are for each fixed population.
Density bounds are expressed as two-sided slope bounds on cumulative
functions, and PPAD and UEOPL use explicit encodings of End-of-Line and
the layered potential-line problem. For seven agents the Lean proof
constructs the crossing curves explicitly. Section~\ref{ueopl:section}
invokes the Jordan Curve hardness theorem of Hollender, Maystre and
Risse~\cite[Theorem~4.1]{HMR} for this step. The rescaling observation after
Theorem~\ref{seven:main} and the asymptotic consequences following
Proposition~\ref{prop:monotone-to-additive} also remain outside the
verified statements.

All checked proofs use only Lean's standard axioms. The development
includes a proof of Brouwer's fixed-point theorem from
\texttt{harfe/fixed-point-theorems-lean4}, unchanged except for import
paths;\footnote{\url{https://github.com/harfe/fixed-point-theorems-lean4}.}
it supplies the existence argument used in the formal development.
The Lean development, including the proof sources, statement-to-proof correspondence and verification receipts, is available at \url{https://github.com/SimonWMackenzie/connected-envy-free-lean}. An interactive inspector of the formal development is available at \url{https://simonwmackenzie.github.io/connected-envy-free-lean/inspector/}.

\subsection{Related work}\label{sec:related}
\paragraph{Existence and precision.}
Stromquist and Woodall established connected envy-free existence in 1980~\cite{Stromquist,Woodall}. Woodall's proof combines a fixed-point argument with Hall's matching theorem. Su presented the ownership-labelled Sperner construction due to Simmons, giving another topological route to a connected allocation~\cite{Su}. Of the query bounds of Br\^anzei and Nisan recalled above~\cite{BN}, the three-agent upper bound discretely simulates the moving-knife procedure of Barbanel and Brams~\cite[Section~2]{BB}, and the logarithmic lower bound also holds in the model used here, with unrestricted evaluation points and output cuts~\cite[Section~8, Corollary~1]{BN}.
Br\^anzei and Nisan formulate their upper bounds for valuations with
densities. They extend to all nonatomic additive valuations by the quantile
interpolation in Appendix~\ref{four:section}. Running their protocol at
tolerance $\eps/2$ on the interpolants costs a constant number of original
queries per simulated query. With $K=\lceil8/\eps\rceil$ quantile
intervals, transferring the allocation back adds at most $4/K$ to the envy.

\paragraph{Algorithms for small populations.}
Deng, Qi and Saberi gave a squared-logarithmic precision algorithm for three agents with monotone Lipschitz valuations that are positive on every interval of positive length~\cite[preprint, Section~6]{DQS}. Hollender and Rubinstein obtained an $O(\log^3(1/\eps))$ value-query bound for four monotone $1$-Lipschitz valuations and an $O(\log^2(1/\eps))$ Robertson--Webb bound for four additive agents, without a supplied density bound~\cite[Theorems~5.1 and~6.1]{HR}. Our four-agent bound in Appendix~\ref{four:section} uses the same model as their additive bound.

Saberi and Wang~\cite{SW} gave a moving-knife procedure for five agents in which a share may consist of several intervals. They also described how to discretise the procedure with polylogarithmic dependence on the precision under a uniform upper bound on the value densities.

Several preprocessing tools in our algorithms come from this earlier work. Equal-value discretisation underlies Br\^anzei and Nisan's general bound. Hollender and Rubinstein develop lazy quantile interpolation, cache the first answer to a nonunique cut query, and localise a search within affine portions of the interpolant~\cite[Lemmas~6.2 and~6.3]{HR}. Our protocols for additive valuations (Theorems~\ref{five:main} and~\ref{four:main}) use these ideas. For monotone valuations that need not be additive (Theorem~\ref{extension:monotone-five}), we use their combination of a strict monotonicity perturbation and triangular piecewise-linear interpolation~\cite[Lemma~5.2]{HR}, with rational rounding and accounting for the five-agent search.

\paragraph{Topological search and matching.}
Our planar search builds on several more specific precedents. Deng, Qi and Saberi locate monotone changes of favourite pieces by binary search, compute the resulting boundary index, and retain a subregion with nonzero index~\cite[preprint, Section~6, Property~3]{DQS}. Igarashi and Meunier combine interpolated favourite-piece counts, planar bisection and a matching construction in their algorithm for dividing one cake among three groups of prescribed sizes~\cite[Theorem~4 and Sections~4.3--4.4]{IM}. Their approximation controls envy in value for monotone Lipschitz valuations. Hollender, Lawrence and Segal-Halevi recast and extend the preference-count approach through approximate root finding, using Hall's theorem to recover the groups~\cite[Section~5]{HLS}; their near-envy guarantee is formulated in terms of small changes to cut positions. In these three-group applications, members of a group share one of three pieces. Our five-agent search must allocate five distinct pieces, so it keeps two pieces assignable throughout and balances demand for the other three.

\paragraph{Approximation with a variable population.}
A complementary line of work charges computation as well as queries and seeks polynomial-time guarantees for arbitrary populations. Under a restriction on the densities,
Barman and Rathi obtain a complete connected $\eps$-envy-free allocation
in time polynomial in $n$, $\log\lambda$ and $\log(1/\eps)$ in the
Robertson--Webb model~\cite{BR}. Their positive densities $f_i$ admit an
agent order in which $f_j/f_i$ is nondecreasing for every $i<j$, and
$\lambda$ bounds the Lipschitz constants of cut and evaluation queries.
Goldberg, Hollender and Suksompong gave a connected allocation with additive envy at most $1/3$~\cite{GHS}. Barman and Kulkarni improved this guarantee to $1/4+o(1)$ and also obtained an approximation scheme when the number of distinct valuations is bounded~\cite{BK}. Goldberg, Hollender and Suksompong also proved NP-hardness when an envy-free allocation must satisfy prescribed ownership or cut constraints~\cite[Theorem~4.1]{GHS}. Those constraints can destroy existence; the unconstrained total search problem here allows every ownership order.

\paragraph{Computational hardness and additive valuations.}
Earlier hardness results show that general preferences can encode difficult search problems. Deng, Qi and Saberi proved hardness for preferences that can depend on the entire division~\cite{DQS}. Gao, Roghani, Rubinstein and Saberi obtained stronger hardness under substantial relaxations of the requested fairness in that broader preference model~\cite{GRRS}. For values depending only on the interval being valued, Hollender and Rubinstein proved PPAD-completeness of approximate connected division with four identical non-monotone valuations~\cite[Theorem~8.3]{HR}. These preference models lie outside both of our classes: the first two let values depend on the whole division, and the last is not monotone.

Kawase, Roy and Sanpui proved PPAD-completeness for both approximate and exact connected envy-free division with additive valuations given by explicit rational piecewise-constant densities~\cite{KRS}. Their population is part of the input. Approximate hardness holds even for unary accuracy, with positive densities taking two common values and having at most six breakpoints per agent; it therefore also implies binary-accuracy hardness. Our hardness results differ in fixing the population and using succinct cumulative circuits, not in the encoding of the accuracy.

Our additive hardness constructions build on the construction of Kawase, Roy and Sanpui~\cite[Section~3, Lemmas~6--9]{KRS}. They use a physical ordering of regions and a separate cyclic comparison order, give each agent two preferred regions, and sum envy comparisons to control the pieces' prices independently of ownership. Subtracting an average from sums before a cut then turns the price relations into equations for cut positions. Our weighted constructions develop these devices for fixed populations and signed coefficients.

\paragraph{Crossing curves and total search.}
Adler, Daskalakis and Demaine introduced the crossing-curves search problem and proved its membership in PPAD~\cite{ADD}. Hollender, Maystre and Risse proved that the Jordan Curve problem is UEOPL-hard~\cite[Theorem~4.1]{HMR}. They used it to obtain UEOPL-hardness and a polynomial inverse-precision value-query lower bound for three identical non-monotone valuations~\cite[Theorem~3.1]{HMR}. For the Robertson--Webb query lower bound we adapt their routing through layered checkpoints and show how to simulate cut queries. For computational hardness we apply their Theorem~4.1 directly and encode the two walks using seven additive valuations (Section~\ref{ueopl:section}). Our thirteen-agent PPAD-hardness reduction instead starts from Chen and Deng's two-dimensional discrete Brouwer problem~\cite{CD}.

\paragraph{Explicit and succinct inputs.}
Alijani, Farhadi, Ghodsi, Seddighin and Tajik gave an exact connected
envy-free algorithm for every fixed number of agents with explicitly
listed piecewise-constant densities: enumerate the cut cells and
ownership orders and solve the resulting linear programs~\cite[Theorem~5]{AFGST}.
Our hardness constructions instead give succinct descriptions, which
can represent exponentially many breakpoints. The computational model
allows polynomial-time evaluation procedures, as in Hollender and
Rubinstein~\cite[Section~3]{HR} and Hollender, Maystre and
Risse~\cite[Section~2.1]{HMR}; rational piecewise-linear cumulative
circuits are a restricted class sufficient for both of our hardness
results. Membership of approximate cake cutting in PPAD follows from
the standard discretisation argument~\cite{DQS,HR}. For the distinct
problem of exact division, Filos-Ratsikas, Hansen, H\o gh and Hollender
prove PPAD membership for hungry agents with piece utilities represented
by piecewise-linear pseudo-circuits~\cite[Section~8.1, Theorem~8.2]{FRHH}.

\section{Model and preliminaries}\label{sec:models}

\subsection{Valuations and allocations}
There are $n$ agents, indexed by $[n]=\{1,\ldots,n\}$, and a cake represented by $[0,1]$. Agent $i$ assigns a nonnegative value $v_i(S)$ to each measurable piece $S$. A valuation is \emph{additive} if the value of a union of disjoint pieces is the sum of their values. That is, for any $k\ge1$ and disjoint measurable pieces $S_1,\ldots,S_k \subseteq [0,1]$, the valuation satisfies
\(
 v_i\!\left(\bigcup_{j=1}^k S_j\right)=\sum_{j=1}^k v_i(S_j).
\)
Unless stated otherwise, valuations are nonatomic and additive measures. We normalise these valuations so that $v_i([0,1])=1$, and write $v_i(a,b)=v_i([a,b])$ for $a\le b$. 

Every normalized, nonatomic and additive valuation can be represented by its \emph{cumulative value
function} $F_i:[0,1]\to[0,1]$, which records the value of the cake to the
left of a given point:
\[
    F_i(t):=v_i([0,t])
    \qquad\text{for all }t\in[0,1].
\]
The function $F_i$ is nondecreasing and continuous, with $F_i(0)=0$
and $F_i(1)=1$. It determines the value of every interval through
\[
    v_i([a,b])=F_i(b)-F_i(a)
    \qquad\text{for all }0\le a\le b\le1.
\]
We will use the cumulative functions $F_i$ as a convenient representation
of the valuations $v_i$. 
A valuation is \emph{hungry} if every interval of positive length has positive value, equivalently if $F_i$ is strictly increasing.

Many of our results hold for a more general class of valuations that need not be additive. In this case, $v_i(a,b)\in[0,1]$ is a continuous function of the endpoints, extended by zero when $b\le a$. The value of the whole cake need not equal one; envy is measured on this absolute scale. We call a valuation \emph{monotone}, that is,
inclusion-monotone: $v_i(a,b)\le v_i(a',b')$ whenever
$[a,b]\subseteq[a',b']$, so enlarging an interval cannot lower its
value. A valuation is \emph{$L$-Lipschitz} if, for all endpoints,
\[
 |v_i(a,b)-v_i(a',b')|\le L(|a-a'|+|b-b'|).
\]

A \emph{complete connected (contiguous) allocation} consists of cuts $0=y_0\le y_1\le\cdots\le y_n=1$ and a bijection assigning intervals to agents. Empty pieces are permitted. Writing $A_i$ for agent $i$'s piece, the allocation is \emph{$\eps$-envy-free} if
\[
 v_i(A_i)\ge v_i(A_j)-\eps\qquad\text{for all }i,j\in[n].
\]
When $\eps=0$, the allocation is \emph{exactly envy-free}.

\subsection{Robertson--Webb queries}\label{sec:rw-model}
A protocol accesses valuations through
\begin{itemize}
\item $\operatorname{Eval}_i(a,b)$, which returns $v_i(a,b)$;
\item $\operatorname{Cut}_i(a,t)$, which returns $b\in[a,1]$ with
      $v_i(a,b)=t$, or reports that no such $b$ exists.
\end{itemize}
Parameters and answers are exact real numbers. We allow arbitrary query
coordinates and arbitrary output cuts. Any valid answer to a nonunique
cut query is permitted. For additive valuations, \emph{prefix queries}
have left endpoint zero. They simulate general queries at constant cost:
\[
 v_i(a,b)=F_i(b)-F_i(a),\qquad
 \operatorname{Cut}_i(a,t)
 =\operatorname{Cut}_i(0,F_i(a)+t)
\]
whenever the requested level is feasible, with a valid answer chosen in
$[a,1]$. At target $t=0$ one may simply return $a$.

The literature uses several Robertson--Webb conventions. The model above
is often called $\mathrm{RW}^{+}$; in \emph{restricted RW}, evaluations,
cut-query starting points and output cuts must be endpoints or previously
returned cut points. For each fixed population, the existence of a
uniform deterministic polylogarithmic query bound for additive
$\eps$-envy-free division is the same in these two models and with
prefix cut queries alone. The simulations may introduce a logarithmic
factor; they need not preserve the precise exponent. With value queries
alone, a known Lipschitz bound is required. Appendix~\ref{sec:other-models}
states these qualifications and proves the transfers. Our five-agent
additive protocol already works in restricted RW, while our lower bound
holds in the more permissive model above.

In query complexity only oracle calls are charged; arithmetic and other
computation between them are free. Bounds are uniform over the indicated
valuation class. Unless stated otherwise, $n$ is fixed, constants may
depend on $n$, and $0<\eps\le1$. For nonadditive monotone
valuations, our upper bounds use only evaluation queries and a supplied
Lipschitz bound. In complexity bounds, logarithmic factors are
understood to be at least one.

\subsection{Valuation descriptions and computation}\label{sec:circuit-model}
For computational complexity, the valuations are given by finite
descriptions rather than accessed only through queries. We follow the
evaluation model of Hollender and Rubinstein~\cite[Section~3]{HR}.
Fix a representation with a uniform algorithm that, given a valuation
description and rational endpoints $a,b$, returns the exact rational
value $v(a,b)$ in time polynomial in their combined encoding length.
Polynomial-time Turing-machine programs with a fixed polynomial clock
are one such representation. The polynomial evaluation bound is fixed
as part of the representation, not an arbitrary exponent in the input.
Our algorithms use only this evaluation guarantee; an algorithm for the
problem may also inspect the descriptions.

\begin{definition}[Computational connected envy-free division]
\label{circuit:problem}
Fix $n\ge2$. An instance of $\eps$-$\ConnEF_n$ consists of descriptions
of $n$ monotone valuations $v_i:[0,1]^2\to[0,1]$, a common rational
Lipschitz bound $L\ge1$, and a rational tolerance $0<\eps\le1$.
The valuations are promised to be $L$-Lipschitz and to satisfy
$v_i(a,b)=0$ for $b\le a$; they need not be additive or strictly
monotone. A solution is a complete connected $\eps$-envy-free
allocation, specified by rational cuts and an assignment $\pi$.
\end{definition}

All rationals are encoded as binary numerator--denominator pairs.
All bit operations are charged, and running time is measured in the
total input length, including the descriptions, $L$ and $\eps$.
Thus polynomial time in $1/\eps$ would not suffice. A supplied $L$
may be replaced by the next power of two without changing the
polynomial bounds. Whole-cake values need not equal one in this
model. One can instead scale values and $\eps$ by $L$ to obtain the
one-Lipschitz convention of~\cite{HR}.

Our hardness constructions use the following restricted descriptions.
A \emph{rational piecewise-linear circuit} has one real input, one
output, and gates for rational constants, addition, subtraction,
multiplication by rational constants, minimum and maximum. This is the gate
set used by Etessami and Yannakakis for linear fixed-point circuits~\cite{EY}. The circuit size
counts the gates and the bit lengths of their data; it may describe
exponentially many affine pieces. A nondecreasing circuit $F_i$ with
$F_i(0)=0$ and $F_i(1)=1$ defines an additive valuation by
$v_i(a,b)=F_i(b)-F_i(a)$ for $a\le b$, and zero otherwise.
Such circuits have polynomial-time rational evaluators and a
computable Lipschitz bound of polynomial bit length, as shown in
Section~\ref{circuit:section}. They therefore provide valid inputs to
the computational problem. The hardness results apply to the
program-based representation and to any representation admitting
these circuit evaluators with polynomial overhead.

\subsection{Relations between valuation classes}\label{sec:valuation-classes}
Every additive valuation is monotone. Thus an upper bound for monotone
$L$-Lipschitz valuations applies to additive valuations with densities
at most $L$, and a lower bound using such additive valuations applies
to the larger monotone class. Our seven-agent construction has densities
in $[\ell_n,u_n]$, so its lower bound holds even when the bound $u_n$
is supplied.

There is also a transfer when the original additive valuations have no
density bound. Replacing spatial coordinates by the merged order of the
agents' quantiles gives nearby additive $n$-Lipschitz valuations whose
values can be queried with logarithmic overhead
(Proposition~\ref{prop:monotone-to-additive}). A polylogarithmic
value-query algorithm for monotone valuations therefore gives one for
arbitrary additive valuations using both cuts and evaluations. This
preserves the distinction between polylogarithmic and polynomial
precision dependence, but not necessarily the best exponent. In
particular, the direct five-agent additive algorithm is sharper than
this transfer. Appendix~\ref{sec:other-models} gives the bounds and
explains why neither removing cut queries without a Lipschitz bound nor
extending an additive algorithm to nonadditive valuations is automatic.

\section{Five agents}\label{five:section}

One geometric construction underlies both the query-efficient algorithm
and the polynomial-time algorithm for valuation descriptions. We first arrange
two adjacent pieces so that they can be assigned to two distinct agents
without envy. This leaves three agents competing for the other three
pieces. A planar search balances those competing demands. The two
algorithms differ in how they evaluate the partitions examined by this
search: through a small number of quantile queries, or by rational
interpolation of polynomial-time computable interval values.

\begin{theorem}\label{five:main}
For five agents with additive valuations and every $0<\eps\leq1$, a
deterministic Robertson--Webb protocol finds a complete connected
$\eps$-envy-free allocation using $O(\log^2(1/\eps))$ queries.
\end{theorem}

\begin{theorem}\label{circuit:five}
There is a deterministic polynomial-time algorithm for
$\eps$-$\ConnEF_5$ (Definition~\ref{circuit:problem}). Thus five agents
with monotone Lipschitz valuations given by polynomial-time evaluation
procedures admit a complete connected $\eps$-envy-free allocation in
time polynomial in the total input length.
\end{theorem}

The same implementation also gives the following value-query bound.

\begin{theorem}\label{extension:monotone-five}
Given $L\ge1$, for five monotone $L$-Lipschitz valuations, which need
not be additive, and every $0<\eps<1$, a deterministic value-query
protocol finds a complete connected $\eps$-envy-free allocation using
$O(\log^3(L/\eps))$ queries. All queries have dyadic endpoints.
The valuations need not be strictly monotone: enlarging an interval may
leave its value unchanged.
\end{theorem}

The additive query bound requires no density or Lipschitz bound. The
monotone bound uses only evaluation queries, but requires a supplied
Lipschitz bound; neither query theorem therefore subsumes the other.
The transfer in Proposition~\ref{prop:monotone-to-additive} would give
$O(\log^4(1/\eps))$ queries for additive valuations, so it does not
recover our sharper bound.

Sections~\ref{five:abstract-section} and~\ref{five:chart-section}
contain the common argument: a search principle and a two-parameter
family of calibrated partitions to which it applies.
Sections~\ref{five:quantiles-section}--\ref{five:protocol-section}
implement it in the query model. The essential saving is to locate all
preference changes on an inspected row together, rather than repeatedly
evaluate partitions during separate binary searches.
Appendix~\ref{extension:monotone-section} gives the monotone-valuation
implementation. Section~\ref{circuit:finite-precision} bounds its bit
complexity, proving the computational theorem.

\subsection{A search principle for three competing pieces}\label{five:abstract-section}

Each vertex $v$ of $\{0,\ldots,H\}^2$, where $H=2^\lambda\geq2$,
carries a matrix $S_v\in\R^{5\times5}$ of agents' scores for five
pieces $L,P,U,M,R$; its entry $S_v[a,k]$, also written $S_{a,k}$, is
agent $a$'s score for $k$. Rows and columns of the grid are indexed
by $0,\ldots,H$ from the bottom and from the left.
Call $P,U$ \emph{private}: we will assign their owners at
each vertex. Call $L,M,R$ \emph{public}: their prospective owners
remain to be balanced.

At a vertex, a \emph{public favourite} of an agent is a label that
maximises her score among $L,M,R$; a \emph{global favourite}
maximises it among all five labels. We assume the following
properties.
\begin{enumerate}[label=(\roman*)]
\item\label{five:h-assignment}
At every vertex there are distinct agents who can receive $P$ and $U$
as global favourites. Each of the three remaining agents has a public
piece that is a global favourite. An assignment of these two private
owners and one such public favourite to each remaining agent is called
\emph{legal}. Public pieces may have several prospective owners.
\item\label{five:h-boundary}
On the left boundary $L$ is nobody's public favourite; on the right
boundary $M$ is nobody's public favourite; on the top boundary $R$ is
nobody's public favourite. On the bottom boundary neither $L$ nor $M$
is anybody's public favourite.
\item\label{five:h-rows}
On every row other than the bottom row, as the column increases, each
agent's score for $L$ strictly increases, her score for $M$ strictly
decreases, and her score for $R$ stays constant.
\end{enumerate}
Fix rules depending only on the score matrix for choosing a legal
assignment and each agent's public favourite. The latter choice is her
\emph{vote}. The rules may resolve ties in any fixed way.
Let the \emph{load} $D_k(v)$ count the agents sent to public label $k$
by the legal assignment, and let $N_k(v)$ count the votes for $k$
among all five agents. Thus $\sum_kD_k=3$ and $\sum_kN_k=5$.

The target load for a public piece depends slightly on the votes:
\begin{equation}\label{five:shift}
 C_k(N)=1+\frac{5/3-N_k}{100},\qquad
 T(v)=D(v)-C(N(v)).
\end{equation}
Thus $T(v)$ measures how far the loads at $v$ are from their targets.
Both vectors $T(v)$ and the origin belong to the plane
$\Pi=\{z\in\R^3:z_L+z_M+z_R=0\}$. The small correction will ensure
that a piece with at most one vote stays strictly to one side of the
target. We take winding numbers about a nearby generic rational point
$q\in\Pi$, with $|q_k|<1/1000$, rather than the origin, to avoid
degeneracies.

\begin{lemma}\label{five:generic}
A rational point $q$ with $|q_k|<1/1000$ can be fixed, independently of
the profile and the grid, such that no segment between two possible
values of $T$ contains $q$.
\end{lemma}
\begin{proof}
There are finitely many possible integer pairs $(D,N)$. In the first
two coordinates of $\Pi$, the parabola $(t,t^2)$ meets each line
through two distinct possible values of $T$ at most twice. Hence we
can choose a positive rational $t<10^{-4}$ avoiding these lines and
the finitely many possible values themselves. The point
$q=(t,t^2,-t-t^2)$ has the required properties and can be found by
enumerating rational candidates, without valuation queries.
\end{proof}

For a grid rectangle $Q$, let $w(Q)$ be the winding number about $q$
of the polygon through its boundary vectors $T$, in counter-clockwise
order. Lemma~\ref{five:generic} makes this well defined. We use three
elementary facts: opposite common edges cancel; a polygon in a convex
set missing $q$ has winding zero; and a deformation avoiding $q$
preserves winding. In particular, a path in an open half-plane
missing $q$ can be replaced by its chord.

A \emph{vertex evaluation} computes $S_v$, hence the legal assignment,
the votes and $T(v)$. Following $T$ along a row needs few of them: on
a stretch where none of the fifteen \emph{public comparisons}
(differences between an agent's scores for two public pieces) vanishes
or changes sign, the votes are fixed and some public piece keeps at
most one vote, so $T$ stays in a half-plane missing $q$ whoever owns
the private pieces.

\begin{lemma}\label{five:abstract}\label{five:abstract-search}\label{five:square-search}
Under properties \ref{five:h-assignment}--\ref{five:h-rows}, the whole
grid has winding number $1$ or $-1$. A unit cell $Q$ with $w(Q)\ne0$
can be found using the following information:
\begin{enumerate}[label=(\alph*)]
\item at most $\lambda+1$ non-bottom rows, with the thirty integers on
each such row that give, for each agent and each of the increasing
differences $S_{a,L}-S_{a,M}$, $S_{a,L}-S_{a,R}$,
$S_{a,R}-S_{a,M}$, its first nonnegative and first positive column
index; an absent index is recorded as $H+1$;
\item at most $47(\lambda+1)+2\lambda$ vertex evaluations.
\end{enumerate}
Without supplied row indices, the same search uses $O(\log^2 H)$
vertex evaluations and reads at most $\lambda+1$ non-bottom rows.
Let $\omega(Q)$ be the maximum oscillation of any individual score over
the four corners of $Q$. From the four corner matrices one can find a
bijection $\pi$ of agents to piece labels such that, at every corner $v$,
\begin{equation}\label{five:corner-envy}
 S_v[a,\pi(a)]\geq S_v[a,k]-2\omega(Q)
 \qquad\text{for every agent $a$ and label $k$}.
\end{equation}
\end{lemma}
\begin{proof}
The proof bounds the targets, finds the winding of the grid,
compresses rows, bisects and extracts the bijection.
The target correction satisfies
\begin{equation}\label{five:target-bounds}
 \frac{29}{30}\leq C_k\leq\frac{61}{60}.
\end{equation}
Thus $D_k=0$ implies $T_k<q_k$, and
\begin{equation}\label{five:scarce}
 D_k\leq N_k\leq1\quad\Longrightarrow\quad
 T_k\leq N_k-1-\frac{5/3-N_k}{100}\leq-\frac1{150}<q_k.
\end{equation}

By property~\ref{five:h-boundary}, on the left, right, and top
boundaries respectively,
$D_L=N_L=0$, $D_M=N_M=0$, and $D_R=N_R=0$.
On the bottom, $D=(0,0,3)$ and $N=(0,0,5)$. Deform the boundary
vectors $T-q$ linearly to $D-(1,1,1)$. Each side stays in the open
half-plane with the corresponding coordinate negative; the bottom
stays in the negative-$L$ half-plane. No edge therefore crosses zero.
The resulting bottom is constant at $e_R=(-1,-1,2)$, while the right,
top, and left join successively
$e_R$, $e_L=(2,-1,-1)$, $e_M=(-1,2,-1)$, and $e_R$.
Replace these paths by their chords within the same half-planes.
The resulting triangle contains zero in its interior and winds once.

Next we compress a non-bottom row. By property~\ref{five:h-rows}, its
fifteen public comparisons are strictly monotone. Their crossing
locations, including zeros at vertices, split the row into at most
fifteen isolated \emph{tie vertices}, at which some public comparison
vanishes, and at most sixteen maximal consecutive \emph{runs} of tie-free
vertices with fixed comparison signs. On a run every public favourite
is unique, so $N$ is constant. Five votes leave some label $k$ with $N_k\leq1$.
Every agent legally assigned to $k$ votes for it, since $k$ is then
her unique public favourite, giving
$D_k\leq N_k$ even if the private owners change. By
\eqref{five:scarce}, the whole run lies in $T_k<q_k$ and may be
replaced by a chord. The supplied indices specify the runs and tie
vertices; their endpoints require at most $2\cdot16+15=47$
evaluations. Without supplied indices, thirty binary searches cost
$O(\log H)$ additional evaluations per row. The bottom vector is
constant and requires none.
Monotone preference thresholds also underlie the boundary compression
of Deng, Qi and Saberi~\cite[preprint, Section~6, Property~3]{DQS}; here the
half-plane argument accommodates changing private owners.

We now run the search. Bisect the whole grid horizontally, keeping a
half with nonzero winding. Such a half exists because opposite common
edges cancel. The kept half is a full-width \emph{slab}. A slab's
vertical sides lie on the left and right boundaries, in $T_L<q_L$ and
$T_M<q_M$ respectively, so replace them by chords and compute its
winding from its two compressed rows.
After $\lambda$ bisections its height is one, and only the top and
at most $\lambda$ intermediate rows have been read. Now bisect
vertically. Each new vertical side is a single edge; truncating the
known horizontal runs requires only the two new endpoints. Another
$\lambda$ steps yield a unit cell, using $2\lambda$ further
evaluations. This needs no monotonicity down columns or of the loads.

Finally we extract the bijection. We combine favourite assignments
from the corners into a fractional assignment, as in Igarashi and
Meunier~\cite[Lemma~4 and Section~4.4]{IM},
and use Hall's theorem on its support, as in Hollender, Lawrence and
Segal-Halevi~\cite[Section~5, proof of Property~2]{HLS}.
Nonzero winding implies that $q$ is in the
convex hull of the four corner vectors. Write
$q=\sum_r\alpha_rT(v_r)$, where $\alpha_r\geq0$ and
$\sum_r\alpha_r=1$, and average the four legal zero-one assignment
matrices with these weights to obtain $X$. Every row sums to one,
each private column sums to one, and each public column sums to
\[
 q_k+\sum_r\alpha_rC_k(N(v_r))<1.02.
\]
For any agent set $A$ and its neighbourhood $B$ in the support of $X$,
\[
 |A|\leq\sum_{k\in B}\sum_aX[a,k]<|B|+3(0.02)<|B|+1.
\]
Integrality gives Hall's condition. Hence the union of the four legal
assignments contains a perfect matching, found by trying all $120$
bijections without computing $\alpha_r$. Each matched piece is a
global favourite at some corner. Moving its score and a competing
score to another corner loses at most $2\omega(Q)$, proving
\eqref{five:corner-envy}.
\end{proof}

\subsection{A two-parameter family of partitions}\label{five:chart-section}

We next construct scores satisfying the search principle: each grid
vertex will carry a partition of the cake into the five
pieces~\eqref{five:piece-order}, and an agent's score for a piece is her
value of it under the cumulative functions introduced next. For this
common construction, let
$G_1,\ldots,G_5$ be continuous, strictly increasing cumulative functions
with $G_i(0)=0$, $G_i(1)=1$, extended by $G_i(z)=z$ outside $[0,1]$;
in the query implementation they will be the auxiliary functions of
Section~\ref{five:quantiles-section}. Appendix~\ref{extension:monotone-section}
gives the corresponding construction for nonadditive valuations, used
in the computational implementation.
Assume that for all $y<z$,
\begin{equation}\label{five:density-bounds}
 m_-(z-y)\leq G_i(z)-G_i(y)\leq m_+(z-y),
 \qquad 0<m_-\leq1\leq m_+,
 \qquad B=m_+/m_-.
\end{equation}
The extension permits inverses at levels outside $[0,1]$; it does not
add cake to the allocation. For cuts $0\leq x\leq c\leq b\leq d\leq1$
the physical order of labels will be
\begin{equation}\label{five:piece-order}
 L=[0,x],\quad P=[x,c],\quad U=[c,b],\quad
 M=[b,d],\quad R=[d,1].
\end{equation}
The constants $m_-,m_+$ bound changes in cumulative value; the capital
letters in \eqref{five:piece-order} are piece labels.

In their four-agent construction, Hollender and Rubinstein~\cite[Section~4]{HR}
use the median of three indifference marks to define their paths. To calibrate the two
private pieces here, we choose $c$ by equating two order statistics, one
for each piece. Given
$(x,b,d)\in[0,1]^3$, without imposing an order for the moment, define
\begin{align}
 f_i&=\max\{G_i(x),G_i(d)-G_i(b),1-G_i(d)\},\notag\\
 \ell_i&=G_i^{-1}(G_i(x)+f_i),\qquad
 u_i=G_i^{-1}(G_i(b)-f_i),\label{five:calibration}\\
 \mathcal A(x,b,d)&=\operatorname{secondmin}_i\ell_i
          -\operatorname{secondmax}_iu_i.\notag
\end{align}
For ordered cuts, $f_i$ is agent $i$'s largest public value. The
\emph{calibration mark} $\ell_i$ is where the left private piece,
growing from $x$, reaches that value, and $u_i$ is where the right
private piece, growing leftwards from $b$, reaches it. Equalising the
indicated order statistics, that is, taking a zero of the
\emph{calibration function} $\mathcal A$ with $c$ their common value, will give
each private piece at least two possible owners and at most one agent
who strictly requires it; we call such partitions \emph{calibrated}.

\begin{lemma}\label{five:calibration-rates}
On the whole cube $[0,1]^3$, whenever both sides are defined and $h>0$,
\begin{align*}
 \mathcal A(x+h,b,d)&\geq \mathcal A(x,b,d)+h/B,\\
 \mathcal A(x,b+h,d)&\leq \mathcal A(x,b,d)-h/B,\\
 \mathcal A(x,d+h,d+h)&\leq \mathcal A(x,d,d)-h/B.
\end{align*}
The function $\mathcal A$ is $6B$-Lipschitz in maximum norm. Moreover
$\mathcal A(0,0,0)=2$, $\mathcal A(0,1,1)=-1$, and $\mathcal A(x,x,d)>0$ whenever
$0\leq x\leq d\leq1$.
\end{lemma}
\begin{proof}
An inverse $G_i^{-1}$ changes at rates between $1/m_+$ and $1/m_-$.
Increasing $x$ by $h$ raises $G_i(x)$ by at least $m_-h$ and does not
lower $f_i$. Thus every $\ell_i$ rises by at least $h/B$, while no
$u_i$ rises. Increasing $b$ does not raise $f_i$; every $u_i$ rises
by at least $h/B$, while no $\ell_i$ rises. With $b=d$, the expression
for $f_i$ is $\max\{G_i(x),1-G_i(d)\}$, so increasing $b=d$ gives
the same latter conclusion. Order statistics preserve coordinatewise
lower bounds on increases. These arguments establish all three rates,
even when the maximising branch or the agent attaining an order
statistic changes.

Under a displacement of maximum size $\delta$, $f_i$ changes by at
most $2m_+\delta$. The arguments of both inverses change by at most
$3m_+\delta$, so each mark changes by at most $3B\delta$. The difference
of the two order statistics changes by at most $6B\delta$. At
$(0,0,0)$ the marks are $1,-1$, and at $(0,1,1)$ they are $0,1$.
When $b=x\leq d$, the three public values sum to one, so
$f_i\geq1/3$ and $\ell_i>x>u_i$ for every agent. This gives the last
assertion.
\end{proof}

\begin{lemma}\label{five:private-owners}
Suppose $0\leq x\leq b\leq d\leq1$ and $\mathcal A(x,b,d)=0$. Let $c$ be
the common order statistic in \eqref{five:calibration}. Then
$x<c<b$. There is a legal assignment in the sense of
\ref{five:h-assignment} for the five pieces in
\eqref{five:piece-order}.
\end{lemma}
\begin{proof}
If $f_i=0$, then $(x,b,d)=(0,1,1)$, where $\mathcal A=-1$. Thus every
$f_i>0$, giving $\ell_i>x$, $u_i<b$, and $x<c<b$.
Call an agent a weak (strict) claimant of a private piece when she
values it at least (more than) her largest public value $f_i$. For
$P$ this means $\ell_i\leq c$ ($\ell_i<c$); for $U$ it means
$u_i\geq c$ ($u_i>c$). The order statistics give each piece at least two weak
claimants and at most one strict claimant.

If the strict claimants are distinct, assign each her claimed piece;
she is not strict for the other piece, so this is a global favourite.
If one agent strictly claims both, give her the one she prefers and
give the other to a different weak claimant, who is strict for
neither. If just one piece has a strict claimant, give it to her and
choose a different weak claimant for the other. If neither has a
strict claimant, choose distinct weak claimants, possible since both
sets have size at least two. Every chosen piece is a global favourite.
All strict claimants are chosen, so every remaining agent has a
public global favourite.
\end{proof}

By Lemma~\ref{five:calibration-rates}, there is a unique
$d_0\in(0,1)$ with $\mathcal A(0,d_0,d_0)=0$. For each $d\in[d_0,1]$ there
is a unique $a(d)\in[0,d)$ with $\mathcal A(a(d),d,d)=0$; it equals zero only
at $d=d_0$. For $0\leq x\leq a(d)$ there is a unique
$b(x,d)\in(x,d]$ with $\mathcal A(x,b(x,d),d)=0$, and $b(x,d)=d$ exactly
when $x=a(d)$. These assertions follow from the rates, continuity,
and the positive sign at $b=x$ in that lemma. Let $c(x,d)$
be the common order statistic at this zero. The promised family, which
we call the \emph{chart}, is
\begin{equation}\label{five:chart}
 d=d_0+(1-d_0)t,\qquad x=s\,a(d),\qquad
 b=b(x,d),\qquad c=c(x,d),\qquad (s,t)\in[0,1]^2.
\end{equation}
On a grid of mesh $1/H$, column $i$ and row $j$ meet at
$(s,t)=(i/H,j/H)$, so a row fixes $d$; since $a(d_0)=0$, the bottom
row carries a single partition.

\begin{lemma}\label{five:quantitative-chart}
The score matrices of the partitions \eqref{five:chart} satisfy
\ref{five:h-assignment}--\ref{five:h-rows}. Every score is
$W$-Lipschitz in maximum norm on the square, where
\begin{equation}\label{five:W}
 W=252m_+B^5.
\end{equation}
Consequently a grid of mesh $1/H$ yields, by
Lemma~\ref{five:abstract}, a corner partition and a bijection with
envy at most $2W/H$.
\end{lemma}
\begin{proof}
Lemma~\ref{five:private-owners} supplies the legal assignment of
property~\ref{five:h-assignment}.
The pieces $L,M,R$ are respectively empty on the left, right, and top
boundaries; $L,M$ are both empty on the bottom. They cannot be public
favourites there because $f_i>0$, which gives
property~\ref{five:h-boundary}. On any other row, $a(d)>0$ and
$x$ strictly increases with $s$. The first two calibration rates
imply that $b(x,d)$ strictly increases with $x$. Hence $L$ grows in
value, $M$ shrinks, and $R$ stays fixed, which is
property~\ref{five:h-rows}.

For the Lipschitz bounds, note that a function increasing at rate at
least $1/B$ moves its zero by at most $B$ times a perturbation in its
values. Together with the
$6B$-Lipschitz bound for $\mathcal A$, this gives a $6B^2$-Lipschitz bound for
$a(d)$ and for $b(x,d)$ as a function of $(x,d)$. On the square,
$d$ is $1$-Lipschitz, $x=s\,a(d)$ is $7B^2$-Lipschitz, and $b$
is consequently $42B^4$-Lipschitz. These zero comparisons may involve
unordered trial points, which is why the calibration rates were
proved on the whole cube. Each $\ell_i$ is $3B$-Lipschitz in
$(x,b,d)$, so $c$ is $126B^5$-Lipschitz on the square. A score is a
difference of two prefix values, giving $W=252m_+B^5$.
Cell corners are within $1/H$ in maximum norm, so
\eqref{five:corner-envy} gives envy at most $2W/H$.
\end{proof}

\subsection{The Robertson--Webb implementation}\label{five:oracle-implementation}

The common search leaves one task to its implementation: obtain exact
scores and preference thresholds on demand. We do this without learning
the full valuations. A nearby piecewise-linear profile can be described
by the order of quantile marks; only the marks needed by the search are
queried. The following steps construct this profile, give access
to it, locate its calibration roots and preference thresholds, and
complete the query accounting.

\subsubsection{Quantiles and an auxiliary profile}\label{five:quantiles-section}

The given valuations need not satisfy \eqref{five:density-bounds}, and
the preceding construction uses exact zeros. Approximating those zeros
without further care would lose the strict boundary exclusions or the
legal private assignment. We therefore build an auxiliary profile that
satisfies \eqref{five:density-bounds} and in which every zero can be
computed exactly, while querying only a small part of its description.
Quantile interpolation, deferred queries, and consistent cut choices on
flat intervals are established tools~\cite[Lemma~6.2]{HR}. We
implement them in the order of the agents' quantile marks, so that the
auxiliary functions have rational coefficients.

Return to the original additive valuations $v_i$ and their cumulative
functions $F_i$, which may have flat intervals. We construct a nearby
auxiliary profile satisfying the assumptions of the common construction.
Let $K=2^\kappa$ be
the power of two with
\begin{equation}\label{five:K}
 256/\eps\leq K<512/\eps,
 \qquad J=5(K-1)+1.
\end{equation}
For each $i$ and $1\leq j\leq K-1$, a query
$\operatorname{Cut}_i(0,j/K)$ returns a \emph{quantile mark} $y_{i,j}$
(not to be confused with the calibration marks $\ell_i,u_i$) satisfying
$F_i(y_{i,j})=j/K$; there are $J-1$ quantile marks in all. For the
definition only, imagine that every such mark has been chosen. The
algorithm will ask for very few of them. Within one agent the marks
strictly increase, even if her valuation has flat intervals. Call the
pairs $(i,j)$ \emph{entries}, and order them by the lexicographic key
$(y_{i,j},i)$; this is the \emph{merged order}.
Let $r(i,j)$ be the \emph{rank} in this order, from $1$ to
$J-1$, and let $p_k$ be the physical position of the entry of rank $k$.
Set $p_0=0$ and $p_J=1$. Write $c_h(k)$ for the number of agent $h$'s
entries among the first $k$ ranks, with $c_h(0)=0$ and $c_h(J)=K$.
The last convention adds the common endpoint as a final mark for each
agent.

Define strictly increasing, piecewise-affine cumulative functions
$G_h$ by linear interpolation between
\begin{equation}\label{five:virtual-values}
 G_h(k/J)=Q_{h,k}
 =\frac{(K-1)c_h(k)}{K^2}+\frac{k}{JK},
 \qquad 0\leq k\leq J.
\end{equation}
Extend them as the identity outside $[0,1]$. The points $k/J$ are
\emph{knots}, $G_h$ is affine on each \emph{knot cell}
$[k/J,(k+1)/J]$, and for $0<k<J$ the knot $k/J$ stands for the entry
of rank $k$. These functions form the \emph{auxiliary profile} and are
the cumulative functions used in the preceding construction; the
variable in this auxiliary cake records the order of quantile marks,
rather than their physical distance. Thus
$Q_{h,k}=(1-1/K)\,c_h(k)/K+(1/K)\,k/J$ mixes the normalised count
$c_h(k)/K$, which approximates $F_h(p_k)$
(Lemma~\ref{five:virtual-transfer}), with the uniform value $k/J$. The
uniform term makes every auxiliary valuation strictly positive on
nonempty intervals.

\begin{lemma}\label{five:virtual-transfer}
The auxiliary functions satisfy \eqref{five:density-bounds} with
$m_-=1/K$, $m_+=5$, $B=5K$. For
\begin{equation}\label{five:projection}
 \rho(t)=p_{\lfloor tJ\rfloor},\qquad 0\leq t\leq1,
\end{equation}
we have $\rho$ nondecreasing and
$|G_h(t)-F_h(\rho(t))|\leq3/K$. Applying $\rho$ to each cut changes
the envy of any fixed assignment, measured with $G_h$ at the cuts and with $F_h$ at their images, by at most $12/K$.
\end{lemma}
\begin{proof}
On the cell $[k/J,(k+1)/J]$ put
$\Delta=c_h(k+1)-c_h(k)\in\{0,1\}$. On the final cell $\Delta=1$
for every agent. Directly from \eqref{five:virtual-values},
\begin{equation}\label{five:integer-affine}
 K^2G_h(t)=s t+r,\qquad
 s=(K-1)\Delta J+K,\qquad
 r=(K-1)(c_h(k)-\Delta k).
\end{equation}
Here $r,s$ are integers and $K\leq s\leq5K^2$. The slopes are
therefore between $1/K$ and $5$, with strict upper inequality.

For a rank $k$ between $1$ and $J-1$, let $p=p_k$, and let $a$ be the
number of agent $h$'s marks strictly left of $p$. Then
$a/K\leq F_h(p)\leq(a+1)/K$. The number $c_h(k)$ is either $a$ or
$a+1$, and it is $a+1$ only if $h$ has a mark at $p$, in which case
$F_h(p)=(a+1)/K$. Thus
$|c_h(k)/K-F_h(p_k)|\leq1/K$, including ranks inside a group of
coincident marks. The endpoint cases also hold. Within a cell the
interpolated count changes by at most $1/K$; the mixing with the
uniform cumulative function changes a value by at most another $1/K$.
This proves the prefix bound. A piece value is a difference of two
prefixes and an envy comparison a difference of two piece values,
giving $12/K$.
\end{proof}

For the chart of the auxiliary profile, take as grid side the least
power of two $H=2^\lambda$ with $H\geq8W/\eps$, where
$W=252\cdot5\cdot(5K)^5$ is \eqref{five:W} with $m_+=5$ and $B=5K$. Then
\begin{equation}\label{five:mesh-bound}
 \lambda=O(\kappa)=O(\log(1/\eps)).
\end{equation}
Indeed, $W=O(K^5)$ and $1/\eps\leq K/256$. The auxiliary envy
$2W/H\leq\eps/4$ leaves enough room for the $12/K$ projection
error. We must now find the cell using $O(\kappa)$ queries per row.

\subsubsection{Exact access to the quantile order}\label{five:access-section}

The auxiliary functions depend on the merged order of all $J-1$
quantile marks, which we access one entry at a time. We call the
following operation a \emph{probe}: given an entry $(i,j)$,
obtain its physical position $y_{i,j}$, its rank, and the five counts
$c_h$ at its rank. All query answers are cached for the entire
protocol.

\begin{lemma}\label{five:rank}
A probe needs at most nine Robertson--Webb queries.
\end{lemma}
\begin{proof}
The rank is $j$ plus the number of other agents' entries preceding
$(i,j)$, which their values at $y_{i,j}$ almost determine.
Ask for $z=y_{i,j}$ and ask the four other agents for
$\alpha_h=F_h(z)$. Let $n_h$ be the number of $h$'s entries preceding
$(z,i)$ in the lexicographic order. If $K\alpha_h$ is not an integer,
then $n_h=\lfloor K\alpha_h\rfloor$: the quantile levels below
$\alpha_h$ have marks strictly left of $z$, and those above it have
marks strictly right of $z$. At levels $0$ and $K$ the counts are
$0$ and $K-1$, respectively. If $K\alpha_h=j'$ is an integer with
$1\leq j'\leq K-1$, ask for the actual stored mark $y_{h,j'}$ and put
\begin{equation}\label{five:plateau-rule}
 n_h=j'-1+\mathbf{1}\{(y_{h,j'},h)<(z,i)\}.
\end{equation}
This extra query is essential: a cut oracle may put $y_{h,j'}$ on
either side of $z$ within a flat interval of $F_h$. We have
$r(i,j)=j+\sum_{h\ne i}n_h$, and the counts at that rank are $j$ for
agent $i$ and $n_h$ for each other agent. There are at most four extra
cuts, giving nine queries.
\end{proof}

We use \emph{Locate} for the operation that, given a rational
$t\in[0,1)$, identifies the entries of ranks $k=\lfloor tJ\rfloor$
and $k+1$, with endpoint conventions, and hence its knot cell
$[k/J,(k+1)/J]$ and all five affine forms \eqref{five:integer-affine}
and values on that knot cell. At $t=1$ the endpoint and its values are
known without queries.

\begin{lemma}\label{five:flattened-access}
Selecting an entry of a given rank, or locating a rational auxiliary
point, needs $O(\kappa)$ probes.
\end{lemma}
\begin{proof}
For Locate, binary-search each agent's own indices for the largest $j$
with $r(h,j)\leq\lfloor tJ\rfloor$. This costs at most $\kappa$
probes per agent. The largest of the five resulting ranks is $k$.
The entry of rank $k+1$ is one of the five successors of these entries
in their respective agents' lists, so at most five further probes find
it. Endpoint ranks and values are known for free. Formula
\eqref{five:integer-affine} supplies the affine forms. Entry selection
is Locate at $t=k/J$. The total is at most $5\kappa+5$ probes;
there is no nested search over ranks.
\end{proof}

An inverse point alone can be obtained even more cheaply. This will
let us evaluate the calibration function at constant cost once its
\emph{forward values} $G_h(x)$, $G_h(b)$, $G_h(d)$ are known.

Agent $h$'s \emph{own} entries are the $(h,c)$; here and in the next
two proofs $c$ is such an index, not a cut.
Write $r_{h,c}=r(h,c)$, and set $r_{h,0}=0$, $r_{h,K}=J$.
The knots at which $G_h$ changes slope occur only immediately before
or at $h$'s own marks:
\begin{equation}\label{five:own-bends}
 (r_{h,c}-1)/J,\quad r_{h,c}/J,\qquad 1\leq c\leq K.
\end{equation}
Moreover, for $r_{h,c}\leq k<r_{h,c+1}$,
\begin{equation}\label{five:ladder}
 Q_{h,k}=\frac{(K-1)c}{K^2}+\frac{k}{JK},
 \qquad c\leq r_{h,c}\leq J-K+c.
\end{equation}
The slope on a cell is $1/K$, except when the entry at its right
endpoint belongs to $h$ or is the final endpoint, in which case the
slope is $((K-1)J+K)/K^2<5$. These statements follow directly from
\eqref{five:integer-affine}; the rank bounds count the $c$ preceding
own entries and the $K-1-c$ following own entries. In particular,
\begin{equation}\label{five:own-levels}
 \frac{(K-1)c}{K^2}\leq Q_{h,r_{h,c}}
 \leq\frac{(K-1)c}{K^2}+\frac1K,
 \qquad
 Q_{h,r_{h,c+1}}-Q_{h,r_{h,c}}>\frac{K-1}{K^2}.
\end{equation}

\begin{lemma}\label{five:two-probe-inverse}
For rational $0\leq z<1$ the exact position $G_h^{-1}(z)$, its knot
cell, and $G_h$'s slope on that cell are obtained using at most two
probes, both to agent $h$'s own entries. For $z$ outside $[0,1)$ the
inverse is $z$ without queries.
\end{lemma}
\begin{proof}
Put $c_0=\lfloor zK^2/(K-1)\rfloor$ and let $c^*$ be the largest
$c\in\{0,\ldots,K-1\}$ with $Q_{h,r_{h,c}}\leq z$. From
\eqref{five:own-levels},
\[
 c^*\leq c_0,\qquad
 z<Q_{h,r_{h,c^*+1}}\leq\frac{(K-1)(c^*+1)}{K^2}+\frac1K,
\]
so $c^*\geq c_0-2$ for $K\geq4$. Test $c=c_0$. This is free and
true at $c=0$, free and false at $c\geq K$, and otherwise requires one
probe, which gives $r_{h,c}$ and therefore $Q_{h,r_{h,c}}$. If true,
$c^*=c_0$ and a second probe gives $r_{h,c^*+1}$, unless this is the
known rank $J$. If false, test $c=c_0-1$ in the same way. A true
answer gives $c^*=c_0-1$; otherwise $c^*=c_0-2$. In both cases the
rank $r_{h,c^*+1}$ is already known from a failed test or is $J$.
Tests at out-of-range endpoint indices use the endpoint conventions.

By \eqref{five:ladder}, the largest knot index $k^*$ with
$Q_{h,k^*}\leq z$ is
\begin{equation}\label{five:inverse-knot}
 k^*=\min\left\{r_{h,c^*+1}-1,
 \left\lfloor JKz-\frac{(K-1)c^*J}{K}\right\rfloor\right\}.
\end{equation}
The value at this knot is
$(K-1)c^*/K^2+k^*/(JK)$. The slope numerator $s$ is
$(K-1)J+K$ if $k^*+1=r_{h,c^*+1}$ and is $K$ otherwise. Linear
interpolation gives
\[
 G_h^{-1}(z)=\frac{k^*}{J}
       +\frac{K^2(z-Q_{h,k^*})}{s}.
\]
Only the ranks at indices among $c_0-1,c_0,c_0+1$ were needed.
\end{proof}

Consequently an exact evaluation of $\mathcal A(x,b,d)$ at rational arguments
costs $O(\kappa)$ probes: locate the three arguments and compute its
ten calibration marks $\ell_i,u_i$ by inversion. If all agents' forward
values at an inverse point are needed, they require a separate Locate.

We say a point is \emph{recorded} if all five forward values there
are known, the numbers of each agent's quantile marks strictly to its
left in the auxiliary order are known, and, if it is not a knot, the five
affine forms on its cell are known. Both endpoints, every probed knot,
and every point processed by Locate are recorded. At a probed knot
the owning entry is known, so the strict-left counts are obtained
from the inclusive counts by subtracting its one entry. This convention
will permit searches over open intervals without ambiguity at knots.

Part~(b) of the next lemma makes $\mathcal A$ explicit within knot cells, so
that its zeros there can be found \emph{offline}, without probes.

\begin{lemma}\label{five:local-functions}
Fix a recorded $d$.
\begin{enumerate}[label=(\alph*)]
\item At recorded $x,\beta$, all ten marks in
\eqref{five:calibration} and $\mathcal A(x,\beta,d)$ are computed with at most
twenty probes.
\item Suppose $(x(\tau),\beta(\tau))$, $0\leq\tau\leq1$, is affine,
each coordinate either constant and recorded, or confined to one
closed knot cell with known affine forms. With at most sixty probes,
all ten marks and $\mathcal A(x(\tau),\beta(\tau),d)$ become explicit
continuous piecewise-affine functions of $\tau$.
\end{enumerate}
\end{lemma}
\begin{proof}
For (a), the ten inverse levels $G_i(x)+f_i$ and $G_i(\beta)-f_i$ are
known from the forward values, so
Lemma~\ref{five:two-probe-inverse} applies ten times. For (b), each
varying forward value moves by less than $5/J$. Every branch of $f_i$
therefore varies by less than $5/J$, and each inverse level ranges
over an interval $[z_1,z_2]$ of length less than $10/J$. Inverting
any level in this interval only needs own indices among
$c_0(z)-1,c_0(z),c_0(z)+1$, by the preceding proof. Since $K\geq8$,
\[
 c_0(z_2)-c_0(z_1)
 \leq\left\lfloor\frac{(z_2-z_1)K^2}{K-1}\right\rfloor+1\leq3,
 \qquad \frac{10K^2}{(5K-4)(K-1)}<3.
\]
Thus at most six own entries cover the entire inverse-level interval.
Querying them gives an explicit piecewise-affine inverse on that
interval: by \eqref{five:own-bends} there are no other changes of
slope relevant to it. The identity extension handles levels outside
$[0,1]$. Ten levels cost at most sixty probes. Compositions, maxima,
and finite order statistics of explicit continuous piecewise-affine
functions remain explicit continuous piecewise-affine functions.
All further operations in (b) are consequently offline.
\end{proof}

\subsubsection{Exact scalar and crossing searches}\label{five:roots-section}

We now compute the chart roots $d_0$, $a(d)$, $b(x,d)$ and $c(x,d)$
exactly: $d_0$, the right cut of the bottom row, once by rational
reconstruction, and the others by discarding quantile marks until only
known affine cells remain.

\begin{lemma}\label{five:rational-roots}
For the auxiliary profile, $d_0$ is rational of denominator at most
$75K^4$ and can be computed with $O(\kappa^2)$ probes. The remaining
chart roots are rational whenever their arguments are rational.
\end{lemma}
\begin{proof}
We first show that at each chart root, eliminating $c$ from the two
calibration equations leaves one linear equation for the remaining
unknown, with a positive integer coefficient (at most $75K^4$ for
$d_0$); this gives rationality and the denominator bound. Then we find
$d_0$ by exact bisection and rational reconstruction.
At a zero write $b^*=b(x,d)$, $c^*=c(x,d)$. Choose agents $h,h'$
attaining the two order statistics, so that their calibration marks
satisfy $\ell_h=c^*=u_{h'}$. On fixed closed knot cells write
$K^2G_i(z)=s_i^zz+r_i^z$, with the integer coefficients of
\eqref{five:integer-affine}. The two equalities are
\begin{align*}
 s_h^cc^*+r_h^c&=s_h^xx+r_h^x+K^2f_h,\\
 s_{h'}^cc^*+r_{h'}^c&=s_{h'}^bb^*+r_{h'}^b-K^2f_{h'}.
\end{align*}
Each $K^2f_i$ is one of its three integer-affine branches in $x,b^*,d$.
Multiply by the opposite $c$ coefficient and subtract to eliminate
$c^*$. The positive coefficient of $b^*$ is
\[
 A_2=s_h^cs_{h'}^b(1\text{ or }2)
          +s_{h'}^cs_h^b(0\text{ or }1).
\]
It is between $K^2$ and $3(5K^2)^2=75K^4$. The other terms are
integer combinations of $x,d,1$, proving rationality of $b^*,c^*$
and nonsingularity of this two-equation system. For $a(d)$ set
$b=d$ and regard $x$ as unknown. Its coefficient after eliminating $c$
is
$s_{h'}^cs_h^x(1\text{ or }2)+s_h^cs_{h'}^x(0\text{ or }1)$,
which is positive. For $d_0$, set $x=0$, $b=d$, so
$f_i=1-G_i(d)$. The coefficient is
$s_{h'}^cs_h^d+2s_h^cs_{h'}^d$, again positive and at most $75K^4$.
If a root is on a knot or branch boundary, choose either adjacent
closed cell or active branch; the equalities remain valid.

It remains to compute $d_0$. A rational zero of denominator at most
$Q=75K^4$ is uniquely
determined by a closed interval of width less than $1/Q^2$, since two
distinct rationals with such denominators differ by at least $1/Q^2$.
Bisect the strictly decreasing function $d\mapsto \mathcal A(0,d,d)$ using
exact signs, stopping at an exact zero if encountered. Since
$\log Q=O(\kappa)$, this takes $O(\kappa)$ steps. Recover the zero as
the unique fraction of denominator at most $Q$ in the bracket, for
example by continued fractions. Each sign costs $O(\kappa)$ probes
by Lemmas~\ref{five:flattened-access}
and~\ref{five:two-probe-inverse}, giving $O(\kappa^2)$ in total.
\end{proof}

\begin{lemma}\label{five:x-root}
Let $d,\beta,X>0$ be recorded, with $d,\beta,X\leq1$. For
$g(x)=\mathcal A(x,\beta,d)$ one can report $g(0)>0$, report $g(X)<0$, or
compute its exact zero in $[0,X]$, using
$O(\kappa)$ probes. A returned zero is recorded. The same
conclusion holds for $\beta=0$.
\end{lemma}
\begin{proof}
Evaluate both endpoints at constant cost by
Lemma~\ref{five:local-functions}(a). The stated exterior-sign cases
are then known, and an endpoint zero is returned directly. Otherwise
keep an open bracket $(X_0,X_1)$ with
$g(X_0)<0<g(X_1)$, initially $(0,X)$. Its ends are recorded, and its
left end is always a knot. The own marks strictly inside form known
intervals of indices. If there is any such mark, take the first agent
who has one, probe her median mark in the bracket, and evaluate $g$
there using Lemma~\ref{five:local-functions}(a). An exact zero ends
the search; otherwise its sign moves one bracket end to that knot. The number of
that agent's marks inside at least halves. Each agent begins with
$K-1=2^\kappa-1$ marks, so after at most $\kappa$ such moves she has
none. Thus there are at most $5\kappa$ rounds of constant cost.

When no mark remains, the bracket has no interior knot and lies in a
single closed knot cell. Its five affine forms are known. If both
ends are knots they are consecutive and their forward values determine
these forms; if the right end is not a knot, it is the original $X$
and its recorded cell supplies them. Apply
Lemma~\ref{five:local-functions}(b) on this interval with $\beta$
constant. At most sixty probes make $g$ an explicit continuous
piecewise-affine function, strictly increasing by
Lemma~\ref{five:calibration-rates}. Find its unique zero
offline. Its forward values and cell forms are known, and its
strict-left counts equal those at the left bracket end including that
end's entry, so it is recorded. The total cost is $O(\kappa)$ probes.
\end{proof}

In particular, after locating $d$, apply this lemma with $\beta=X=d$
to compute $a(d)$, the cut $x$ at the right end of the row of $d$.

We also need to evaluate the score matrix at a prescribed grid vertex,
that is, to compute $b(x,d)$ and $c(x,d)$.
For fixed recorded or rational $x<d$, first consider the interior case
$\mathcal A(x,d,d)<0$, that is, $x<a(d)$. Set
\[
 \ell(\beta)=\operatorname{secondmin}_i\ell_i(x,\beta,d),\qquad
 u(\beta)=\operatorname{secondmax}_iu_i(x,\beta,d).
\]
The first is nonincreasing and the second strictly increasing.
Their unique crossing is $(b^*,c^*)=(b(x,d),c(x,d))$ with
$x<c^*<b^*<d$. For any trial value $\beta\in[x,d]$ of the cut $b$ and
any trial value $\gamma$ of the cut $c$, the signs
of $\gamma-\ell(\beta)$ and $\gamma-u(\beta)$ give the following
tests:
\begin{equation}\label{five:vertex-signs}
\begin{array}{c|c}
\text{sign condition}&\text{conclusion}\\\hline
 \gamma>\ell(\beta),\ \gamma>u(\beta)&\gamma>c^*\\
 \gamma<\ell(\beta),\ \gamma<u(\beta)&\gamma<c^*\\
 \ell(\beta)\geq\gamma\geq u(\beta),\ \ell(\beta)>u(\beta)
       &\beta<b^*\\
 \ell(\beta)\leq\gamma\leq u(\beta),\ \ell(\beta)<u(\beta)
       &\beta>b^*\\
 \ell(\beta)=\gamma=u(\beta)&(\beta,\gamma)=(b^*,c^*) .
\end{array}
\end{equation}
The five rows cover all nine pairs of signs. To verify the first two tests,
$c^*$ lies between $\ell(\beta)$ and $u(\beta)$ for every $\beta$:
before the crossing $\ell\geq c^*\geq u$, and afterwards
$\ell\leq c^*\leq u$. The next two tests express the sign of
$\mathcal A(x,\beta,d)=\ell(\beta)-u(\beta)$. All tests use only forward
values at $x,d,\beta,\gamma$. For example
$\gamma>\ell(\beta)$ exactly when at least two agents satisfy
$G_i(\gamma)>G_i(x)+f_i$; weak inequalities give the weak version.
Likewise compare $G_i(\gamma)$ with $G_i(\beta)-f_i$ for $u$.

Recall that a legal assignment (property~\ref{five:h-assignment})
gives $P,U$ to distinct agents as global favourites and every other
agent a public global favourite.

\begin{lemma}\label{five:vertex-crossing}
For a vertex in a non-bottom row, its complete score matrix and a legal
assignment can be obtained using $O(\kappa)$ probes, assuming
that the row's $d$ and $a(d)$ are already recorded. In the interior
case the procedure computes $b(x,d)$ and $c(x,d)$ exactly. The bound
also covers the right boundary $x=a(d)$.
\end{lemma}
\begin{proof}
The search shrinks brackets for $b^*$ and $c^*$ in phases; the right
boundary is treated at the end.
In the interior case locate $x$, at cost $O(\kappa)$; $d$ is already
recorded. Maintain an open $b$-bracket and an open $c$-bracket, both
initially $(x,d)$, containing $b^*,c^*$. All bracket ends are recorded. Thus each agent's
marks strictly inside each bracket form a known interval of indices.

If both brackets contain marks, probe a median own mark of the first
agent with marks in the $b$-bracket, and do the same for the
$c$-bracket. Call their knot positions $\beta,\gamma$. The tests
\eqref{five:vertex-signs} either find the crossing or move one bracket
end to the corresponding median. This costs two probes and halves one
agent's marks in one bracket.

Suppose instead that the $b$-bracket contains no marks. It then lies
in one closed knot cell with known affine forms. Probe a median mark
$\gamma$ in the $c$-bracket. We determine without further queries
whether $c^*>\gamma$, $c^*<\gamma$, or $c^*=\gamma$. The first holds
exactly when some $\beta$ in the open $b$-bracket has
$\gamma<\ell(\beta),u(\beta)$; the second has the reversed strict
inequalities; the third holds exactly when some $\beta$ has
$\ell(\beta)=\gamma=u(\beta)$. Necessity follows by taking
$\beta=b^*$, and sufficiency follows from
\eqref{five:vertex-signs}. For this fixed $\gamma$, the twenty
conditions comparing $G_i(x)+f_i(\beta)$ and
$G_i(\beta)-f_i(\beta)$ with $G_i(\gamma)$, in strict and weak
form, are piecewise-affine inequalities with known rational
coefficients on the bracket. Partition it at all their zeroes and
branch changes, and inspect those points and one point in each
remaining interval. This decides the required existential tests
offline. A strict verdict moves the $c$-bracket end. In the equality
case $c^*=\gamma$ is a probed knot. The five known strictly increasing
functions
$\beta\mapsto G_i(\beta)-f_i(\beta)-G_i(\gamma)$ determine the
unique $\beta$ with $u(\beta)=\gamma$: at least two functions are
nonnegative and at most one is strictly positive there. Compute it
offline from their piecewise-affine forms. This gives $b^*$ and ends
the search.

If the $c$-bracket $(C_0,C_1)$ contains no marks but the $b$-bracket
does, probe a median $\beta$ in the latter. On the former bracket
every $G_i$ is affine. The known levels $G_i(x)+f_i$ and
$G_i(\beta)-f_i$ tell whether each $\ell_i,u_i$ is below $C_0$,
above $C_1$, or inside, where it is known exactly by affine inversion.
The order statistics therefore have the same trichotomy. Since
$C_0<c^*<C_1$, a value $u(\beta)\leq C_0$ gives $\beta<b^*$ and
$u(\beta)\geq C_1$ gives $\beta>b^*$. Otherwise $u(\beta)$ is
known inside the interval: $\ell(\beta)\geq C_1$ or
$\ell(\beta)>u(\beta)$ gives $\beta<b^*$, while
$\ell(\beta)\leq C_0$ or $\ell(\beta)<u(\beta)$ gives
$\beta>b^*$. Equality gives the crossing. Thus this phase also
halves one agent's remaining marks at the cost of one probe.

When neither bracket contains marks, they lie in two closed knot cells
with known affine forms. The two equations in the proof of
Lemma~\ref{five:rational-roots}, for each pair of agents and each
choice of their three public-maximising branches, give at most
$25\cdot9=225$ linear systems in $b,c$. Their determinants are
nonzero by the positive coefficient calculated there. Solve them and
retain a solution in the cells satisfying
$\ell(b)=c=u(b)$. It exists and is unique; the equalities are checked
with the known forward affine forms. This last step costs no probes.
If $b^*$ or $c^*$ is itself a knot, its open bracket cannot become
mark-free and an equality is detected in one of the previous phases.

It remains to count the cost.
Every nonterminal round halves the marks of one agent in one of two
brackets. There are at most $10\kappa$ rounds, each costing at most
two probes. Including the initial Locate, the cost is $O(\kappa)$.
All five forward values at the returned $b,c$ are
known, either from their probed knots or from their final affine
cells; the same is true at $x,d$. The twenty-five scores, the legal
assignment from Lemma~\ref{five:private-owners}, and the votes (each
agent's chosen public favourite) are then obtained offline.

Finally, on the right boundary $x=a(d)$ we already know $b=d$.
The five levels $G_i(x)+f_i$ are known; their inverses cost at most
ten probes by Lemma~\ref{five:two-probe-inverse}. Their second
smallest is $c$. Locate $c$ to obtain all forward values, for a total
of $O(\kappa)$ probes on this boundary too.
\end{proof}

\subsubsection{Preference changes along a row}\label{five:row-section}

To obtain the bound $O(\log^2(1/\eps))$, it is not enough that each
vertex costs $O(\kappa)$ probes: without supplied row indices, the
search of Lemma~\ref{five:abstract-search} might evaluate
$O(\lambda^2)$ vertices, where $H=2^\lambda$ is the grid side and
$\kappa=\log_2K$. We instead compute all thirty preference-change
indices of one row in $O(\kappa)$ probes, namely, for each public
comparison, the first columns where it is nonnegative and where it is
positive, and use the supplied-row form of that lemma.

Fix a non-bottom row, with recorded $d>d_0$ and recorded $a=a(d)>0$.
Write $b(x)=b(x,d)$ for $0\leq x\leq a$. The map $b$ is continuous
and strictly increasing, $x<b(x)\leq d$, and $b(a)=d$. By the strict
decrease of $\mathcal A$ in its second argument on the whole cube
(Lemma~\ref{five:calibration-rates}),
\begin{equation}\label{five:row-calibration-sign}
 \sgn \mathcal A(x,\beta,d)=\sgn(b(x)-\beta),
 \qquad 0\leq x\leq a,\quad0\leq\beta\leq1.
\end{equation}
For a fixed agent $h$, the three increasing public score differences
along this row are
\begin{align}
 \delta_{LM}(x)&=G_h(x)+G_h(b(x))-G_h(d),\notag\\
 \delta_{LR}(x)&=G_h(x)-(1-G_h(d)),\label{five:row-comparisons}\\
 \delta_{RM}(x)&=G_h(b(x))-(2G_h(d)-1).\notag
\end{align}
The first is the only one requiring a simultaneous crossing search,
since it involves both $x$ and $b(x)$.

\begin{lemma}\label{five:row-signs}
There is a unique $x^*\in(0,a)$ with $\delta_{LM}(x^*)=0$. Put
$b^*=b(x^*)\in(0,d)$ and
$\phi(x,\beta)=G_h(x)+G_h(\beta)-G_h(d)$. For
$g=\sgn \mathcal A(x,\beta,d)$ and $p=\sgn\phi(x,\beta)$, with
$0\leq x\leq a$, $0\leq\beta\leq1$, the following conclusions hold:
\begin{equation}\label{five:row-sign-table}
\begin{array}{c|l}
 (g,p)&\text{conclusion}\\\hline
 (+,+)&x>x^*\\
 (-,-)&x<x^*\\
 (+,-)&\beta<b^*\\
 (-,+)&\beta>b^*\\
 (0,+)&x>x^*\text{ and }\beta>b^*\\
 (0,-)&x<x^*\text{ and }\beta<b^*\\
 (+,0)&x>x^*\text{ and }\beta<b^*\\
 (-,0)&x<x^*\text{ and }\beta>b^*\\
 (0,0)&(x,\beta)=(x^*,b^*).
\end{array}
\end{equation}
\end{lemma}
\begin{proof}
The function $\delta_{LM}(x)=\phi(x,b(x))$ is continuous and strictly
increasing. At $x=0$ its value is $G_h(b(0))-G_h(d)<0$, because
$b(0)<d$; at $x=a$ it equals $G_h(a)>0$. This proves existence,
uniqueness, and the stated strict bounds.

For $(+,+)$, if $x\leq x^*$ then
$\beta<b(x)\leq b^*$ by \eqref{five:row-calibration-sign}, forcing
$\phi(x,\beta)<\phi(x^*,b^*)=0$, a contradiction. The $(-,-)$ case
is symmetric. For $(+,-)$, if $\beta\geq b^*$ then
$b(x)>\beta\geq b^*$ gives $x>x^*$ and hence $\phi(x,\beta)>0$,
again a contradiction; $(-,+)$ is symmetric. If $g=0$, then
$\beta=b(x)$ and $p$ is precisely the sign of $x-x^*$, and also of
$\beta-b^*$. For $(+,0)$ the preceding argument forces $x>x^*$;
the equality $\phi(x,\beta)=\phi(x^*,b^*)$ then forces
$\beta<b^*$. The $(-,0)$ case is symmetric. Finally $g=p=0$ is
exactly the unique crossing.
\end{proof}

\begin{lemma}\label{five:row-crossing}
For each agent, the pair $(x^*,b^*)$ in
Lemma~\ref{five:row-signs} is computed exactly with $O(\kappa)$ probes.
\end{lemma}
\begin{proof}
As in Lemma~\ref{five:vertex-crossing}, we shrink brackets for $x^*$
and $b^*$ in phases.
Keep open brackets $(X_0,X_1)$ and $(B_0,B_1)$ containing $x^*$
and $b^*$, initially $(0,a)$ and $(0,d)$. Their ends are recorded,
their left ends are knots, and the own marks strictly inside form
known intervals of indices. All updates below use strict conclusions,
so the unknown crossing stays strictly inside both brackets.

If both brackets contain marks, probe a median mark of the first
agent with marks in each bracket, with knot positions $\xi,\beta$.
Compute $\mathcal A(\xi,\beta,d)$ at constant cost by
Lemma~\ref{five:local-functions}(a). The value of $\phi(\xi,\beta)$
is already known. Table~\eqref{five:row-sign-table} either returns
the crossing or moves an end of at least one bracket to its probed
median.

Suppose the $x$-bracket has no marks but the $b$-bracket does. The
$x$-bracket lies in one closed knot cell with known affine forms, as
in Lemma~\ref{five:x-root}. Probe a median $\beta$ from the
$b$-bracket. By Lemma~\ref{five:local-functions}(b), sixty further
probes make $x\mapsto \mathcal A(x,\beta,d)$ explicit on $[X_0,X_1]$.
If $\mathcal A(X_0,\beta,d)\geq0$, strict increase and $x^*>X_0$ imply
$\mathcal A(x^*,\beta,d)>0$, so $\beta<b^*$. If
$\mathcal A(X_1,\beta,d)\leq0$, the symmetric argument gives $\beta>b^*$.
Otherwise compute its unique zero $x'$ in the bracket offline.
Then $b(x')=\beta$, and
\[
 \sgn\phi(x',\beta)=\sgn\delta_{LM}(x')
                  =\sgn(x'-x^*)=\sgn(\beta-b^*).
\]
This either determines which $b$-bracket end to move or finds
$(x^*,b^*)=(x',\beta)$.

If instead the $b$-bracket has no marks and the $x$-bracket does,
probe a median $\xi$ from the latter and make
$\beta\mapsto \mathcal A(\xi,\beta,d)$ explicit on $[B_0,B_1]$, again using
at most sixty further probes. If $\mathcal A(\xi,B_1,d)\geq0$, then
$b(\xi)\geq B_1>b^*$, giving $\xi>x^*$. If
$\mathcal A(\xi,B_0,d)\leq0$, then $b(\xi)\leq B_0<b^*$, giving
$\xi<x^*$. Otherwise compute its zero $\beta'=b(\xi)$ offline;
$\sgn\phi(\xi,\beta')=\sgn(\xi-x^*)$ gives the update or the
exact crossing.

When both brackets are mark-free, all forward functions are affine
on their two known closed cells. Write
$G_h(x)=\sigma x+\eta$ and
$G_h(\beta)=\sigma'\beta+\eta'$, with positive slopes. The equation
$\phi=0$ is the decreasing line
\[
 \beta=\beta(x)=\frac{G_h(d)-\eta-\eta'-\sigma x}{\sigma'}.
\]
The values $x\in[X_0,X_1]$ for which
$\beta(x)\in[B_0,B_1]$ form a closed interval $I$ containing $x^*$.
The function $x\mapsto \mathcal A(x,\beta(x),d)$ is strictly increasing:
increasing its first argument and decreasing its second both increase
$\mathcal A$ by Lemma~\ref{five:calibration-rates}. Sixty probes make this
function explicit by Lemma~\ref{five:local-functions}(b). Its
unique zero is $x^*$, found offline, and then $b^*=\beta(x^*)$.

It remains to count the rounds and to deal with knots.
Every nonterminal round halves the marks of one agent in one bracket.
Each bracket has at most $5\kappa$ such halvings, so there are at
most $10\kappa$ rounds, each of constant cost, followed by sixty
probes in the final step. If either crossing
coordinate is a knot, its open bracket always contains that mark and
cannot become mark-free. When that mark is tested, the table either
ends the search or moves the other bracket. The one-cell phases
also decide equality exactly. Thus knots cause no extra iterations
or loss of exactness.
\end{proof}

\begin{lemma}\label{five:row-data}
For a non-bottom row with recorded $d$ and $a=a(d)$, all thirty indices
required in Lemma~\ref{five:square-search} can be computed using
$O(\kappa)$ probes.
\end{lemma}
\begin{proof}
For an increasing comparison $\delta$ on $[0,a]$, let its
\emph{change point} be its exact zero in $[0,a]$ if one exists,
$-\infty$ if it is everywhere positive, and $+\infty$ if it is
everywhere negative. The three change points for agent $h$ are found
as follows.

For $L$ versus $R$, compute
$z=G_h^{-1}(1-G_h(d))$ with at most two probes. If $z\leq a$, this
is the change point; otherwise the comparison is negative everywhere
and the change point is $+\infty$.

For $R$ versus $M$, set $z=2G_h(d)-1$. If $z\leq0$,
$\delta_{RM}$ is strictly positive because $b(x)>0$, so the change
point is $-\infty$. Otherwise compute
$\widehat b=G_h^{-1}(z)\in(0,d]$ and locate it, unless it is the
already recorded $d$.
Apply Lemma~\ref{five:x-root} to
$g(x)=\mathcal A(x,\widehat b,d)$ on $[0,a]$. These operations cost
$O(\kappa)$ probes. The signs of this function and of
$\delta_{RM}(x)=G_h(b(x))-z$ agree by
\eqref{five:row-calibration-sign}. Moreover
$g(a)\geq0$ because $\widehat b\leq d=b(a)$. Thus the scalar search
either reports $g(0)>0$, giving change point $-\infty$, or returns
the exact zero, including a possible endpoint zero.

For $L$ versus $M$, use Lemma~\ref{five:row-crossing}, also costing
$O(\kappa)$ probes.

The grid columns $i=0,\ldots,H$ on the row have $x_i=ia/H$, since
$s=i/H$. For a finite change point $z\in[0,a]$, the first nonnegative
and first positive indices are exactly
\begin{equation}\label{five:index-formulas}
 i_{\geq}=\left\lceil\frac{Hz}{a}\right\rceil,\qquad
 i_{>}=\left\lfloor\frac{Hz}{a}\right\rfloor+1.
\end{equation}
For change point $-\infty$ both indices are zero; for $+\infty$
both are $H+1$. Formula~\eqref{five:index-formulas} also handles a
zero at $0$ or $a$. All numbers are exact rationals, so whether the
two indices differ is decided exactly. They differ precisely when a
comparison vanishes at a vertex, that is, at a tie vertex, which must
be kept separate from the tie-free runs. All fifteen change points,
and hence all thirty indices, cost $O(\kappa)$ probes.
\end{proof}

\subsubsection{The protocol and its query bound}\label{five:protocol-section}

The protocol computes $d_0$ by Lemma~\ref{five:rational-roots} and then
runs the supplied-row search of Lemma~\ref{five:square-search}. When a new
non-bottom row $j$ is requested, compute and locate
$d=d_0+(1-d_0)j/H$, compute $a(d)$ by Lemma~\ref{five:x-root},
obtain its thirty comparison indices by Lemma~\ref{five:row-data},
and evaluate by Lemma~\ref{five:vertex-crossing} its run ends and tie
vertices (the ends of the tie-free runs, and the vertices where a
public comparison vanishes; see the proof of
Lemma~\ref{five:abstract}).
Each row costs $O(\kappa)$ probes.
On the top row $d=1$ is known for free. On the bottom row the vector
$T(v)=D(v)-C(N(v))$ of \eqref{five:shift} is constant and needs no
evaluation. During column bisection, apply the
vertex routine only above the bottom row.

Once a cell of nonzero winding is found, obtain its four legal
assignments. If it touches the bottom, the constant bottom vector
alone is insufficient, because the matching needs the legal assignment
there: compute its partition $x=0$, $b=d_0$ by
locating $d_0$, taking $c$ as the second smallest of the five
$G_i^{-1}(f_i)$, and locating $c$. This costs $O(\kappa)$ probes.
Use the same legal assignment at both bottom corners. Find a perfect
matching in the union of the four assignments as in
Lemma~\ref{five:abstract}.

Choose an upper corner of the cell. Locate its four auxiliary cuts
$x,c,b,d$ and obtain their physical images under $\rho$, the
projection $\rho(t)=p_{\lfloor tJ\rfloor}$ of
Lemma~\ref{five:virtual-transfer}, at a further cost of $O(\kappa)$
probes. Output these cuts with the matching.
The cuts are nondecreasing; coincidences produce permitted empty
intervals. Lemmas~\ref{five:quantitative-chart} and
\ref{five:virtual-transfer} give envy at most
$2W/H+12/K\leq\eps/4+12/K<\eps$.

For adaptive cut answers, cache each answer permanently. Complete all
unasked quantile entries by arbitrary admissible answers. Distinct
quantile levels of one agent are necessarily in strict order, so
this gives a consistent completed profile. Lemma~\ref{five:rank}
determines ranks under every such completion: noninteger evaluated
levels force the counts, and at an integer level the uncertain mark
is fetched by \eqref{five:plateau-rule}. All computed affine forms,
roots, indices, and decisions therefore agree with a completed
auxiliary profile. No advance commitment by the cut oracle is needed.

Every probe costs at most nine queries by Lemma~\ref{five:rank}. The
complete accounting is
\[
 \underbrace{O(\kappa^2)}_{d_0}
 +\underbrace{O(\lambda\kappa)}_{\text{rows}}
 +\underbrace{O(\lambda\kappa)}_{\text{column bisection}}
 +\underbrace{O(\kappa)}_{\text{bottom assignment and output}}.
\]
There are at most $\lambda+1$ non-bottom rows and $2\lambda$
additional vertices. Since $\lambda=O(\kappa)$ and
$\kappa=O(\log(1/\eps))$, this proves Theorem~\ref{five:main},
including every sign test, inverse, exact root, and output query.

Five agents enter the argument in two places. Because exactly three
public pieces remain, the vectors $T(v)$ lie in the plane $\Pi$, which
is what makes a planar winding-number search possible; and the
compression step of Lemma~\ref{five:abstract} needs a public piece with
at most one vote, which five votes among three public pieces guarantee.

\subsection{Computation from valuation descriptions}\label{circuit:finite-precision}
For the computational upper bound, we use the monotone implementation
of Appendix~\ref{extension:monotone-section}. It constructs nearby
rational piecewise-linear valuations from lazily evaluated grid values,
so it does not require the input valuations themselves to be piecewise
linear. We show that the work between evaluations also takes polynomial
time.

\begin{proof}[Proof of Theorem~\ref{circuit:five}]
For $\eps=1$, any allocation suffices; hence assume $\eps<1$.
Replace the supplied Lipschitz bound by a power of two and write
$\mathcal L=\max\{1,\log_2(L/\eps)\}$. The parameters
$N,\theta,\eta,E,\Lambda,H$ in
Appendix~\ref{extension:monotone-section} have $O(\mathcal L)$-bit
representations. Every original valuation query has endpoints
$p/N,q/N$, of $O(\mathcal L)$ bits. By the representation assumption,
its exact rational answer can be computed in polynomial time and has
polynomial length. Computing the rounded entry
\[
 w_i(p,q)=\eta\left\lfloor v_i(p/N,q/N)/\eta\right\rfloor
\]
therefore uses polynomial-time integer arithmetic. The entry itself
has $O(\mathcal L)$ bits and is cached. Evaluations of the triangular
interpolant and its linear perturbation then use only rational
arithmetic on the relevant entries.

Lemma~\ref{extension:root-denominators} bounds the denominators of
$d_0,a(d),b(x,d),c(x,d)$ at grid vertices by fixed powers of
$\Lambda$ and $H$. These cuts lie in $[0,1]$, so their numerators
also have $O(\mathcal L)$ bits. Each sign bisection uses
$O(\mathcal L)$ steps on polynomial-length rationals. To recover
the root from a sufficiently short bracket, find its rational of
smallest denominator using continued fractions, which takes
polynomial time in the bit lengths of the bracket endpoints.

The threshold routines in Section~\ref{extension:roots-section}
access their breakpoint lists by index rather than constructing them.
With first endpoint $x$ fixed, the breakpoints of $W_i(x,y)$ in
the cake belong to the two progressions $q/N$ and $x+q/N$, for
integer $q$, restricted to $[0,1]$.
Their sorted union can be indexed by polynomial-bit integer
arithmetic. Binary search therefore takes $O(\log N)$ steps even
though the list may have exponentially many entries. The search with
the second endpoint fixed is symmetric, and the affine tails are
handled directly. The joint bisection for $b,c$ likewise performs
polynomially many operations on polynomial-length rationals.

Finally, Lemma~\ref{five:abstract-search} uses polynomially many
row and vertex computations. The generic point in
Lemma~\ref{five:generic} is fixed independently of the input;
the legal assignments, winding numbers and final matching require
only constant-size combinatorial calculations and rational arithmetic.
There are $O(\mathcal L^3)$ original evaluations and polynomial
additional computation. Appendix~\ref{extension:monotone-section}
proves that the resulting complete connected allocation is
$\eps$-envy-free. Its cuts have polynomial bit length, completing the
proof.
\end{proof}

\section{A query lower bound from seven agents}
\label{seven:section}

For seven or more agents, the number of Robertson--Webb queries cannot be
bounded by any polynomial in $\log(1/\eps)$.  This holds even for
additive valuations with densities bounded above and away from zero by
constants independent of the precision; such valuations are \emph{hungry},
that is, positive on every interval of positive length
(Section~\ref{sec:models}), and Lipschitz.

\begin{theorem}\label{seven:main}
For every fixed integer $n\geq7$ there are constants $c_n,\eps_n>0$ and
$0<\ell_n<u_n<\infty$ with the following property.  Let
$0<\eps<\eps_n$.  Any deterministic Robertson--Webb protocol which, on every
profile of $n$ hungry additive valuations, returns a
complete connected $\eps$-envy-free allocation makes at least
$c_n\eps^{-1/3}$ queries on some profile.
The hard profile can be chosen to have piecewise constant densities in
$[\ell_n,u_n]$.
\end{theorem}

The hard profile may depend on $\eps$. The proof uses the correctness
of the protocol only on profiles with piecewise constant densities in
$[\ell_n,u_n]$, so the bound holds even for protocols that need to
succeed only on such profiles. These valuations are additive and
$u_n$-Lipschitz; note that $u_n\ge1$, since each density integrates to
one. Hence, for every $L\ge u_n$, the bound holds for every protocol, in
particular every value-query protocol, that is given $L$ and succeeds on
all additive, or all monotone, $L$-Lipschitz valuations. For monotone
valuations, whose value of the whole cake need not be one, it holds for
every known Lipschitz bound $L>0$, with constants depending on $n$ and
$L$: if $L<u_n$, run the protocol on the valuations $(L/u_n)v_i$, which
are monotone and $L$-Lipschitz. Each query to them is one query to
$v_i$, and an $\eps$-envy-free allocation for them is
$(u_n\eps/L)$-envy-free for the $v_i$. For additive valuations, which
are normalised, this rescaling is not available. For them we prove the
bound when $L\ge u_n$ and, for value-query protocols, a bound weaker by
a logarithmic factor for every $L\ge n$
(Section~\ref{sec:monotone-to-additive}); some lower limit on $L$ is
necessary (Section~\ref{sec:rw-conventions}).

\subsection{From a hidden path to an envy-free allocation}\label{seven:architecture}

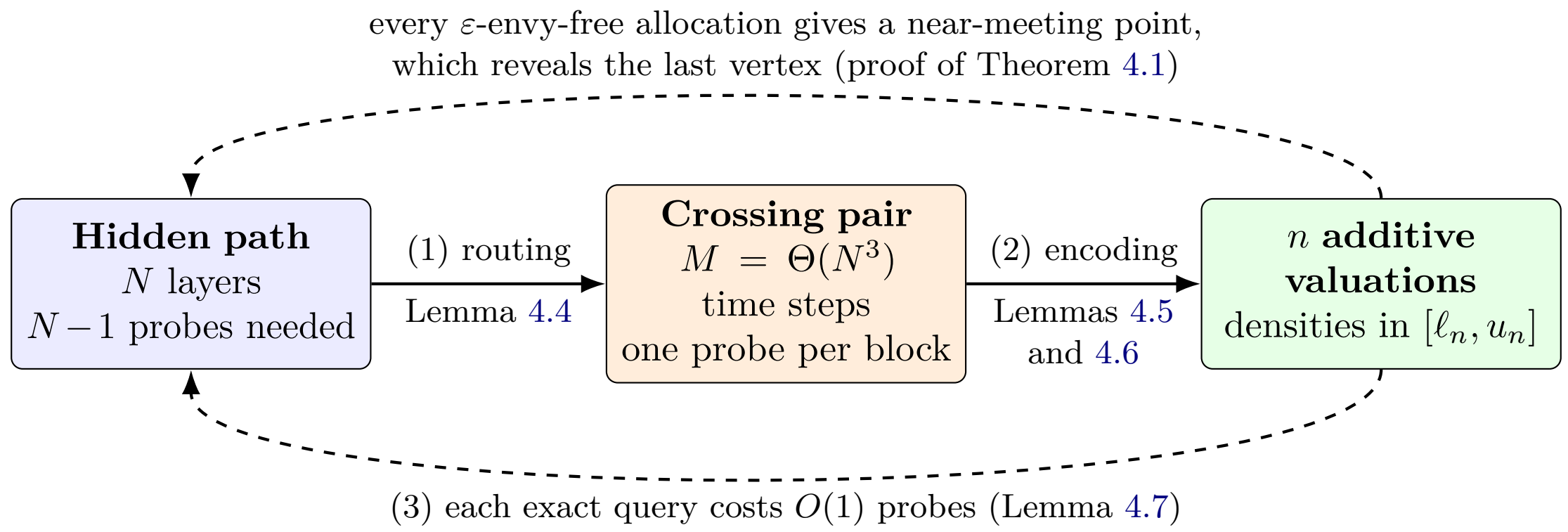
\begin{figure}[ht]
\centering
\begin{tikzpicture}[font=\small,>=Latex,
  stage/.style={draw,rounded corners=3pt,align=center,text width=3.1cm,
                minimum height=1.6cm,inner sep=4pt},
  lab/.style={font=\footnotesize,align=center}]
\node[stage,fill=blue!8] (path) at (0,0)
  {\textbf{Hidden path}\\$N$ layers\\$N-1$ probes needed};
\node[stage,fill=orange!14] (curves) at (5.6,0)
  {\textbf{Crossing pair}\\$M=\Theta(N^3)$ time steps\\one probe per block};
\node[stage,fill=green!10] (cake) at (11.2,0)
  {\textbf{$n$ additive valuations}\\densities in $[\ell_n,u_n]$};
\draw[->,thick] (path) -- node[lab,above]{(1) routing}
  node[lab,below]{Lemma~\ref{seven:source}} (curves);
\draw[->,thick] (curves) -- node[lab,above]{(2) encoding}
  node[lab,below]{Lemmas~\ref{seven:compiler}\\and~\ref{seven:padding}} (cake);
\draw[->,thick,dashed] (cake.south) .. controls +(0,-1.35) and +(0,-1.35) ..
  node[lab,below]{(3) each exact query costs $O(1)$ probes
  (Lemma~\ref{seven:oracle})} (path.south);
\draw[->,thick,dashed] (cake.north) .. controls +(0,1.25) and +(0,1.25) ..
  node[lab,above]{every $\eps$-envy-free allocation gives a
  near-meeting point,\\ which reveals the last vertex
  (proof of Theorem~\ref{seven:main})} (path.north);
\end{tikzpicture}
\caption{Architecture of the query lower bound.  Solid arrows build the
instance; dashed arrows show how a protocol for the cake yields a probing
algorithm for the hidden path.}
\label{seven:pipeline}
\end{figure}

We encode the endpoint of a hidden path in every approximately envy-free
allocation.  The essential requirement is that each exact cake query reveal
only constantly many vertices of the path.  Figure~\ref{seven:pipeline}
summarises the reduction.  In this section a \emph{probe} means a query to
the hidden-path instance, rather than a cake query.

\paragraph{The source problem.}
The \emph{hidden path} problem has $N$ layers, each containing $N$
labelled vertices.  A path visits one vertex in each layer, starting at a
known vertex of the first layer.  A probe at a vertex reveals its
predecessor and successor on the path, or reveals that the vertex is not
on the path.  The task is to find the label of the last vertex; every
deterministic algorithm needs $N-1$ probes (Lemma~\ref{seven:hidden-path}).

\paragraph{Crossing curves.}
The path is encoded in two curves in the unit square.

\begin{definition}\label{seven:crossing-pair}
A \emph{crossing pair} consists of two continuous curves
$R,D:[0,1]\to[0,1]^2$, the \emph{red} and the \emph{blue} curve, with
\begin{equation}\label{seven:endpoints}
 R(0)=(0,0),\ R(1)=(1,1),\qquad
 D(0)=(0,1),\ D(1)=(1,0).
\end{equation}
We call the common parameter \emph{time}.
\end{definition}

The red curve joins two opposite corners of the unit square and the blue
curve joins the other two, so the curves meet.  The crossing-curves problem
of Adler, Daskalakis and Demaine asks for such a meeting point~\cite{ADD}.
We need only an approximate version, for curves that can be inspected at
chosen times: a \emph{near-meeting point}, that is, times $s,t$ with
$\|R(s)-D(t)\|_\infty$ small.  Hollender, Maystre and
Risse~\cite[Section~4.1]{HMR} route two such curves through checkpoints
in a layered graph so that a meeting point yields a solution of the
underlying path problem. We adapt
their routing to a single hidden path with a cubic number of time steps.
After encoding the curves as additive valuations, we show how to answer
each exact cut query with a constant number of path probes.

\paragraph{The reduction.}
\begin{enumerate}[label=(\arabic*)]
\item \emph{Path to curves} (Lemma~\ref{seven:source}).
Route the two curves on an integer grid of width $W=3N^2$.
Each vertex has a short horizontal interval, its \emph{checkpoint
interval}, and a time block.  Red handles odd layers and blue handles
even layers.  Normally the curves follow distinct horizontal lines.
At an internal path vertex, the responsible curve detours through the
checkpoint intervals of its predecessor and successor.  Those neighbours
have removed the relevant part of their own line, so the curves miss
each other there.  Only the last vertex retains its line, leaving a
single intersection whose horizontal coordinate identifies the endpoint.
The routes take $M=\Theta(N^3)$ unit steps.  One vertex probe determines
an entire time block, and hence a curve value at any specified time.
\item \emph{Curves to cake} (Lemmas~\ref{seven:compiler}
and~\ref{seven:padding}).  From a crossing pair of polygonal
curves whose coordinates are $\Lambda$-Lipschitz, and a gain $\delta$ with
$\delta\Lambda\le1/1000$, we build $n$ hungry additive valuations with
densities bounded above and below by positive constants.  The idea is a
fixed-point equation.  The cake is divided into nominal pieces called
\emph{slots}.  Two cuts, near two designated slot boundaries, have
displacements $s$ and $t$ from their nominal positions, and clipping these
to $[0,1]$ gives times $S$ and $T$.  Near those boundaries the valuations
are perturbed by $\delta$ times the coordinates of $R(S)$ and $D(T)$, and
the envy inequalities force
$(s-S,t-T)=\delta\,(D(T)-R(S))$ up to an error $O_n(\eps)$
(equation~\eqref{seven:residual}).  When $s,t\in[0,1]$ the left side
vanishes, so $R(S)$ and $D(T)$ are close; at the ends of the range the
endpoint conditions~\eqref{seven:endpoints} give the same conclusion.
Thus in every complete connected $\eps$-envy-free allocation, whatever its
ownership order, $\|R(S)-D(T)\|_\infty=O_n(\eps/\delta)$.  The valuations
depend on the curves only in two narrow bands, fixed in advance next to
the two designated boundaries.
\item \emph{Queries to probes} (Lemma~\ref{seven:oracle}).  After one
initial probe, which reveals the first red block, every evaluation query
and every exact cut query to these valuations can be answered with a
constant number of probes.  A prefix value depends on at most one point
of one curve, so an evaluation depends on at most two such points.  For a
cut query, a reference curve, which keeps the true first block and follows
the default route in every later block, gives cumulative values within
$O(N/M)$ of the true ones; the exact answer therefore lies in a short
bracket that meets at most two time blocks, and probing their vertices
determines it.
\end{enumerate}

\paragraph{Parameters.}
After normalisation the curves of step~(1) have coordinate Lipschitz
constant $\Lambda=M/W=\Theta(N)$, and we take $\delta=W/(1000M)$.  A
near-meeting point with $\|R(S)-D(T)\|_\infty\le1/(2W)$ already identifies
the unique common point, because two disjoint unit segments of the grid
are at distance at least $1/W$ after normalisation.  Step~(2) achieves this
accuracy when $\eps\le c_n'/M$ for a constant $c_n'>0$ depending only on
$n$, that is, for $N$ of order $\eps^{-1/3}$.  By step~(3), a protocol
making $Q$ queries yields a path algorithm making $O(Q)$ probes, so
$Q=\Omega(N)=\Omega(\eps^{-1/3})$.  The exponent $1/3$ reflects the
duration $M=\Theta(N^3)$ of the curves; the other two steps lose only
constant factors.

\subsection{A locally accessible pair of crossing curves}\label{seven:source-section}

We first specify the hidden-path instance and its two curves.
The source problem is a single-path restriction of the
layered model of G\"o\"os et al.~\cite[Section~3.2]{GHH+}, in the form used by
Hollender, Maystre and Risse~\cite[Section~2.3]{HMR}.  There are $N$ layers,
each containing $N$ labelled vertices.  One vertex in each layer is active:
its label is $L_i\in\{1,\ldots,N\}$, with $L_1=1$.  Consecutive active
vertices point to each other.  Probing a vertex reveals its predecessor
and successor, or null pointers if it is inactive.  The first and last
active vertices have one null pointer.  The task is to output $L_N$.

\begin{lemma}\label{seven:hidden-path}
Every deterministic algorithm that always outputs $L_N$ requires
at least $N-1$ vertex probes in the worst case.
\end{lemma}

\begin{proof}
Maintain an exposed prefix $L_1,\ldots,L_k$, initially $k=1$.
Answer probes outside it with null pointers.  When its final vertex
is probed and $k<N$, choose $L_{k+1}$ among labels not previously probed in
layer $k+1$, reveal this successor and extend the prefix.  Replies
at previously probed vertices stay unchanged.

After $q\leq N-2$ probes, at most $q+1<N$ labels have been exposed.
Each unassigned layer retains at least $N-q\geq2$ unqueried labels.
Completing the path with such labels agrees with every reply and
allows at least two choices of $L_N$.  Thus any output made by then
is wrong for one consistent input.
\end{proof}

For the rest of this section, \emph{public} means independent of the
hidden path or curves, hence known without probes; \emph{private} means
dependent on them.  These terms describe access to the source instance,
not ownership of cake pieces.  The \emph{clock} is the raw time interval
$[0,M]$, divided into one block per vertex and a final block.

\begin{lemma}\label{seven:source}
For $N\geq3$ the hidden path determines two curves satisfying
\eqref{seven:endpoints}, obtained by normalising axis-aligned unit grid
walks on $[0,W]^2$, where $W=3N^2$.  They have a public clock of length
$M=\Theta(N^3)$ and coordinate Lipschitz constant at most $M/W$.
Their only common physical point has raw-grid abscissa
$3((N-1)N+L_N-1)+1$ (and normalised abscissa equal to this number divided
by $W$).  A whole ordinary clock block (the block of a vertex other than
the first), and hence a curve
value at any specified time, is determined by one vertex probe.  The
exceptional first block requires one probe in advance.
\end{lemma}

\begin{proof}
We place the checkpoint intervals and levels, route every block, show
that the walks meet only at the last vertex, and then time the blocks.
Place vertex $(i,j)$ at abscissa
$X_{i,j}=3((i-1)N+j-1)$ and write $X_i=X_{i,L_i}$.
Choose $h=\lfloor W/2\rfloor-2$.  The red \emph{baseline} (its default
line in step~(1)) is $h+4$, its inner level $h+3$ and its outer level $h$; the blue levels are,
respectively, $h+1,h+2,h+5$.  All six levels lie strictly between
$0$ and $W$.

Each walk has a block for each lexicographically ordered vertex.  Normally
a block follows its own baseline across its vertex's checkpoint interval
$[X,X+3]$.  Red is responsible
for odd layers and blue for even layers.  Only at an active, internal
vertex of its own parity does a walk replace this baseline segment by a
\emph{detour}.
Writing $y$ for its baseline, $y'$ for the opposite baseline, and
$\sigma=-1$ for red and $+1$ for blue, the successive waypoints are
\begin{align}\label{seven:route}
 &(X_i,y),\ (X_i,y+\sigma),\ (X_{i-1}+2,y+\sigma),\nonumber\\
 &(X_{i-1}+2,y'+\sigma),\ (X_{i+1}+1,y'+\sigma),\nonumber\\
 &(X_{i+1}+1,y+\sigma),\ (X_i+3,y+\sigma),\ (X_i+3,y).
\end{align}
The detour's long verticals, at $X_{i-1}+2$ and $X_{i+1}+1$, are its
predecessor and successor \emph{ports}, and its
horizontals at the inner level $y+\sigma$ are its \emph{shoulders}.
The final active vertex follows the baseline, like an inactive vertex.
At the first red vertex use instead
\[
 (0,0),\ (X_2+1,0),\ (X_2+1,h+3),\ (3,h+3),\ (3,h+4).
\]
Blue starts with its public descent from $(0,W)$ to $(0,h+1)$ and
then its baseline segment to $(3,h+1)$.  At the end, the public red
connector at $x=W$ goes up to $(W,W)$ and the blue connector goes down
to $(W,0)$.

We now show that the walks have only one common point.  The two
baselines differ.
A detour crosses the opposite baseline only at $X_{i-1}+2$ or
$X_{i+1}+1$, strictly inside an active neighbour's checkpoint interval.
At each nonterminal neighbour the opposite walk has removed its
baseline from $(X_j,X_j+3)$, including $(0,3)$ at the red start.
Thus the only baseline contact is the penultimate vertex's successor
port, at the terminal vertex.  The special first leg obeys the same
rule at $X_2+1$.

Off the baselines, opposite-colour detours at layer distance at least
three have disjoint horizontal spans.  At adjacent layers they belong
to consecutive path vertices.  The shoulder horizontals at $X_i$ occupy
\[
 [X_{i-1}+2,X_i]\quad\hbox{and}\quad[X_i+3,X_{i+1}+1].
\]
The opposite port targeting $i$ lies at $X_i+1$ or $X_i+2$, in the
omitted gap; its other port is outside these ranges.  This excludes
contacts at either shoulder.  Long verticals cannot coincide: ports of
the same role (predecessor or successor) target different parities, and
those of different roles differ modulo three.  The red outer level $h$
lies below every ordinary blue point and the blue outer level $h+5$
above every ordinary red point.  The unit verticals joining a detour to
its own baseline, at $X_i$ and $X_i+3$, do not reach the opposite
shoulder.

Finally the initial red shoulder $[3,X_2+1]$ misses the blue port at
abscissa $2$ (the predecessor port of the layer-$2$ detour) and all
later odd-layer ports.  Its bottom segment is below every
ordinary blue point and ends before the final connector at $W$.
The blue initial connector stops
above the red origin; all later red segments have positive abscissa.
Private ports lie in $0<x<W$, and the final connectors have disjoint
height ranges.  These observations also exclude endpoint contacts.
The unique common point is
$(X_N+1,h+1)$ for even $N$ and $(X_N+1,h+4)$ for odd $N$.
Its abscissa identifies $L_N$ directly.

It remains to time the walks.  Follow each listed segment at unit grid
speed and wait at the last waypoint for the rest of its public block.
An ordinary block has length at most $24N$: its horizontal travel is at
most $18N-11$ and its vertical travel at most eight.
Give the exceptional first block, which combines the initial connector
with the first vertex's route, length $2W$, every remaining
vertex block length $24N$, and the final connectors length $W$.
The first route fits in $2W$ for $N\geq3$, and so
\[
 M=3W+24N(N^2-1)=\Theta(N^3).
\]
Every ordinary block is determined by the probed vertex's two pointers
and parity; the first is known after probing the public start vertex.
Normalise space by $W$ and time by $M$, and linearly interpolate at
unit timestamps.  This gives the endpoints and Lipschitz bound claimed.
\end{proof}

The non-crossing argument relies on one feature of the routing: the
predecessor port in~\eqref{seven:route} lies at $X_{i-1}+2$, inside the
predecessor's checkpoint interval, where the opposite walk has left its
baseline.

\subsection{Encoding two curves in seven additive valuations}\label{seven:encoding-section}

We now encode an arbitrary crossing pair in seven valuations.
We first work on a longer interval, divided into seven
slots (nominal pieces), numbered $0,\ldots,6$ in physical order.  Each
agent values the centres of two slots especially highly.
These \emph{bonuses} restrict ownership; small cumulative-value changes
near the boundaries encode the curves.  We normalise the measures and
interval at the end.

Hollender, Maystre and Risse~\cite[Section~3, proof of Theorem~3.2]{HMR}
use two cuts as curve parameters and encode agreement of the two coordinates
in the envy comparisons of three agents with identical non-monotone
valuations. To implement this idea with additive valuations, we use the
separation of physical slot order from cyclic utility comparisons
introduced by Kawase, Roy and Sanpui~\cite[Section~3]{KRS}.
The \emph{prices} are numbers $q_j$, one
per slot and determined by the cuts, whose differences are the agents' utility
differences between their two favoured slots (see \eqref{seven:prices}).
The weights and four coefficient columns below encode two independently
parametrised curves with seven agents.

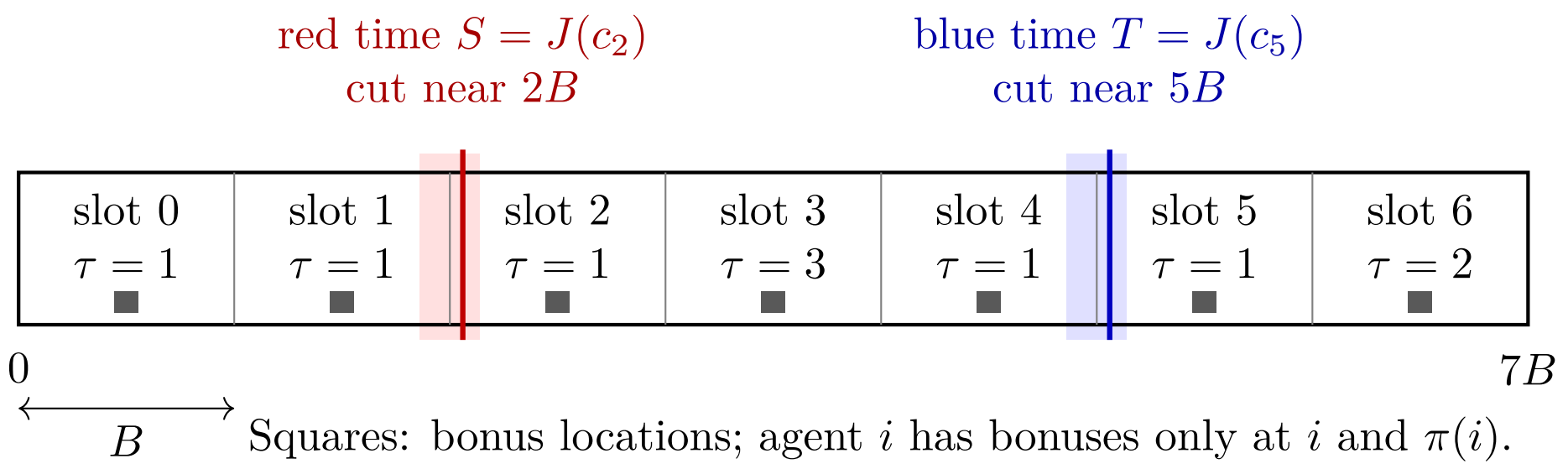
\begin{figure}[htbp]
\centering
\begin{tikzpicture}[x=0.85cm,y=1.2cm,font=\small]
 \fill[red!12] (3.72,-0.10) rectangle (4.28,1.12);
 \fill[blue!12] (9.72,-0.10) rectangle (10.28,1.12);
 \draw[thick] (0,0) rectangle (14,1);
 \foreach \k in {1,...,6} {\draw[gray] (2*\k,0)--(2*\k,1);}
 \foreach \j in {0,...,6} {
   \node at (2*\j+1,0.76) {slot $\j$};
   \fill[black!65] (2*\j+0.89,0.08) rectangle (2*\j+1.11,0.22);
 }
 \foreach \j/\t in {0/1,1/1,2/1,3/3,4/1,5/1,6/2} {
   \node at (2*\j+1,0.40) {$\tau=\t$};
 }
 \draw[red!75!black,very thick] (4.12,-0.10)--(4.12,1.15);
 \draw[blue!75!black,very thick] (10.12,-0.10)--(10.12,1.15);
 \node[red!65!black,align=center] at (4.12,1.75)
   {red time $S=J(c_2)$\\cut near $2B$};
 \node[blue!65!black,align=center] at (10.12,1.75)
   {blue time $T=J(c_5)$\\cut near $5B$};
 \node[below] at (0,-0.08) {$0$};
 \node[below] at (14,-0.08) {$7B$};
 \draw[<->] (0,-0.55)--(2,-0.55);
 \node[below] at (1,-0.55) {$B$};
 \node[align=center] at (8,-0.75)
   {Squares: bonus locations; agent $i$ has bonuses only at $i$ and $\pi(i)$.};
\end{tikzpicture}
\caption{The seven-slot encoding on the raw interval.  Only the bands
near the second and fifth boundaries contain curve information.  The
other bands implement public increasing functions.  Here $B$ is the raw
slot length, $c_k$ is the displacement of the $k$th cut from its nominal
position $kB$, and $J$ clips to $[0,1]$.  The source bands, bonus widths
and small cut displacements are schematic and not to scale.}
\label{seven:layout}
\end{figure}

The two slots eligible for agent $i$ are $i$ and $\pi(i)$, where $\pi$ is the
single cycle
\begin{equation}\label{seven:cycle}
 (0,2,3,4,6,1,5).
\end{equation}
We call slot $i$ agent $i$'s \emph{own} slot and $\pi(i)$ its
\emph{successor} slot.  Assign weights
\[
 \tau=(1,1,1,3,1,1,2)
\]
to the slots, so $\sum_j\tau_j=10$.  They convert prices into
cut displacements.  Physical order determines how cuts affect lengths;
the cycle \eqref{seven:cycle} restricts ownership
(Figure~\ref{seven:layout}).

For a vector $v\in\R^7$, define the two weighted centred prefixes
\begin{equation}\label{seven:projection}
 \mathcal P(v)_k=
 \sum_{j<k}\tau_jv_j-
 \frac{\sum_{j<k}\tau_j}{10}\sum_j\tau_jv_j,
 \qquad k\in\{2,5\}.
\end{equation}
The centring removes a value common to every slot.  These prefixes will
give the displacements of the second and fifth cuts
(equation~\eqref{seven:prefix-equation}).  We use the four columns
\begin{align*}
 a&=(-4,3,-7,2,2,-4,2),&
 A&=(0,2/3,8/3,0,0,0,0),\\
 b&=(-2,2,-2,-2,7,-3,2),&
 C&=(0,0,0,0,-7/2,-3/2,0).
\end{align*}
Direct substitution in \eqref{seven:projection} gives
\begin{equation}\label{seven:four-identities}
 \mathcal P(a)=(-1,0),\quad \mathcal P(A)=(0,1),\quad
 \mathcal P(b)=(0,-1),\quad \mathcal P(C)=(1,0).
\end{equation}

Let $R,D$ be a crossing pair of polygonal curves
(Definition~\ref{seven:crossing-pair}), and let $\Lambda$ bound the
Lipschitz constant of each coordinate.  Write
$J(x)=\max\{0,\min\{1,x\}\}$, and choose a \emph{gain} $\delta$, the
amplitude with which the curves enter the measures, with
\begin{equation}\label{seven:gain}
 0<\delta\leq1/1000,\qquad \delta\Lambda\leq1/1000.
\end{equation}
The second condition will keep the densities positive.  Define two
vector-valued functions, extended by clipping outside $[0,1]$:
\begin{align}
 f^2(x)&=a\bigl(J(x)-\delta R_1(J(x))\bigr)
                    +A\delta R_2(J(x)),\label{seven:f2}\\
 f^5(x)&=b\bigl(J(x)+\delta D_2(J(x))\bigr)
                    -C\delta D_1(J(x)).\label{seven:f5}
\end{align}
Set $f^k=0$ at the other interior boundaries.  The entry $f^k_j(x)$ is
the source term that a displacement $x$ of the cut at boundary $k$ adds
to the price of slot $j$ (see \eqref{seven:prices}).
The coefficients are chosen both for \eqref{seven:four-identities} and for
positive density.

\begin{lemma}[Positive-density encoding]\label{seven:compiler}
For the curves above there are seven hungry additive valuations
and a public constant $Z$, independent of the curves and $\delta$, such that
for every $\eps>0$ with $Z\eps<1/7$, every complete connected
$\eps$-envy-free allocation, in any owner order,
determines $S,T\in[0,1]$ satisfying
\begin{equation}\label{seven:decoder-bound}
 \|R(S)-D(T)\|_\infty\leq 70Z\eps/\delta,
 \qquad Z\eps<1/7.
\end{equation}
These valuations have piecewise constant densities bounded above and below by
positive constants independent of $\Lambda$ and $\delta$.  For rational
polygonal curves and rational $\delta$ the construction is rational
piecewise linear at the level of cumulative values.
\end{lemma}

In \eqref{seven:decoder-bound}, $Z$ is the common raw total mass, and the
hypothesis $Z\eps<1/7$ is repeated alongside the conclusion: it keeps
the envy tolerance, measured in raw units, small enough to localise the
cuts.  In the construction below, the cumulative
values in a band of half-width $L$ around each nominal boundary are
adjusted so that utility differences become differences of prices; the
functions $G_j$ build the weights $\tau_j$ into these values (slope
$1/\tau_j$ on $[-H,H]$, where all displacements will lie), and their clip
at $H$ bounds every adjustment by $A_0$.  The proof builds the measures,
localises the cuts, shows that only two assignments occur, and solves the
resulting near-equalities for the cuts.

\begin{proof}
Put $H=2000$ and
\[
 G_j(x)=x+(1/\tau_j-1)\clip(x,-H,H).
\]
Each $G_j$ is strictly increasing and equals $x/\tau_j$ on $[-H,H]$.
Choose public constants
\begin{equation}\label{seven:physical-constants}
 A_0=9H+50,\quad K=10nA_0,\quad
 L=2n(2K+4A_0+2)+H,\quad
 B=10(L+4A_0+2),\qquad n=7.
\end{equation}
Use the raw interval $[0,nB]$.  Each row, that is, each agent's raw
measure, starts with density one and receives
a mass $K$, uniformly distributed on the middle unit interval of each of
its two eligible slots.  At nominal boundary $kB$, let
\[
 e^k_j=\mathbf1_{j=k-1}-\mathbf1_{j=k},\qquad
 D^k_i=e^k_i-e^k_{\pi(i)}.
\]
Thus the \emph{incidence} $e^k_j$ is $1$ if slot $j$ ends at boundary
$k$, $-1$ if it starts there, and $0$ otherwise, and $D^k_i$ is a
coefficient (not the blue curve $D$).
For displacement $x$ from that boundary define
\begin{equation}\label{seven:local-prefix}
 F_{ik}(x)=
 \begin{cases}
 \displaystyle\frac{e^k_iG_i(x)-e^k_{\pi(i)}G_{\pi(i)}(x)
                      +f^k_i(x)-f^k_{\pi(i)}(x)}{D^k_i},&D^k_i\ne0,\\
 x,&D^k_i=0.
 \end{cases}
\end{equation}
Add $F_{ik}(x)-x$ to the cumulative value on $|x|\leq L$.
Continue this adjustment linearly to zero on each adjoining interval of
length $4A_0$ (a \emph{taper}), and use zero outside.  Boundary bands and
central bonuses are pairwise disjoint by the choice of $B$.

If $D^k_i=0$, the two relevant entries of each column at boundary $k$
agree; we use this \emph{equal-entry check} after \eqref{seven:prices}.
We now bound the slopes, which are the raw densities.  For $k=2$ and $5$,
the nonzero ratios
\[
 \frac{e^k_i/\tau_i-e^k_{\pi(i)}/\tau_{\pi(i)}
             +g_i-g_{\pi(i)}}{D^k_i},
 \qquad g=a\ (k=2),\quad g=b\ (k=5),
\]
are, respectively, $(4,8,10,2)$ in rows $(0,1,2,6)$ and
$(6,10,6,2)$ in rows $(1,3,4,5)$.  On $0<x<1$ these are the public
parts of the slopes in \eqref{seven:local-prefix}.  The remaining slope
has absolute value at most
$\delta\Lambda(14+7)\leq21/1000$.  Outside this interval, the source
terms $f^k$ are constant and the slope is $G_i'$, $G_{\pi(i)}'$, or their
arithmetic mean, and is at least $1/3$.  The same is true at the
other interior boundaries, where $f^k=0$.
The correction has absolute value at most $A_0$: indeed
$|G_j(x)-x|\leq H$ here and $|f_i^k-f_{\pi(i)}^k|<40$.
Every taper therefore retains slope at least $3/4$.
Thus the raw densities are at least $1/3$ and bounded above by a public
constant, including on the bonuses.  Every correction returns to zero,
so each row has total mass
\[
 Z=nB+2K.
\]
Divide by $Z$ and rescale the interval to $[0,1]$.

Now fix an allocation as in the lemma, and let $\eta=Z\eps<1/n$, its
envy tolerance in raw units.  The raw value of any interval differs from
its length by $[-2A_0,2K+2A_0]$.  Comparing the shortest output piece, in its owner's
row, with the longest shows that their lengths differ by at most
$2K+4A_0+\eta$.  As the average length is $B$ and the left endpoint is
fixed, every ordered cut is within $n(2K+4A_0+1)<L$ of its nominal
boundary.  In particular all central bonuses lie intact in their
respective output pieces.

Next, the bonuses force eligibility.  Compare the actual assignment with
the assignment giving slot $i$ to agent $i$.  The length terms cancel
after summation over agents.  Every agent assigned outside its eligible
pair loses $K$ in bonus value.
The corrections at the two ends of each piece, each at most $A_0$ in
absolute value, can favour the actual assignment by at most $4nA_0$, whereas approximate envy-freeness allows a total gain of only
$n\eta$ from the comparison assignment.  Since $K>4nA_0+n\eta$,
every owner is eligible.  The eligibility graph defined by
\eqref{seven:cycle} has exactly two perfect matchings: if any agent takes
its successor slot, the displaced agent must do likewise around the
whole cycle.  Thus either all agents receive their own slots or all
receive their successor slots.

Write the cuts as $kB+c_k$, with $c_0=c_n=0$ (this $c_n$ is not the
constant of Theorem~\ref{seven:main}), and put
\begin{equation}\label{seven:prices}
 d_j=f^2_j(c_2)+f^5_j(c_5),\qquad
 q_j=G_j(c_{j+1})-G_j(c_j)+d_j.
\end{equation}
These are the prices: by \eqref{seven:local-prefix}, agent
$i$'s utility for its own slot minus that for its successor slot
is exactly $q_i-q_{\pi(i)}$.  Constant lengths and equal bonuses cancel; the coefficient of boundary $k$ is
$D_i^k$, including its zero case by the equal-entry check.
In the two matchings, respectively,
$q_i-q_{\pi(i)}\geq-\eta$ for all $i$ or
$q_i-q_{\pi(i)}\leq\eta$ for all $i$.  Following the directed cycle gives
\begin{equation}\label{seven:spread}
 \max_jq_j-\min_jq_j\leq n\eta.
\end{equation}

It remains to solve for the cuts.  Both vector functions in
\eqref{seven:f2}--\eqref{seven:f5} together give $|d_j|\leq20$.  We
cannot have all $q_j>20$, since then strict increase of every $G_j$
in \eqref{seven:prices} would imply
$c_0<c_1<\cdots<c_n=c_0$.  Similarly not all $q_j<-20$.
It follows from \eqref{seven:spread} that $|q_j|\leq20+n\eta$.
Inductively every cut is in $[-H,H]$: the candidate next cut is
$c_j+\tau_j(q_j-d_j)$, and all partial sums of these increments have
absolute value at most $10(40+n\eta)<H$.  On this central interval it
solves \eqref{seven:prices}; strict increase gives uniqueness globally.
We have therefore proved
\[
 c_{j+1}-c_j=\tau_j(q_j-d_j).
\]
Let $\bar q=\sum_j\tau_jq_j/10$.  Summing the last identity and using
$c_n=c_0$ gives $\bar q=\sum_j\tau_jd_j/10$.  Taking the two prefixes
then yields
\begin{equation}\label{seven:prefix-equation}
 (c_2,c_5)=-\mathcal P(d)+E,
 \qquad \|E\|_\infty\leq10n\eta,
\end{equation}
because $|q_j-\bar q|\leq n\eta$.

Set $s=c_2$, $t=c_5$, $S=J(s)$ and $T=J(t)$.  Substituting
\eqref{seven:f2}--\eqref{seven:f5} and \eqref{seven:four-identities}
into \eqref{seven:prefix-equation} gives
\begin{equation}\label{seven:residual}
 (s-S,t-T)=\delta\bigl(D(T)-R(S)\bigr)+E.
\end{equation}
This is the fixed-point equation of step~(2) in
Section~\ref{seven:architecture}.  If $s<0$, then $s-S<0$ and
$D_1(T)-R_1(0)\geq0$; if $s>1$, these signs are reversed.  If
$s\in[0,1]$, the left side $s-S$ is zero.  In every case the first coordinate of $\delta(D-R)$ has absolute value
at most $\|E\|_\infty$.  For the second coordinate use
$D_2(0)=1$ and $D_2(1)=0$ in the same way.  This proves
\eqref{seven:decoder-bound}.  All construction operations preserve
rational piecewise linearity, completing the proof.
\end{proof}

\subsection{Larger fixed populations}\label{seven:padding-section}

The same encoding works for every fixed $n\geq7$.
Adding arbitrary agents to a connected
division need not preserve a lower bound, but here we can increase the
population by splitting a slot that is remote from the two source
boundaries (the second and fifth, which carry the curves).

\begin{lemma}[Padding]\label{seven:padding}
For every fixed $n\geq7$, Lemma~\ref{seven:compiler} holds with the
right-hand side $10nZ\eps/\delta$ and the condition $Z\eps<1/n$,
for a public constant $Z$ depending on $n$.  There remain exactly two
source boundaries, and at any physical point a cumulative value depends
on at most one of the two curves at one time.
\end{lemma}

\begin{proof}
We split slot $0$ and check that each step of the proof of
Lemma~\ref{seven:compiler} survives.
Replace slot $0$ by $r=n-6$ consecutive copies, each of weight $1/r$
and with the old slot's four coefficients.  Replace its occurrence in
the cycle by this chain in physical order.  Total weight remains $10$,
and the two distinguished boundaries become $r+1,r+4$.  Their weighted
prefixes and totals are unchanged, preserving
\eqref{seven:four-identities}.

At either source boundary the copies have zero spatial incidence.
Cycle edges between copies have equal coefficients, and those entering
or leaving the chain retain the old differences and incidences.
Thus all source slope checks are unchanged.  At other boundaries only
positive slopes of the functions $G_j$ occur, bounded below by $1/3$ since $\tau_j\leq3$.
Use the constants \eqref{seven:physical-constants} for the present $n$,
with $A_0=(n+2)H+50$; this bounds the larger correction because
$|G_j(x)-x|\leq\max\{n-7,1\}H$.

Length localisation and owner eligibility follow as before.  The two
matchings again bound the spread of the prices by $n\eta$,
while $|d_j|\leq20$.  Total weight $10$ still places every displacement in $[-H,H]$,
and then \eqref{seven:prefix-equation} has error $10n\eta$.  Equation
\eqref{seven:residual} proves the stated decoder bound.  The two source
bands remain disjoint, all new boundaries are public, and corrections
return to zero, preserving equal total mass; the density bounds now
depend only on $n$.
\end{proof}

\subsection{Exact queries and the precision accounting}\label{seven:oracle-section}

To complete the lower bound, we simulate every allowed cake query.
The clock of
Lemma~\ref{seven:source} has $\Theta(N^3)$ steps, but each Cut query must
be answered with only constantly many probes.  Here $W=3N^2$ and
$M=\Theta(N^3)$ are the grid width and clock length of
Lemma~\ref{seven:source}.

\begin{lemma}\label{seven:oracle}
Apply Lemma~\ref{seven:padding} to the curves of
Lemma~\ref{seven:source}, with $\delta=W/(1000M)$.
After one initial source probe, each Eval query and each exact Cut
query from an arbitrary starting point can be answered with a constant
number of source probes, independent of $N$.
\end{lemma}

\begin{proof}
We treat Eval queries first, and then Cut queries, whose answers we
bracket with the reference curve of step~(3).
Normalisation is public.  Write $V_i$ for agent $i$'s raw cumulative
function.  Each raw prefix $V_i(y)$ is public unless $y$ lies in one of
the two source bands, where its unknown part is $\delta$ times a linear
combination of one curve's coordinates at one time.  Each coefficient
has magnitude at most $14$, including division by $D_i^k$
in~\eqref{seven:local-prefix}.
One vertex probe supplies the whole block containing that time.
Clipped tails and tapers use public endpoint values.  Thus a prefix,
and an Eval using its two endpoints, costs $O(1)$ probes.

Cut queries invert the functions $V_i$, and for them we need a more
accurate reference than deleting the curve altogether.  Keep the true
exceptional initial block, obtained with the one initial probe, and the
public final block.  In every ordinary block replace the detour by the
direct baseline segment, traversed at
unit speed and then followed by waiting, with the same clock endpoints.
Call this the reference curve.  At each time in an ordinary block its
horizontal discrepancy from the true curve is at
most $(6N+3)/W$, and its vertical discrepancy is at most $4/W$.
The first bound follows because both predecessors and successors are
in adjacent layers; the second follows from the six-row strip of levels
$h,\ldots,h+5$.
The reference has the same endpoints and coordinate Lipschitz bound
$M/W$.  Replacing the source curve in each raw prefix by its reference
therefore gives a completely known piecewise linear function
$\widehat V_i$ such that, uniformly in position,
\begin{equation}\label{seven:prefix-reference}
 |V_i-\widehat V_i|\leq
 d_*:=\frac{14(6N+7)}{1000M}.
\end{equation}
Both the true and reference full prefixes have slopes at least $1/4$.
Indeed the positivity check in the proof of Lemma~\ref{seven:compiler}
uses only the endpoint conditions and the coordinate Lipschitz bound,
both preserved by the reference; the smaller constant $1/4$ also
accommodates all joins.
No uncertainty accumulates from earlier bands, since every correction
has public zero total mass.

To answer $\operatorname{Cut}_i(x,\alpha)$, first evaluate the actual
starting value $V_i(x)$ and put $v=V_i(x)+Z\alpha$ in raw mass units.
Rescale $x$ to the raw interval when necessary.  If the target is
outside the total mass this is detected publicly after this evaluation.
Otherwise the exact inverse lies in
\[
 [\widehat V_i^{-1}(v-d_*),\widehat V_i^{-1}(v+d_*)],
\]
with inverse arguments clipped to the total mass range.  This bracket
has physical length at most $8d_*$.  Inside a source band, where the
physical local variable is the curve's normalised time, its raw clock
length is at most
\[
 8Md_*=112(6N+7)/1000<24N.
\]
It consequently meets at most two ordinary blocks, together with the
already known exceptional blocks.  If it meets a band endpoint, its
extension outside the source-time range is public.  The two source
bands are separated by much more than this bracket length, so no
second collection of private blocks is needed.  Probe the at most two
ordinary blocks it meets.  The entire true prefix in the bracket is
then known piecewise linearly; inspect its breakpoints and interpolate
on the appropriate affine piece to obtain the exact inverse.
This procedure includes equality at a breakpoint or block boundary.
It also includes arbitrary starting points, because the actual starting
value $V_i(x)$, rather than its reference value $\widehat V_i(x)$, was
used to form $v$.  Computation on known pieces is free in the query
model.
\end{proof}

\begin{proof}[Proof of Theorem~\ref{seven:main}]
We first assume $\eps$ small compared with $1/M$, so that the decoded
times reveal $L_N$, the label of the last vertex; then we count probes
and choose $N$.
Fix $n$ and let $Z$ be the public raw total mass from
Lemma~\ref{seven:padding}.  Take the hidden path problem with $N\geq3$
layers, the curves of Lemma~\ref{seven:source} and their encoding by
Lemma~\ref{seven:padding}, with $\delta=W/(1000M)$, and suppose
\begin{equation}\label{seven:epsilon-scale}
 \eps\leq\frac{1}{20000nZM}.
\end{equation}
The gain conditions \eqref{seven:gain} hold and $Z\eps<1/n$.
Every successful allocation gives times $S,T$ with
\[
 \|R(S)-D(T)\|_\infty\leq\frac{10nZ\eps}{\delta}
 \leq\frac1{2W}.
\]
Fetch the unit grid edges traversed at these two times, using a constant
number of vertex probes.  Two disjoint closed horizontal or vertical
unit lattice edges, including stationary points, have supremum distance
at least one in raw units.  Thus these two edges intersect.  Their
intersection contains a common lattice vertex, which is the unique
common point from Lemma~\ref{seven:source}; its abscissa gives $L_N$.
This argument covers times at clock endpoints and stationary steps.

By Lemma~\ref{seven:oracle}, a protocol making $Q$ queries gives a
deterministic path algorithm making at most $a_nQ+b_n$ probes, including
the initial probe and output decoding.  The constants are independent of
$N$.  Lemma~\ref{seven:hidden-path} therefore forces
$Q\geq(N-1-b_n)/a_n=\Omega_n(N)$ on some profile.

Finally $M=\Theta(N^3)$.  For every sufficiently small $\eps$, choose
an integer $N\geq3$ of order $\eps^{-1/3}$, with a sufficiently small
constant prefactor that \eqref{seven:epsilon-scale} holds.  The raw
construction constants depend only on $n$, so normalisation gives
positive upper and lower density bounds independent of $N$ and $\eps$.
This proves the theorem.
\end{proof}

\begin{proposition}[Randomized lower bound]\label{seven:randomized}
For every fixed integer $n\ge7$, there are constants $c_n,\eps_n>0$
and $0<\ell_n<u_n<\infty$ with the following property. For every
$0<\eps<\eps_n$, any randomized Robertson--Webb protocol that returns
a complete connected $\eps$-envy-free allocation with probability at
least $2/3$ on every profile of $n$ hungry additive valuations has
expected query cost at least $c_n\eps^{-1/3}$ on some profile with
piecewise constant densities in $[\ell_n,u_n]$.
\end{proposition}

Here the internal random seed is independent of the input, and all
randomized decisions are measurable; exact uniform-real samples are
allowed. Appendix~\ref{app:randomized-lower} proves the proposition
using the same construction and exact-query simulation as
Theorem~\ref{seven:main}.

\section{Computational hardness}\label{circuit:section}

The query lower bound does not itself prove computational hardness:
an algorithm given valuation descriptions may inspect them. We give
polynomial-time reductions for the problem of
Definition~\ref{circuit:problem}. In the program-based representation,
the problem is UEOPL-hard for every fixed
$n\ge7$ and PPAD-complete for every fixed $n\ge13$, even for
additive valuations represented by rational piecewise-linear cumulative
circuits. Both reductions allow every ownership order and decode
\emph{every} allocation meeting the prescribed tolerance.

Recall that a total search problem has a solution on every instance.
A polynomial-time reduction constructs a target instance and converts
any target solution to a source solution in polynomial time.
PPAD consists of problems reducible to End-of-Line: given a directed
graph of indegree and outdegree at most one, represented by circuits,
and one starting endpoint, find another endpoint~\cite{Papadimitriou}.
UEOPL is defined through Unique End of Potential Line, where a potential
increases along paths and violations of the intended uniqueness also
count as answers~\cite{FGMS}. A problem is hard for a class if every
problem in that class reduces to it, and complete if it also belongs
to the class.

\paragraph{Promises and finite answers.}
The problem is total on inputs satisfying its valuation and evaluation
promises. An algorithm need not check these promises. A reduction to
the cake problem must produce a promised instance; a reduction from
it need only work on promised instances. PPAD membership is understood
in this sense.

Short rational answers exist: round the cuts of an exact envy-free
allocation down to a dyadic grid of mesh at most $\eps/(8L)$.
This preserves their order and changes each interval value by at most
$\eps/4$, hence each envy comparison by at most $\eps/2$.
The grid coordinates have $O(\log(L/\eps))$ bits. We may therefore
restrict outputs to polynomial-length rational cuts, as required for
an NP search problem. A proposed answer is verified in polynomial time
by evaluating its interval values and checking the envy inequalities.
Our hardness decoders also work for longer rational answers, in time
polynomial in their length.

\paragraph{Circuit descriptions of the hard instances.}
A rational piecewise-linear circuit of size $S$ can be evaluated at a
rational point of bit length $b$ in time polynomial in $S+b$. To see
this, choose an attaining predecessor at each min/max gate. The
resulting affine circuit has coefficients whose denominators divide
the product of all constant denominators and whose magnitudes are
at most $2^{O(S)}$: expanding into paths gives at most exponentially
many terms, each with total factor bit length $O(S)$. The same bound
applies to every intermediate gate value. It also bounds the absolute
slopes on every affine piece, supplying a Lipschitz bound with
$O(S)$ bits. Thus each cumulative circuit constructed below gives
a polynomial-time interval evaluator, together with a valid bound $L$,
with polynomial description length. The valuation and all-output
decoding arguments are unchanged by this representation.

\subsection{Membership in PPAD}\label{circuit:membership-section}

\begin{proposition}\label{circuit:membership}
For every fixed $n\ge2$, $\eps$-$\ConnEF_n$
(Definition~\ref{circuit:problem}) belongs to PPAD.
\end{proposition}
\begin{proof}
Use the standard discretised Sperner reduction for approximate
connected cake cutting~\cite{DQS,HR}. Its colouring evaluates interval
values at rational grid points and compares them to choose favourite
pieces. For a supplied Lipschitz bound, a mesh of size polynomial
in $\eps/L$ suffices; its coordinates have polynomial bit length,
and the colouring is polynomial-time computable under our evaluation
assumption. A fully labelled cell yields an assignment with envy at
most the prescribed tolerance. This is the PPAD-membership argument
for the computational model of~\cite[Section~3]{HR}.

For the boundary colouring, we can ensure strict positivity on intervals
of positive length using the perturbation of Hollender and
Rubinstein~\cite[Section~4]{HR}. On legal intervals, replace $v_i(a,b)$ by
\[
 \widetilde v_i(a,b)=(1-\delta)v_i(a,b)+\delta(b-a),
 \qquad \delta=\eps/4.
\]
These valuations are hungry, have a supplied polynomial-bit Lipschitz
bound, and differ from the originals by at most $\delta$.
Apply the reduction with tolerance $\eps/2$. Its output has envy at
most $\eps/2+2\delta=\eps$ for the original valuations. Every
evaluation and every grid operation has polynomial bit cost.
\end{proof}

\subsection{UEOPL-hardness from seven agents}\label{ueopl:section}

Hollender, Maystre and Risse proved that the Jordan Curve problem is
UEOPL-hard~\cite[Theorem~4.1]{HMR}. We reduce that problem to connected
cake cutting. Lemma~\ref{seven:compiler} encodes the two grid walks in
seven additive valuations with densities bounded above and away from zero.
Every sufficiently accurate envy-free allocation, in any ownership order,
determines a vertex visited by both walks. Lemma~\ref{seven:padding}
extends the construction to every fixed population larger than seven.

\begin{theorem}\label{ueopl:hardness}
In the program-based representation, for every fixed $n\ge7$,
$\eps$-$\ConnEF_n$
(Definition~\ref{circuit:problem}) is UEOPL-hard, even for additive
valuations given by rational piecewise-linear cumulative circuits.
The instances constructed by the reduction have densities bounded above and below by
positive constants that depend only on $n$.
\end{theorem}

\paragraph{The source problem.}
We use the form of Jordan Curve with prescribed endpoints and valid
grid steps, which suffices for hardness by~\cite[Section~2.2 and
Theorem~4.1]{HMR}. Two Boolean circuits give the positions of a red
and a blue walk at each integer time in $\{0,\ldots,M-1\}$. Both walks
stay in $\{0,\ldots,W\}^2$, with $W\ge1$. The red walk goes from
$(0,0)$ to $(W,W)$ and the blue walk from $(0,W)$ to $(W,0)$.
At each step a walk stays in place or moves one unit along a grid edge.
A solution is a pair of timestamps, one for each walk, at which their
positions agree.

Following Hollender, Maystre and Risse~\cite[Section~3 and Remark~3.7]{HMR},
we join consecutive positions by straight segments. To represent the resulting
walks as rational piecewise-linear circuits, we need to interpolate a table
given by a Boolean circuit. The one-dimensional interpolation lemma below
follows by specialising the construction of Fearnley, Goldberg, Hollender
and Savani~\cite[preprint, Section~1.2 and proof of Lemma~5.1]{FGHS}.
Filos-Ratsikas, Hansen, H\o gh and Hollender give a related construction
using piecewise-linear pseudo-circuits~\cite[preprint, Proposition~3.2]{FRHH}.
We give a direct proof for the signed tables used in both reductions.

\begin{lemma}\label{ueopl:table-circuit}
Let a Boolean circuit, on input $i\in\{0,\ldots,2^a\}$, output a
signed $q$-bit integer $A_i$, and let $\rho$ be a rational number.
In polynomial time one can construct a rational piecewise-linear
circuit whose value at $i$ is $\rho A_i$ and which is affine on every
$[i,i+1]$.  Its size and constant bit lengths are polynomial in
$a,q$, the input circuit size, and the bit length of $\rho$.
The statement allows several output coordinates, affine
rescaling of the domain, and constant continuation outside the
domain by clipping.
\end{lemma}

\begin{proof}
The idea is to treat even and odd sample points separately and add the
two results.  For each parity at most one interpolation weight is
positive at a given point, and we obtain the weighted sample value by
running the Boolean circuit,
with minimum, maximum and subtraction in place of its gates, on the
address bits of that sample point, each multiplied by the weight.

Set $K=2^a$ and first take $\rho=1$.  The triangular function
$\max\{0,1-|u-k|\}$ is the interpolation weight of sample $k$.
Group these functions by the parity $p\in\{0,1\}$ of their centres:
\[
 z_p(u)=1-\operatorname{dist}(u-p,2\mathbb Z),
 \qquad 0\le u\le K.
\]
When $z_p(u)>0$, a unique centre $p+2h$ is within distance one of $u$.
For noninteger $u$, the two parities select its neighbouring samples,
with weights summing to one.

The periodic distance functions have small circuits.  For a dyadic
integer period $P$ and a bounded argument $v$, add a fixed multiple of
$P$ to put $v$ in $[0,2^tP]$, where $t=O(a+1)$ below.
For $j=t,t-1,\ldots,1$, replace $v$ by
$\min\{v,2^jP-v\}$.  The current interval is halved at each step,
and reflection about an integer multiple of $P/2$ preserves
distance to $P\mathbb Z$.  The result lies in $[0,P]$, where its
distance is $\min\{v,P-v\}$.  Thus these distances use $O(a+1)$
rational piecewise-linear gates and polynomial-bit constants.

Fix a parity $p$.  We now construct, for each address bit, its
Boolean value multiplied by the weight $z=z_p(u)$ of the selected
centre.  The least significant bit is $pz$.
For address bit $j+1$, where $0\le j<a$, let $m=2^j$.
This is bit $j$ of $h$ in the address $p+2h$.
Its one-runs comprise $m$ successive
centres, with first centre $p+2m$ and last centre $p+4m-2$,
repeated with period $P=4m$.  The corresponding triangular supports form the
interval from $p+2m-1$ to $p+4m-1$, modulo $P$.
Set
\[
 c=p+3m-1,\qquad
 H(u)=\max\{0,m-\operatorname{dist}(u-c,P\mathbb Z)\}.
\]
Outside these support intervals $H=0$.  Inside, $H$ is distance to the
nearer endpoint, and hence at least the triangular weight $z$.
It follows that $\min\{z,H\}$ is exactly the desired address bit
multiplied by $z$.  This argument also covers points where $z=0$.
Use $a+1$ address bits to include the right endpoint $K$.

Convert the table circuit to AND, OR and NOT gates.
For this parity replace the Boolean constant one by $z$,
AND by minimum, OR by maximum, and NOT of a value $v$ by $z-v$.
At a point with $z>0$, induction over the gates says that every
gate equals its Boolean value on the selected address times $z$.
At $z=0$, every gate is zero.  Signed output bits can therefore
be assembled using their usual signed binary place values.
This gives $z_p(u)$ times the sample value at the selected centre.
Adding the two parity outputs gives the exact polygonal
interpolant, including at integer endpoints.

There are $O(a^2)$ gates for the scaled address bits and a constant
number of copies of the given Boolean circuit, followed by $O(q)$
output gates per coordinate.  All powers of two used have
polynomial bit length.  Finally multiply by $\rho$.
An affine change of input and the map
$u\mapsto\min\{K,\max\{0,u\}\}$ give the other assertions.
\end{proof}

\paragraph{Recovering a shared grid vertex.}
Let $T$ be the smallest power of two at least $\max\{W,M-1\}$.
Pad both walks to time $T$ by evaluating their position circuits at
$\min\{i,M-1\}$ on input $i$. The comparison and selection use
polynomially many Boolean gates, since $\log T$ is bounded by the
bit lengths of the timestamps and coordinates. Apply
Lemma~\ref{ueopl:table-circuit} to the padded coordinates, divide by
$W$, and rescale time to $[0,1]$. The resulting red and blue curves
$R,D$ form a crossing pair (Definition~\ref{seven:crossing-pair}).
Every coordinate is $T/W$-Lipschitz, and $1/W$ is one grid unit.

\begin{lemma}\label{ueopl:near-crossing}
Given rational $s,t\in[0,1]$ with
$\|R(s)-D(t)\|_\infty<1/W$, one can find in polynomial time a pair of timestamps
at which the original walks visit the same vertex.
\end{lemma}

\begin{proof}
Locate the two unit-time intervals containing $Ts$ and $Tt$,
using the final interval if one of these times equals $T$.
Evaluate the two endpoints of each interval in the integer grid.
Each interval traces a closed axis-aligned unit lattice segment,
possibly a stationary point.  Disjoint such segments have
$\ell_\infty$ distance at least one: in some coordinate their
closed projections are disjoint integer-endpoint intervals, so
the gap is at least one.  The assumed distance is smaller than
one in grid units, so the two segments intersect.
Two intersecting axis-aligned unit lattice segments share a
lattice endpoint. Test the at most four endpoint pairs to find equal
positions. Replacing each resulting timestamp $i$ by $\min\{i,M-1\}$
gives timestamps in the original walks with those same positions.
\end{proof}

\begin{proof}[Proof of Theorem~\ref{ueopl:hardness}]
We choose the parameters so that the decoding bound of the additive
encoding falls below the threshold $1/W$ of
Lemma~\ref{ueopl:near-crossing}.  Fix $n\ge7$, and apply
Lemma~\ref{seven:padding} (for $n=7$ this is Lemma~\ref{seven:compiler}) to
the curves $R,D$ above, with
\[
 \delta=\frac{W}{1000T},\qquad
 \eps=\frac{\delta}{1000nZW},
\]
where $Z$ is the common raw total mass in that construction for
$n$ agents.  These positive rational parameters have polynomial
bit length.  Since $T\ge W$, we have $\delta\le1/1000$ and
$\delta(T/W)=1/1000$, which is condition~\eqref{seven:gain} with
$\Lambda=T/W$, as required for positive densities.
The localisation condition of Lemma~\ref{seven:padding} on raw envy
also holds: $Z\eps=\delta/(1000nW)<1/n$.

Every allowed $\eps$-envy-free allocation, regardless of its
assignment, supplies the two clipped rational times of that lemma,
called $S,T$ there and $s,t$ here.  Explicitly, let $\beta$ be the raw
slot length of the construction for $n$ agents (called $B$ in the proof
of Lemma~\ref{seven:compiler}), and let $J(x)=\max\{0,\min\{1,x\}\}$
clip to $[0,1]$.  The curves are carried by the $(n-5)$th and
$(n-2)$th cuts (the two source boundaries in the proof of
Lemma~\ref{seven:padding}), so cuts $0=y_0\le\cdots\le y_n=1$
give
\[
 s=J\bigl(n\beta y_{n-5}-(n-5)\beta\bigr),\qquad
 t=J\bigl(n\beta y_{n-2}-(n-2)\beta\bigr),
\]
rationals computable in time polynomial in the bit length of the cuts.
By Lemma~\ref{seven:padding},
\[
 \|R(s)-D(t)\|_\infty
 \le \frac{10nZ\eps}{\delta}
 =\frac1{100W}<\frac1W,
\]
and Lemma~\ref{ueopl:near-crossing} then returns a solution of the Jordan Curve instance.

The cumulative circuits have polynomial size: compose the four curve
circuits with the affine operations, minimum, maximum, clipping and
fixed linear extensions of the proof of Lemma~\ref{seven:compiler}.
For $n>7$, Lemma~\ref{seven:padding} replaces slot~$0$ by $n-6$ copies
of the same form, so the number of these operations still depends only
on $n$.  On each taper the correction is a public constant,
computed from $R(0)$, $R(1)$, $D(0)$ and $D(1)$, times a clipped affine
function of the input, so no product of two computed values occurs.
The total mass $Z$ depends only on $n$, so normalisation is rational
scalar multiplication.  For every source instance the lemma also gives
additive valuations with densities bounded above and below by positive constants
that depend only on $n$, so the constructed instance satisfies the
promises of Definition~\ref{circuit:problem} and the density bounds of
the theorem. The construction and the decoder both run in polynomial
time. The claimed UEOPL-hardness follows from~\cite[Theorem~4.1]{HMR}.
\end{proof}

\subsection{PPAD-completeness from thirteen agents}\label{circuit:ppad-section}

To obtain PPAD-hardness, we replace the Jordan Curve source by a
two-dimensional discrete Brouwer problem,
whose instance is a succinct three-colouring of an exponentially large
grid.  The reduction has three steps.  First, we turn the colouring
into a field of small signed direction vectors on a refined grid, whose
zero labels mark solutions of the source problem.  Second, we construct thirteen
valuations so that every approximately envy-free allocation forces two
numbers $X,Y$, read off from the displacements of two marked cuts, to
satisfy two feedback equations: the point $(X,Y)$ is
nearly unchanged by one clipped step along the field, which is read at
three sample addresses carried by further cuts
(Lemma~\ref{circuit:thirteen-module} and the proof of
Theorem~\ref{circuit:ppad13}).  Third, such a nearly unchanged
point lies in a refined square with a zero corner, which yields a
solution of the source (Lemma~\ref{circuit:thirteen-source}).  As in the
seven-agent encoding of Section~\ref{seven:encoding-section}, the
physical slots, separate cyclic comparison order, and near-equal prices
(the quantities $q_j$ of \eqref{seven:prices})
build on Kawase, Roy and Sanpui~\cite[Section~3]{KRS}.

\begin{theorem}\label{circuit:ppad13}
In the program-based representation, for every fixed $n\ge13$,
$\eps$-$\ConnEF_n$
(Definition~\ref{circuit:problem}) is PPAD-complete, even for additive
valuations given by rational piecewise-linear cumulative circuits.
The instances constructed by the reduction have positive densities, with
bounds that may depend on the source instance.
\end{theorem}

\paragraph{A planar source with compatible directions.}
For the first step we use the PPAD-complete two-dimensional discrete Brouwer problem of
Chen and Deng \cite[Definition~9 and Theorem~2, ECCC version]{CD}.  Its grid side may
be exponential in the input length; a Boolean circuit returns one of
three colours at a grid vertex, as in Mehta's circuit formulation
of this problem~\cite[preprint, Definition~6]{Mehta}. The prescribed boundary colours are
hardwired. A solution is a unit square containing all three colours.
Deng, Qi and Saberi used this planar source to prove hardness for
three cake-cutting agents whose preferences may depend on the whole
division~\cite[preprint, Section~5, Theorem~3]{DQS}. Here we must encode
the colours with additive interval valuations.

Deng, Qi and Zhang~\cite[Lemma~2]{DQZ} refine a Brouwer grid so that a
zero of a direction-preserving map identifies a tricoloured square
(see also Deng, Qi, Saberi and Zhang~\cite[Lemma~3.2]{DQSZ}).
For our table, we refine the grid by affine interpolation.
Associate the bottom colour with $(0,1)$, the remaining left-edge colour
with $(1,0)$, and the other boundary colour with $(-1,-1)$.  Split every
coarse square along the same diagonal and interpolate these vectors
affinely on the two triangles.  On a $100$-fold refinement label a vertex
by zero if the interpolated vector has sup norm at most $1/10$;
otherwise label it by the signed unit vector along the interpolated vector's coordinate of largest magnitude,
with fixed tie breaking.  Write the refined side as $m$.  Each component
varies by at most $4/100$ across a fine square: each triangle has
coordinate slopes of magnitude at most two, and the common-edge
interpolation is continuous.  Opposed labels on an axis would require
variation greater than $2/10$.  Thus every fine square is
\emph{sign-compatible}: it has no two nonzero labels opposed on one
axis.  No boundary label points out of the grid.

A zero label gives a tricoloured coarse square, because the convex hull
of any two
of the three vectors $(1,0)$, $(0,1)$ and $(-1,-1)$ stays at sup distance
at least $1/3$ from zero, including on a shared edge.
The refined-label circuit uses at most three coarse labels
and a constant-precision interpolation on coordinate residues modulo
$100$.  At the upper edge use the last coarse square and residue $100$.
It has size polynomial in the original circuit size and $\log m$.

\paragraph{Three samples and two feedback equations.}
Daskalakis, Goldberg and Papadimitriou~\cite[Section~5]{DGP} sample nearby
points and average their directions to control errors near grid boundaries.
Chen, Deng and Teng~\cite[preprint, Section~8.1]{CDT} refined this technique, and
Mehta~\cite[preprint, Section~3]{Mehta} used sixteen samples and two
clipped coordinate updates in a planar reduction. For our thirteen-agent construction, we store a point
$(u,v)\in[0,m]^2$ of the refined grid as two numbers $X,Y$. Two selectors
determine three table addresses. As we show below, at least two of the
addresses interpolate directions from the square containing the point. Put
\[
 \tau=\frac{1}{16(m+1)^2},\qquad
 \delta=\frac{\tau}{10^8},\qquad
 \alpha=\eta=\frac{\delta}{10^8}.
\]
Here $\tau$ is the spacing of table addresses, $\delta$ the length of
each sampled direction vector (one step along the field adds three
samples), $\alpha$ a small perturbation of the slot
weights, and $\eta$ the envy tolerance in raw units, as in
Section~\ref{seven:encoding-section}
($\eps=\eta/Z$ in Lemma~\ref{circuit:thirteen-module}).
For $0\le v\le m$, write $v=j+t$, $j=\lfloor v\rfloor$,
$0\le t<1$, and put
$j_k(v)=j+\clip_{[0,1]}(3t-k)$ for $k=0,1,2$;
at $v=m$ set every $j_k=m$.  These \emph{slow indices}, which will select
table rows, are continuous nondecreasing
staircases and $j_0+j_1+j_2=3v$.  In the physical units of cut
displacements, define
\begin{align*}
 X&=\tau u,& Y&=3\tau(m+1)v,&&0\le u,v\le m,\\
 S_0(Y)&=\tau(m+1)j_0(v),&
 S_1(Y)&=\tau(m+1)j_1(v).
\end{align*}
Call $S_0,S_1$ the \emph{selectors}.  Extend them constantly outside the
legal interval $0\le Y\le3\tau(m+1)m$; they have slopes zero or one.  For each coordinate $p$, let $T_p(a)$ be
the linear interpolation of the table
\[
 T_p\bigl(\tau((m+1)r+s)\bigr)
       =\delta\,D_p(s,r),\qquad 0\le r,s\le m,
\]
where $D_p(s,r)$ is the $p$-component of the refined direction label at
vertex $(s,r)$.  Its constant extension is
$2\delta/\tau$-Lipschitz.  The three addresses are
\[
 A=X+S_0(Y),\qquad B=X+S_1(Y),\qquad
 C=X+Y-S_0(Y)-S_1(Y).
\]
Since $j_0+j_1+j_2=3v$, these equal $\tau((m+1)j_k(v)+u)$ for $k=0,1,2$,
so the $k$-th address reads row $j_k(v)$ of the table at column $u$.
At most one slow index is nonintegral, so at least two of these
addresses read horizontal interpolations of directions in the containing
unit square of $(u,v)$.  We call these the \emph{good samples}; the
remaining sample, if any, is \emph{bad}.

\begin{lemma}\label{circuit:thirteen-source}
If the two clipped updates
\[
 X\mapsto\clip_{[0,\tau m]}\bigl(X+T_x(A)+T_x(B)+T_x(C)\bigr),
 \quad
 Y\mapsto\clip_{[0,3\tau(m+1)m]}
                    \bigl(Y+T_y(A)+T_y(B)+T_y(C)\bigr)
\]
have \emph{residual} (absolute difference between the updated and the current value) at most $\delta/100$ in each coordinate, the containing
unit square of $(u,v)$ has a zero direction corner.  Its four corners
can be checked in polynomial time.
\end{lemma}
\begin{proof}
Suppose none of those corners has label zero; we show that the samples
then move one stored coordinate by at least $.47\delta$, which clipping
cannot cancel.  On each axis choose the
sign occurring in the corners, or zero if the axis is absent.
Sign-compatibility, which excludes opposed nonzero labels on one axis,
makes this choice consistent.  Each good sample interpolates two corner
labels, so its scalar product, in units $\delta$, with this sign vector
is one.  The possible bad sample has one-norm at most one.

Discard an axis when its chosen sign points towards a face of the clipping
box within $3\delta$ of the stored coordinate $X$ or $Y$.  Near a horizontal face, the
distance in slow-index units is at most
$3\delta/(3\tau(m+1))<1/3$; every good slow index is then the boundary
row, so it has no outward component on that axis.  Near a vertical
face, a good horizontal interpolation gives at most
$3\delta/\tau\le.03$ weight to the nonboundary column.  The boundary
column has no outward component, so discarding that axis loses at
most $.03$ from a good sample.  Therefore the dot product retained
after discarding dangerous axes is at least $2(.97)-1=.94$.
One retained signed component is at least $.47$.

The corresponding stored coordinate is farther than $3\delta$ from
its favourable face.  Its update, whose magnitude is at most
$3\delta$, is not clipped in that direction.  Thus at least one
residual is at least $.47\delta$, a contradiction.  For points on a grid
line, use the cell starting at that line, except at the upper boundary;
an interpolation endpoint has zero weight on the other horizontal
vertex.  This also covers coincident samples on a boundary.
\end{proof}

\paragraph{The thirteen-slot data.}
We now give the data for the thirteen valuations of the second step.  The
cake will have thirteen consecutive \emph{slots}, indexed $0,\ldots,12$.
As in Section~\ref{seven:encoding-section}, each agent values two slots
so highly that every approximately envy-free allocation gives her one of
them; as these pairs form one cycle, all slot prices are then nearly
equal, which determines each cut displacement, up to a small error,
through weighted centred prefixes of the slots' \emph{charges}, the
readout-dependent parts of the prices.
Seven cuts are \emph{marked}: each stores its displacement from the
boundary between two slots, and a \emph{readout} is a specified function
of that displacement; since the charges depend on readouts, the
equations for the cuts become feedback equations.  The four readouts
$X,Y,S_0,S_1$ use source-independent functions and, as in
Section~\ref{seven:section}, are called \emph{public}; the six table
readouts carry the source colours and are called \emph{private}.
Agent $i$ will compare slots $i$ and $\sigma(i)$,
where $\sigma$ is the single cycle
\[
             (0,3,8,4,1,6,11,9,2,5,12,10,7).
\]
The seven marked cuts, or \emph{markers} (public for $X,Y,S_0,S_1$ and
private for $A,B,C$), indexed by the slot ending there, are
\[
      (a_X,a_Y,a_{S_0},a_{S_1},a_A,a_B,a_C)=(9,4,2,5,7,1,11).
\]
The table gives limiting slot weights $t^0_j$ and public charge
coefficients $s_{jh}$, the coefficient of readout $h$ in the charge of
slot $j$; two coefficients grow as $\alpha^{-1}$, and we call their slots
$6$ and $10$, both of limiting weight zero, \emph{singular}.
\begin{center}
\begin{tabular}{rccccc}
\toprule
$j$&$t^0_j$&$s_{jX}$&$s_{jY}$&$s_{jS_0}$&$s_{jS_1}$\\
\midrule
0&$1/2$&$-1$&$1$&$-1$&$-1$\\
1&$1/2$&$-1$&$-1$&$1$&$-1$\\
2&$1$&$1$&$-1$&$0$&$1$\\
3&$0$&$-1$&$1$&$-2$&$-1$\\
4&$0$&$-1$&$2$&$1$&$-1$\\
5&$0$&$1$&$-2$&$-1$&$2$\\
6&$0$&$-1$&$-1$&$1$&$\alpha^{-1}-2$\\
7&$1$&$-1$&$1$&$-1$&$-1$\\
8&$1/2$&$-1$&$1$&$1$&$-1$\\
9&$1/2$&$1$&$-1$&$1$&$1$\\
10&$0$&$\alpha^{-1}-2$&$1$&$-1$&$-1$\\
11&$1$&$-1$&$-1$&$1$&$1$\\
12&$1$&$1$&$1$&$-1$&$-1$\\
\bottomrule
\end{tabular}
\end{center}
Let $h_j=t^0_j$ if positive, and $h_j=\alpha$ otherwise, and use the
actual weight
\[
        t_j=\frac{h_j}{1+\alpha h_j},\qquad
        G_j(c)=c+(t_j^{-1}-1)\clip_{[-H,H]}(c),\quad H=1000.
\]
The reciprocals $t_j^{-1}=h_j^{-1}+\alpha$ will be local slopes.  Their
added $\alpha$ ensures strict increase at a tight public edge, a cycle
edge at a public marker whose density would otherwise vanish.

For weights $t$, let the centred weighted prefix of a charge vector
$d$, the analogue of \eqref{seven:projection}, be
\[
 P^t_k(d)=\sum_{j=0}^k t_jd_j-
    \frac{\sum_{j=0}^k t_j}{\sum_{j=0}^{12}t_j}
                     \sum_{j=0}^{12}t_jd_j.
\]
The limiting weighted public columns are
$w_{jh}=\lim_{\alpha\downarrow0}t_js_{jh}$; thus they include mass
one at slot $10$ for $X$ and at slot $6$ for $S_1$.
Their full sums vanish.  Direct prefix summation gives the following
matrix of \emph{negative} centred prefixes, with rows in the marker order
$X,Y,S_0,S_1,A,B,C$ and columns in the readout order $X,Y,S_0,S_1$:
\begin{equation}\label{circuit:public-matrix}
 \mathcal R=\begin{pmatrix}
 1&0&0&0\\0&1&0&0\\0&1&0&0\\0&1&0&0\\
 1&0&1&0\\1&0&0&1\\1&1&-1&-1
 \end{pmatrix}.
\end{equation}
Here centred prefixes of weighted limiting columns mean
$\sum_{j\le k}w_{jh}-(\sum_{j\le k}t^0_j/6)\sum_jw_{jh}$,
since $\sum_jt^0_j=6$.  By Lemma~\ref{circuit:thirteen-module}, the rows
of $\mathcal R$ give the public parts of the marker displacements: $X$, $Y$, $Y$,
$Y$, $X+S_0$, $X+S_1$ and $X+Y-S_0-S_1$.

For each private marker use two classes $U,V$ of slots and a matrix $\mathsf L$
as follows:
\begin{center}
\begin{tabular}{ccll}
\toprule
marker&$U$&$V$&$\mathsf L$\\
\midrule
$A$&$\{7\}$&$\{0,3\}$&
$\begin{pmatrix}-2&1\\-2&-2\end{pmatrix}$\\[2mm]
$B$&$\{6,11,9\}$&$\{5,12,10,7,0,3,8,4\}$&
$\begin{pmatrix}2&0\\-2&2\end{pmatrix}$\\[2mm]
$C$&$\{9,2,5\}$&$\{10,7,0,3,8,4,1,6\}$&
$\begin{pmatrix}1/2&-5/2\\-3/2&3/2\end{pmatrix}$\\
\bottomrule
\end{tabular}
\end{center}
For component $p\in\{x,y\}$ set
$r_{jap}=\mathsf L_{1p}\boldsymbol1_{j\in U}+\mathsf L_{2p}\boldsymbol1_{j\in V}$; we call
these coefficients the \emph{private lifts}.
Computing their negative centred prefixes with weights $t^0$ gives
$(1,0)$ at $(a_X,a_Y)$ for $p=x$, and $(0,1)$ for $p=y$.
These equalities follow by substituting the displayed sets, weights,
and matrices.  Each coefficient has magnitude at most $5/2$.

To specify limiting prefixes even at the singular slots, hold the
observed readouts $z$ fixed (their values $z_h$, $z_{ap}$ at the markers,
named in Lemma~\ref{circuit:thirteen-module}) and define
\[
 w_j(z)=\sum_h w_{jh}z_h+
                  \sum_{a,p}t^0_jr_{jap}z_{ap},\qquad
 P^0_k(z)=\sum_{j=0}^k w_j(z)-
          \frac{\sum_{j=0}^k t^0_j}{6}\sum_{j=0}^{12}w_j(z).
\]
Below, $P^0_k(d)$ denotes this prefix with the observed readouts fixed.
This definition retains the nonzero limiting charge at a slot of zero
limiting weight.

\begin{lemma}[Physical realisation]\label{circuit:thirteen-module}
The displayed data produce thirteen normalised hungry additive
valuations such that every $\eps$-envy-free complete connected allocation,
where $\eps=\eta/Z$ and $Z$ is the raw mass defined below, gives the
following relations.  Let $z_X,z_Y$ be the clipped readouts at their
marked cuts, let $z_{S_0},z_{S_1}$ be the selector readouts, and let
$z_{ap}=T_p(c_{a})$ be the six private readouts.  With
\[
 d_j=\sum_{h\in\{X,Y,S_0,S_1\}}s_{jh}z_h
                 +\sum_{a\in\{A,B,C\},p\in\{x,y\}}r_{jap}z_{ap},
\]
the displacement $c_k$ at every marked cut satisfies
\[
            |c_k+P^0_k(d)|\le E,
       \qquad E=3000\alpha+196\eta.
\]
Here $P^0$ uses the limiting weighted columns, including the two
singular weighted limits.  At $X,Y$ these prefixes contain exactly the
corresponding three private direction components; at the remaining
markers their private contribution has magnitude at most $672\delta$.
The conclusion holds for every assignment of pieces to agents.
\end{lemma}
\begin{proof}
We use the same three steps as in Lemma~\ref{seven:compiler}:
construct positive densities, force the cuts and ownership choices, and
recover cut displacements from nearly equal prices.  The larger
coefficients here require a weighted estimate in the last step.  Set
\[
 A_0=\frac{3000}{\alpha},\quad K=10n(A_0+1),\quad
 L=2n(2K+4A_0+2)+H,\quad B_0=10(L+4A_0+3),\quad n=13.
\]
The raw cake is $[0,nB_0]$, with slots $[jB_0,(j+1)B_0]$ and a
unit-width bonus interval at the centre of each slot.  Agent $i$ receives
baseline density one and additional bonus mass $K$ on slots $i$ and $\sigma(i)$.
Her other corrections are supported around the interior boundaries.
At boundary $k$, between slots $k$ and $k+1$, put
$e_{ik}=\boldsymbol1_{i=k}-\boldsymbol1_{i=k+1}$ and
$D_{ik}=e_{ik}-e_{\sigma(i),k}$.  In the local coordinate $c$, the
displacement from boundary $k$, prescribe
the agent's prefix, apart from its constant at that boundary, as
\begin{equation}\label{circuit:local-prefix}
 \frac{e_{ik}G_i(c)-e_{\sigma(i),k}G_{\sigma(i)}(c)
                      +d_i^k(c)-d_{\sigma(i)}^k(c)}{D_{ik}}
                    \quad(D_{ik}\ne0).
\end{equation}
Boundaries with public or private markers are \emph{public} or
\emph{private}; the others are \emph{unused}.  At an unused boundary $d^k=0$.  At a public boundary $a_h$, set
$d_j^k(c)=s_{jh}f_h(c)$, where $f_X,f_Y$ are the clips to the legal
intervals $[0,\tau m]$ and $[0,3\tau(m+1)m]$ of
Lemma~\ref{circuit:thirteen-source}, and $f_{S_0}=S_0$, $f_{S_1}=S_1$ are
the selectors.  At a private boundary $a$,
set $d_j^k(c)=\sum_p r_{jap}T_p(c)$.  If $D_{ik}=0$, prescribe
the unmodified prefix $c$.  Inspection of the displayed columns and
sets shows $d_i^k=d_{\sigma(i)}^k$ whenever $D_{ik}=0$:
the coefficient is constant on each of the two open paths left after
removing the two slots incident to the marker from the cycle.

Use \eqref{circuit:local-prefix} on the boundary band $|c|\le L$, subtract its baseline
$c$, and taper that correction linearly to zero over a further
$4A_0$ on each side.  Its magnitude is below $A_0$: the baseline
correction is at most $H(1/\alpha+2)$ and the readout correction
is below $12/\alpha$.  Outside $|c|\le H$ all readouts and
$G_j(c)-c$ are constant, so these tapers join continuously.  Supports
are disjoint and miss every bonus interval by the choice of $B_0$.

On a public boundary band, write $\lambda\in[0,1]$ for the public readout
slope.  The central density for $D_{ik}\ne0$ is
\[
 \frac{e_{ik}/t_i-e_{\sigma(i),k}/t_{\sigma(i)}
                  +\lambda(s_{ih}-s_{\sigma(i),h})}{D_{ik}}.
\]
At $\lambda=0$ it is a convex combination of reciprocal weights,
all at least one.  At $\lambda=1$ it equals
$\alpha+b/\alpha+d$.  The pairs $(b,d)$, listed in cycle-edge order
for the four edges having nonzero incidence at each public marker,
are
\[
 \begin{array}{c|rrrr}
 X&(0,4)&(0,2)&(0,3)&(0,1)\\
 Y&(1,1)&(1,3)&(1,1)&(1,3)\\
 S_0&(1,1)&(1,3)&(0,0)&(0,2)\\
 S_1&(0,1)&(0,3)&(1,1)&(1,3).
 \end{array}
\]
This is obtained by substituting the thirteen rows of the coefficient
table into the preceding formula.  Every endpoint slope is at least $\alpha$, hence the same
is true for intermediate $\lambda$.  On a private boundary band, each
component loses at most $5(2\delta/\tau)$ from the baseline density;
both together lose at most $20\delta/\tau<1/2$.  The rest of the
band has density one.  Each exterior taper changes the density by
at most $1/4$, and bonuses add only positive density.  Thus every raw
cumulative is strictly increasing.  The total raw mass is the common
public number $Z=nB_0+2K$ because boundary corrections vanish at the
cake endpoints.  A raw cumulative $V_i$ becomes
$F_i(x)=V_i(nB_0x)/Z$ on $[0,1]$.  Its density is
$(nB_0/Z)V_i'(nB_0x)$, so positivity survives both normalisations.

We next force the equations for every $\eps$-envy-free allocation,
whatever its assignment.
For every raw interval, its value minus its physical length lies
between $-2A_0$ and $2K+2A_0$.  Comparing a shortest assigned piece
with a longest piece bounds their length difference by
$2K+4A_0+\eta$.  The mean piece length is $B_0$, so every cut
lies within $n(2K+4A_0+\eta)<L$ of its prescribed boundary, and
all bonus intervals are captured intact.

Compare the total welfare with the assignment in which agent $i$
receives the $i$-th physical piece, containing bonus interval $i$.  Approximate envy-freeness
loses at most $n\eta$ in this comparison.  The physical lengths cancel.
An agent given neither the piece with bonus $i$ nor that with bonus $\sigma(i)$ loses $K$
in bonus value, while the total possible compensating boundary
correction is at most $4nA_0$.  Since $K>4nA_0+n\eta$, there can
be no such agent.  In the eligibility graph $i\mapsto\{i,\sigma(i)\}$,
a perfect matching must be either all identity or all successor:
choosing a successor forces the next successor around the single cycle.

Set $c_{-1}=c_{12}=0$ and, for $j=0,\ldots,12$,
\[
                  q_j=G_j(c_j)-G_j(c_{j-1})+d_j.
\]
These are the slot prices, as in \eqref{seven:prices}.
Here $c_j$ is the displacement of the interior cut ending slot $j$ from
$(j+1)B_0$.  Formula~\eqref{circuit:local-prefix}, multiplied by its
incidence, telescopes to the exact own-minus-successor difference
$q_i-q_{\sigma(i)}$; equal bonuses and slot lengths cancel.
For either full matching, approximate envy gives the appropriate
directed inequality on each cycle edge, and hence
$\max q_j-\min q_j\le n\eta$.

It remains to locate the displacements; the large unweighted singular
coefficients require a weighted argument.  Freeze the observed $d_j$ and put
$e_j=q_j-q_0$, $T=\sum_jt_j$, and
\[
 Q=\frac{\sum_jt_j(d_j-e_j)}{T},\qquad
 c'_k=\sum_{j=0}^k t_j(Q+e_j-d_j).
\]
Then $c'_{12}=0$.  Every true weighted public coefficient differs
from its limiting weighted coefficient by at most $4\alpha$;
at a singular slot this follows directly from
$(1-2\alpha)/(1+\alpha^2)$.  The same bound holds for a private
coefficient.  Since $|z_h|\le1$, $|z_{ap}|\le\delta$, the sum of
absolute weighted observed charges is below $60$, and $1\le T\le14$.
Consequently
\[
        |c'_k|\le120+2n\eta T<1000=H.
\]
Thus $G_j(c)=c/t_j$ at these proposed displacements, and $c'$ is a
\emph{closing chain} with prices $Q+e_j$: started at zero with these
slot prices, it returns to zero at $k=12$.  For fixed
$d,e$ and starting
displacement zero, each recurrence
$c_j=G_j^{-1}(G_j(c_{j-1})+Q+e_j-d_j)$ is strictly increasing in
the common price $Q$.  Its final displacement is strictly increasing
as well.  A closing price is therefore unique.  The observed
displacements form a closing chain with prices $q_j=q_0+e_j$, so they
must be this central chain.  Summing its linear recurrence gives
\[
                    |c_k+P^t_k(d)|\le196\eta.
\]

Finally, the uniform comparison of true and limiting prefixes uses only
short sums.  The total weight error is at most
$14\alpha$, a prefix weight fraction changes by at most
$28\alpha$, and the sum of absolute limiting weighted charges is
below $57$.  Each weighted slot charge changes by at most
$40\alpha$.  Thus the centred-prefix error is less than
\[
       (2\cdot14\cdot40+28\cdot57)\alpha<3000\alpha.
\]
This proves the asserted $E$ bound.  Matrix
\eqref{circuit:public-matrix} and the displayed private lifts give the
claims at $X,Y$.  At any other marker, the absolute private centred
prefix is at most $2\cdot14\cdot24\delta=672\delta$, a loose bound
covering the six coefficients of magnitude at most $5/2$.
\end{proof}

\begin{proof}[Proof of Theorem~\ref{circuit:ppad13}]
We first show that every $\eps$-envy-free allocation decodes to a source
solution, then that the construction and the decoding take polynomial
time, and finally add agents.
Apply Lemma~\ref{circuit:thirteen-module} to the succinct planar source.
Let $X=z_X$, $Y=z_Y$, which already belong to their legal intervals.
At the selector cuts, the actual selector inputs differ from $Y$ by at most
$672\delta+E$.  The selectors are one-Lipschitz.  At the three
private cuts, their public parts are $X+z_{S_0}$,
$X+z_{S_1}$, and $X+Y-z_{S_0}-z_{S_1}$.  Including their own
private contributions, their differences from the ideal addresses
$A,B,C$ are at most $2016\delta+3E$.  Each table component changes by at most
$(2\delta/\tau)(2016\delta+3E)$.  At $X,Y$, the relations are
$c_X=X+\sum_a z_{ax}+O(E)$ and
$c_Y=Y+\sum_a z_{ay}+O(E)$, whereas the readouts equal the respective
clips of $c_X,c_Y$.  Hence the residuals of $X$ and $Y$ under the
clipped updates of Lemma~\ref{circuit:thirteen-source} are bounded by
\[
        E+3(2\delta/\tau)(2016\delta+3E)<\delta/100.
\]
Lemma~\ref{circuit:thirteen-source} recovers a refined zero corner,
which recovers a tricoloured source square.

It remains to check that the reduction constructs small circuits,
rather than an exponentially long table.  Apply the table-interpolation
Lemma~\ref{ueopl:table-circuit} to the
Boolean circuit for $D_p$ at
row-major address $(m+1)r+s$.  Binary division by $m+1$ computes
$r,s$ on integer addresses in polynomial time and is compiled to a
Boolean circuit.  Extend the finite sample table constantly to the
next power-of-two address range and clip the real input to its true
last address.  The lemma constructs exactly the interpolated table
using rational scalar gates, addition/subtraction and min/max.  Clipping
its argument gives the constant extensions.  No floor gate on a real
input and no multiplication of computed wires is used.

The public selectors $S_0,S_1$ also have compact circuits.  Put
$\ell=\tau(m+1)$.  On one period of length $3\ell$, the $k$-th selector,
$k=0,1$, is $\ell\clip_{[0,1]}(Y/\ell-k)$; across periods its ordinate
increases by $\ell$.  Subtract the drift $Y/3$.  The residual is periodic
and affine between integer multiples of $\ell$; its sample values at those
knots form a fixed finite rational alphabet determined by the index
modulo three.  The same table lemma, followed by restoring $Y/3$ and
clipping to the legal input interval, therefore gives an exact
polynomial-size circuit.  The baseline functions $G_j$, clips, bonuses,
and exterior tapers have rational affine/min/max descriptions.
Continuous pieces on public intervals are joined by summing their
clamped-input increments.  On a taper its endpoint correction is a
constant; any source label needed there is evaluated at construction
time.  Thus no variable times variable is hidden in this joining step.
All coefficients and public endpoints have $O(\log m)$ bits, apart
from the given source description, and the number of physical
intervals is constant.  This proves polynomial construction.

With $Z$ the raw total mass of Lemma~\ref{circuit:thirteen-module}, the
normalised tolerance is $\eps=\eta/Z=\Theta(m^{-4})$ since
$\delta=\Theta(m^{-2})$, $\alpha=\eta=\delta/10^8$, and
$Z=\Theta(\alpha^{-1})$.  Its binary encoding is polynomial in
$\log m$, although its inverse need not be polynomial in the source
length.  For a returned rational answer, undo length normalisation:
$c_k=nB_0y_{k+1}-(k+1)B_0$, with the slot width $B_0$ from the proof of
Lemma~\ref{circuit:thirteen-module}.  Compute the two clipped readouts, divide
by $\tau$ and $3\tau(m+1)$ to obtain $u,v$, and check the four
corners of the containing fine cell.  Use the cell starting at a grid
line, except at the upper endpoint.  These are polynomial-time exact
rational operations and constantly many calls to the label circuit.
The decoding argument above applies to every returned assignment and rational
cut vector satisfying the target tolerance. Polynomial-length rational
outputs exist by the grid-rounding argument at the start of this
section; membership is Proposition~\ref{circuit:membership}.

For each larger fixed population, as in
Section~\ref{seven:padding-section}, replace slot zero by $n-12$
consecutive clones in both the physical order and the comparison
cycle.  Divide its limiting weight $1/2$ equally and retain all its
charge coefficients.  It borders no marked cut.  Every old limiting
prefix and the private lifts are preserved; internal clone edges have
equal charges, and both external cycle edges retain their old
differences.  The comparison graph remains one cycle.  At an active
band, one at a marked boundary, every clone has zero incidence, so the
positivity calculations there are unchanged.  Apply the same formula for $t_j$.
Its errors are $O_n(\alpha)$ and are absorbed by choosing the two
constant ratios $\tau/\delta$ and $\delta/\alpha$ sufficiently large
for that fixed $n$.  Rechoose the construction constants $H,A_0,K,L,B_0$
accordingly.  The central
closing-chain argument and all-output decoder then apply verbatim,
with constants depending on $n$.  The exponent and polynomial
encoding bounds are unchanged.
\end{proof}

If instead every cumulative function is given by an explicit list of
breakpoints, every fixed population is solvable exactly in polynomial
time~\cite[Theorem~5]{AFGST}.  The hardness above
therefore uses succinct cumulative circuits; the approximate hardness
additionally uses binary-encoded accuracy.

\appendix
\section{Four agents}\label{four:section}

This appendix proves the four-agent bound of \Cref{tab:oracle}, for
additive valuations.

\begin{theorem}\label{four:main}
For four agents with additive valuations and every $0<\eps\leq1$, a
deterministic Robertson--Webb protocol finds a complete connected
$\eps$-envy-free allocation with $O(\log(1/\eps))$ queries.
\end{theorem}

Together with the deterministic logarithmic lower bound of
Br\^anzei and Nisan~\cite[Section~8, Corollary~1]{BN}, this gives the optimal
dependence on precision for four agents with additive valuations.

We adapt the five-agent calibration idea to four agents, choosing the
middle cut so that the two middle pieces, again called \emph{private}, can go
to distinct agents as \emph{global favourites}, pieces they value at
least as much as every other piece. The calibrated divisions form a
curve $C$, on which envy-freeness fails only when two agents compete for
the same end piece (Lemma~\ref{four:assignment}). Each agent's
preference between the end pieces switches once along $C$; the switches
split the agents into two pairs, whose feasibility boundaries cross
inside a region called the band, and the crossing yields an envy-free
division (Lemmas~\ref{four:bracket} and~\ref{four:crossing-output}).
The search thus reduces to a fixed number of consecutive
one-dimensional searches; placing one search inside another would give
the logarithmic-squared dependence of the previous bound~\cite{HR}. We first
describe the geometry for hungry valuations (strictly increasing
cumulative functions), then carry out each search exactly, with
$O(\log(1/\eps))$ queries, on a lazily accessed quantile interpolation
of arbitrary additive valuations (Lemma~\ref{four:localisation}).

\subsection{Two middle pieces and two end pieces}

Throughout the geometric argument each cumulative function $F_i$ is
strictly increasing. Extend it to the real line by $F_i(t)=t$ outside
$[0,1]$, so that all the inverses below are defined.
Write a division as
\[
 L=[0,x],\qquad P=[x,c],\qquad U=[c,b],\qquad R=[b,1].
\]
The end pieces $L,R$ are called public and the middle pieces $P,U$
private. For each agent put
\[
 f_i=\max\{F_i(x),1-F_i(b)\},\qquad
 \ell_i=F_i^{-1}(F_i(x)+f_i),\qquad
 u_i=F_i^{-1}(F_i(b)-f_i).
\]
Here $f_i$ is agent $i$'s largest public value, and $\ell_i$ and $u_i$
are her \emph{lower} and \emph{upper marks}.
Thus $P$ is worth at least $f_i$ precisely when $c\geq\ell_i$,
and $U$ is worth at least $f_i$ precisely when $c\leq u_i$.
Let $\ell_{(2)}$ denote the second smallest lower mark and $u_{[2]}$
the second largest upper mark, and define
\[
 \mathcal B(x,b)=\ell_{(2)}-u_{[2]}.
\]
This function, the four-agent counterpart of the calibration function
of Section~\ref{five:chart-section}, is continuous, strictly
increasing in $x$ and strictly decreasing in $b$. Indeed, each
$\ell_i$ strictly increases in $x$ and weakly decreases in $b$, whereas
each $u_i$ has the opposite monotonicities, with strict increase in $b$.
Since $\mathcal B(0,1)=-1$ and $\mathcal B(x,x)>0$, its zero set in $0\leq x\leq b\leq1$
is a strictly increasing continuous curve, denoted $C$ and called the
\emph{calibrated curve}, from the side
$x=0$ to the side $b=1$. At a point of $C$ set
$c=\ell_{(2)}=u_{[2]}$. The inequalities $\ell_i\geq x$ and $u_i\leq b$
ensure that this is a physical division.

\begin{lemma}\label{four:assignment}
At every point of $C$, the private pieces can be given to distinct agents
as global favourites, leaving each other agent with a public piece as a
global favourite. If the division is not envy-free, those two remaining
agents must both strictly prefer the same public piece.
\end{lemma}
\begin{proof}
Since $c=\ell_{(2)}=u_{[2]}$, each of $P$ and $U$ has at least two
\emph{weak claimants}, agents $i$ who value it at least $f_i$, and at
most one \emph{strict claimant}, who values it more. The
case analysis in the proof of Lemma~\ref{five:private-owners} uses only
these counts, so it applies verbatim: the private pieces go to distinct
agents as global favourites, and every strict claimant receives one.
Hence each of the two remaining agents values both private pieces at
most her largest public value, so her public favourites are global
favourites. If their public favourites include different pieces, a
complete envy-free assignment follows. Otherwise both strictly prefer
the same end piece, since a public tie would permit assigning the ends
distinctly.
\end{proof}

Call the two possible failures in this lemma a \emph{right collision},
where both remaining agents strictly prefer $R$, and a \emph{left
collision}, where both strictly prefer $L$. We do not assume that
these types change monotonically along $C$; instead,
$F_i(x)+F_i(b)$ strictly increases along $C$, from below $1$ to above
$1$. Agent $i$ therefore switches from preferring $R$ to preferring $L$
at one point $t_i$, where $b=F_i^{-1}(1-F_i(x))$. It is found by solving
\begin{equation}\label{four:switch}
 \mathcal B\bigl(x,F_i^{-1}(1-F_i(x))\bigr)=0,
 \qquad 0\leq x\leq F_i^{-1}(1/2).
\end{equation}
The endpoint signs are negative and positive, respectively.

Order these four switch points along $C$ as $t_{(1)},\dots,t_{(4)}$,
breaking ties by agent index. Let $A$ be the first two agents and $B$
the last two. The \emph{band} is the closed region
where both agents in $A$ weakly prefer $L$ to $R$, and both agents in $B$
weakly prefer $R$ to $L$; it meets $C$ in the arc from $t_{(2)}$ to
$t_{(3)}$.

At a non-envy-free point of this band, a right collision must leave
exactly $B$ as the \emph{public pair} (the agents left for the end
pieces): an agent in $A$ who favoured $R$ would
be tied between the end pieces and could take $L$. Similarly, a left
collision must leave $A$ as the public pair. At $t_{(2)}$ an agent in
$A$ is tied between the end pieces, so a left collision is impossible
there: that agent does not strictly prefer $L$; the symmetric statement holds at
$t_{(3)}$.
Test these two divisions and stop if either is envy-free. Otherwise,
writing $t_{(j)}=(x_j,b_j)$, we have a right collision at $t_{(2)}$,
a left collision at $t_{(3)}$, and
\[
 x_2<x_3,\qquad b_2<b_3.
\]
Coincident switch points would permit both end pieces to be assigned.

\subsection{A crossing inside the band}

In the remaining case the private pieces go to $A$ at $t_{(2)}$ and to
$B$ at $t_{(3)}$; we show that the feasibility boundaries of the two
pairs, defined below, cross inside the band, and that the crossing
yields an envy-free division. Inside the band the marks simplify to
\[
\begin{array}{lll}
 a\in A:&\ell_a=F_a^{-1}(2F_a(x)),
 &u_a=F_a^{-1}(F_a(b)-F_a(x)),\\
 e\in B:&\ell_e=F_e^{-1}(F_e(x)+1-F_e(b)),
 &u_e=F_e^{-1}(2F_e(b)-1).
\end{array}
\]
Use these formulas also outside the band to define two explicit curves:
\begin{align*}
 c_A(x)&=\min_{a\in A}F_a^{-1}(2F_a(x)),&
 b_A(x)&=\min_{a\in A}F_a^{-1}(F_a(c_A(x))+F_a(x)),\\
 c_B(b)&=\max_{e\in B}F_e^{-1}(2F_e(b)-1),&
 x_B(b)&=\max_{e\in B}F_e^{-1}(F_e(c_B(b))-1+F_e(b)).
\end{align*}
Both $b_A$ and $x_B$ are continuous and strictly increasing. Their
meaning comes from the equivalences
\begin{align}
 \max_{a\in A}u_a\leq\min_{a\in A}\ell_a
 &\quad\Longleftrightarrow\quad b\leq b_A(x),\label{four:A-feasible}\\
 \max_{e\in B}u_e\leq\min_{e\in B}\ell_e
 &\quad\Longleftrightarrow\quad x\geq x_B(b).\label{four:B-feasible}
\end{align}
In the band the first condition says that some middle cut makes $L$
a global favourite for both agents in $A$; the second says the same
for $R$ and $B$. Strict inequality in either display prevents that
pair of agents from taking both private pieces as favourites. We call
the curves $b=b_A(x)$ and $x=x_B(b)$ the \emph{feasibility boundaries}
of $A$ and $B$.

To find the part of the boundary $b=b_A(x)$ that lies in the band, let
$x_-$ be the unique zero of
\[
 \min_{a\in A}\{F_a(x)+F_a(b_A(x))\}-1,
\]
and let $x_+$ be the unique zero of $\max_{e\in B}\{F_e(x)+F_e(b_A(x))\}-1$.
These two searches are bracketed by $0$ and
$M=\max_iF_i^{-1}(1/2)$. At $0$ each expression is $-1$; at $M$ it
is positive because $b_A(M)>M$. A minimising or maximising agent is
tied between the end pieces at the corresponding zero.

\begin{lemma}\label{four:bracket}
If neither $t_{(2)}$ nor $t_{(3)}$ is envy-free, then $x_-<x_+$ and
\[
 x_- - x_B(b_A(x_-))\geq0,\qquad
 x_+ - x_B(b_A(x_+))\leq0.
\]
Every point $(x,b_A(x))$ with $x\in[x_-,x_+]$ lies in the band.
\end{lemma}
\begin{proof}
The collisions at $t_{(2)}$ and $t_{(3)}$ give the signs at $x_-$ and
$x_+$, respectively. At $t_{(2)}$, the private owners are $A$ and the
public pair is $B$.
Thus \eqref{four:A-feasible} cannot hold strictly and
\eqref{four:B-feasible} holds:
$b_2\geq b_A(x_2)$ and $x_2\geq x_B(b_2)$.
If $x_-<x_2$, the agent in $A$ tied at $t_{(2)}$ would strictly prefer
$R$ at $(x_-,b_A(x_-))$, a contradiction. Hence $x_-\geq x_2$.
If $b_A(x_-)>b_2$, the agent in $A$ tied at $x_-$ would strictly
prefer $R$ at $t_{(2)}$, again a contradiction. Consequently
\[
 x_B(b_A(x_-))\leq x_B(b_2)\leq x_2\leq x_-.
\]
At $t_{(3)}$ the symmetric argument gives
$x_+\leq x_3$, $b_A(x_+)\geq b_3$ and
$x_B(b_A(x_+))\geq x_B(b_3)\geq x_3\geq x_+$.
In particular, $b_A(x_-)\leq b_2<b_3\leq b_A(x_+)$, proving
$x_-<x_+$. The definitions of these endpoints give the band conditions
throughout the interval. They also imply $0\leq x\leq b_A(x)\leq1$:
$b_A(x)\geq c_A(x)\geq x$, and an agent in $B$ tied at $x_+$
certifies $b_A(x_+)\leq1$.
\end{proof}

We can therefore find a zero $x_*$ of the continuous function
$x-x_B(b_A(x))$ on this bracket. No monotonicity of this function is
needed. Write $z_*=(x_*,b_*)$, where $b_*=b_A(x_*)$, for the resulting
intersection of the two feasibility boundaries, and
abbreviate $c_A=c_A(x_*)$ and $c_B=c_B(b_*)$. At $z_*$,
\[
 c_A=\min_A\ell_a=\max_Au_a,\qquad
 c_B=\max_Bu_e=\min_B\ell_e,
 \qquad x_*\leq c_A,c_B\leq b_*.
\]

\begin{lemma}\label{four:crossing-output}
If $\mathcal B(z_*)\geq0$, a constant number of comparisons gives an envy-free
division with outer cuts $x_*,b_*$. If $\mathcal B(z_*)<0$, the unique point of
$C$ on the horizontal line $b=b_*$ is envy-free.
\end{lemma}
\begin{proof}
At $z_*$, which lies in the band by Lemma~\ref{four:bracket}, the middle
cut $c_A$ makes $L$ a global favourite for both agents in $A$, and
$c_B$ does the same for $R$ and $B$.
If $c_A=c_B$, set $c$ to their common value. Give $P$ to an agent in
$A$ attaining $\min_A\ell_a$, $L$ to the other agent in $A$, $U$ to
an agent in $B$ attaining $\max_Bu_e$, and $R$ to the other agent.
All receive global favourites.

Suppose $c_B<c_A$. Choose $a_1\in A$ with $u_{a_1}=c_A$ and
$e_1\in B$ with $\ell_{e_1}=c_B$, and denote the other two agents
by $a_0,e_0$. For $c\in[c_B,c_A]$, agent $a_1$ favours $U$ and
agent $e_1$ favours $P$. Their partners can take $L,R$ precisely
when $u_{a_0}\leq c\leq\ell_{e_0}$. If
$u_{a_0}\leq\ell_{e_0}$, choose $c=\max\{c_B,u_{a_0}\}$.
Otherwise the second smallest lower mark is $\ell_{e_0}$ and the
second largest upper mark is $u_{a_0}$, so $\mathcal B(z_*)<0$.
When $c_A<c_B$, choose an agent in $A$ attaining $\ell_a=c_A$ to
receive $P$, and an agent in $B$ attaining $u_e=c_B$ to receive $U$.
Writing $a_0,e_0$ for their partners, the feasible middle cuts satisfy
$u_{e_0}\leq c\leq\ell_{a_0}$ within $[c_A,c_B]$.
The choice $c=\max\{c_A,u_{e_0}\}$ works unless
$\mathcal B(z_*)=\ell_{a_0}-u_{e_0}<0$. This proves the first assertion.

Now suppose $\mathcal B(z_*)<0$. Let $q_h$ and $q_v$ be the intersections of
$C$ with $b=b_*$ and $x=x_*$, respectively. These exist because
$\mathcal B(x_*,b_*)<0<\mathcal B(b_*,b_*)$ and
$\mathcal B(x_*,x_*)>0$; they satisfy $x(q_h)>x_*$ and $b(q_v)<b_*$.
On the open arc of $C$ between $q_v$ and $q_h$, every lower mark in the
band formulas strictly increases from its value at $z_*$,
and every upper mark strictly decreases. Thus both
\eqref{four:A-feasible} and \eqref{four:B-feasible} hold strictly.
Where this arc lies in the band, neither pair $A$ nor pair $B$ can
own both private pieces. By Lemma~\ref{four:assignment} and the
analysis of collisions in the band, a collision there is impossible.

Both agents in $A$ strictly prefer $L$ at $q_h$, so $q_h$ occurs
after $t_{(2)}$. Both agents in $B$ strictly prefer $R$ at $q_v$,
so $q_v$ occurs before $t_{(3)}$. Hence the open arc from
$\max\{q_v,t_{(2)}\}$ to $\min\{q_h,t_{(3)}\}$ is nonempty
and consists of envy-free divisions. Its endpoints are also envy-free
by continuity: among the finitely many ownership permutations, one
works along a sequence converging to either endpoint.
Since $t_{(3)}$ was assumed not envy-free, the right endpoint must
be $q_h$.
\end{proof}

The construction uses eight scalar searches: four for the switches,
two for the band endpoints, one for $z_*$, and, if necessary, one for
$q_h$. The last search is bracketed by $[x_*,b_*]$ and uses
$\mathcal B(x,b_*)$. Each search target is an explicit fixed expression in
cumulative functions, their inverses, affine operations, minima,
maxima and order statistics. Earlier searches supply constants to
later ones; no search evaluation invokes another search. In particular,
no evaluation needs a point of $C$, which would itself require a search.

\subsection{Exact scalar searches for arbitrary additive valuations}

From now on $F_i$ may have flat intervals, and the geometric
construction is run on hungry interpolants.
We use the quantile interpolation underlying
Hollender and Rubinstein's preprocessing~\cite[Lemma~6.2]{HR}.
For an integer $K\geq2$, cache one
answer
\[
 q_{i,k}=\operatorname{Cut}_i(0,k/K),\qquad 0<k<K,
 \qquad q_{i,0}=0,\quad q_{i,K}=1,
\]
whenever it is needed. Let $\widehat F_i$ linearly interpolate
$(q_{i,k},k/K)$, with the identity extension outside $[0,1]$.
Distinct levels have strictly increasing locations, so $\widehat F_i$
is hungry, and
\[
 \|\widehat F_i-F_i\|_\infty\leq1/K.
\]
Every interval value changes by at most $2/K$, and an exact envy-free
division for the interpolants has envy at most $4/K$ for the original
valuations.

An inverse evaluation of $\widehat F_i$ needs at most two original
cut queries. A forward evaluation at $y$ needs one original evaluation
and at most two cuts. Indeed, the value $F_i(y)$ locates its quantile
cell. If $F_i(y)=k/K$, compare $y$ with the cached $q_{i,k}$ to choose
the adjacent cell; this handles all flat-interval answers. All revealed
quantiles can be completed afterwards to one fixed collection of grids,
and every computed value depends only on already fetched adjacent
knots. Adaptive oracle choices therefore cause no inconsistency.

The following adapts the affine-cell localisation of Hollender and
Rubinstein~\cite[Lemma~6.3]{HR} to the fixed expressions used by our
searches.

\begin{lemma}[Scalar localisation]\label{four:localisation}
Let $T(s)$ be a continuous function given by a fixed finite sequence of
operations using $\widehat F_i$, $\widehat F_i^{-1}$, affine operations
whose coefficients do not depend on $s$, and fixed-arity minima, maxima
and order statistics. Given $a\leq b$ with $T(a)T(b)\leq0$, a zero
can be found exactly with $O(\log K)$ original Robertson--Webb queries.
The constant depends only on the sequence of operations.
\end{lemma}
\begin{proof}
The idea is to shrink the bracket until $T$ is affine on it.
Maintain a bracket with opposite endpoint signs, returning immediately
at an endpoint zero. Inspect the operations in their evaluation order.
An interpolation operation (an application of $\widehat F_i$ or
$\widehat F_i^{-1}$) is resolved when its inputs at
the two endpoints contain no breakpoint strictly between them. A
comparison operation is resolved when no pair of its arguments has
strictly opposite order at the two endpoints.

If all earlier operations are resolved, each earlier quantity is affine
throughout the bracket. This follows inductively: an affine input that
does not cross an interpolation breakpoint stays in one affine piece;
affine arguments with no order reversal give the same selected affine
function throughout the interior, and continuity handles endpoint ties.

At the first unresolved interpolation operation, split at the preimage
of the median interior breakpoint. Its input is already affine, so
this preimage is explicit; a forward-operation breakpoint is a quantile
location obtained by at most one extra cut. The endpoint levels identify
the relevant integer quantile indices, so the median is located without
enumerating the intervening knots. Evaluate $T$ there and keep
a subinterval with opposite endpoint signs, or return a zero. The number
of interior breakpoints for this operation at least halves. Thus it
requires $O(\log K)$ splits. At an unresolved comparison operation,
split at a crossing of two affine arguments. That pair cannot cross
again on the retained bracket, so an operation with $m$ arguments
requires at most $\binom m2$ splits. Earlier resolved operations remain
resolved under restriction. When all operations are resolved, $T$ is
affine and its zero is explicit. Each evaluation costs a constant
number of original queries, proving the bound.
\end{proof}

\begin{proof}[Proof of Theorem~\ref{four:main}]
Choose $K=8\cdot2^{\lceil\log_2(1/\eps)\rceil}$ and apply the
geometric construction to the interpolants. Every scalar target has
the fixed expression required by Lemma~\ref{four:localisation}, and all
brackets were established above. Hence a constant number of searches
and assignment checks gives an allocation that is exactly envy-free for
the interpolants, with $O(\log K)$ original queries. Its envy for the
original valuations is at most $4/K\le\eps/2$.
\end{proof}

\section{Monotone valuations with value queries}
\label{extension:monotone-section}

This appendix proves Theorem~\ref{extension:monotone-five}, whose
protocol makes value queries only, as Hollender and Rubinstein do for
monotone valuations~\cite[Section~3]{HR}. The protocol of
Theorem~\ref{five:main} cannot be used: it obtains $v_i(a,b)$ by
subtracting two prefix values, which requires additivity, and it makes
cut queries; without additivity Proposition~\ref{prop:value-transfer}
does not apply either. Br\^anzei and Nisan~\cite[Theorem~8.2]{BN} use bisection to approximate
cut queries for additive valuations with densities bounded above. Approximate
cut answers need not be consistent with any single profile
(Section~\ref{sec:rw-conventions}), and the assignment argument needs
exact calibration at vertices with ties. Instead the protocol queries
each valuation only on intervals with endpoints on a dyadic grid and
replaces it by a strictly monotone surrogate $W_i$ with rational
coefficients, within about $\eps/4$ of it
(Lemma~\ref{extension:surrogate-properties}). It then runs the search of
Section~\ref{five:abstract-section} on the surrogates, whose calibrated
cuts it computes exactly (Section~\ref{extension:roots-section}). The
planar search of Section~\ref{five:abstract-section} uses only scores
satisfying properties \ref{five:h-assignment}--\ref{five:h-rows}, and
the assignment rule of Lemma~\ref{five:private-owners} uses only
claimant counts, so neither needs additivity; we verify their hypotheses
below for the strictly monotone surrogates.

We apply the \emph{abstract score search} of
Lemma~\ref{five:abstract-search}.
Suppose that each vertex of a grid of side $H$ carries the agents'
scores for two \emph{private} pieces $P,U$ and three \emph{public}
pieces $L,M,R$, satisfying properties
\ref{five:h-assignment}--\ref{five:h-rows}, which we verify below for the surrogate scores.
Then the search evaluates $O(\log^2H)$ vertices and finds a unit cell
and a bijection of agents to pieces whose envy in scores at each
corner is at most twice the oscillation of the scores over the cell.

\subsection{A strictly monotone rational approximation}

We combine a strictly monotone tilt with triangular interpolation,
as in~\cite[Lemma~5.2]{HR}, and round table entries to obtain rational
coefficients.  Let $\theta=2^{-t}$ be the largest
power of two at most $\eps/8$, and let $N=2^r$ be the least power of two
at least $16L/\eps$.  Put
\[
 \eta=\frac{\theta}{8N},\quad E=\frac{2^t}{\eta},\quad
 \Lambda=\lceil E(L+1)\rceil,\quad
 \kappa=\frac{7\theta}{8},\quad K=L+2\theta,\quad B=K/\kappa.
\]
Here $\theta$ weights the tilt, $1/N$ is the table mesh and $\eta$ its
rounding unit. Multiplied by $E$, the surrogate $W_i$ below has
integer coefficients on each affine region, with endpoint coefficients
of size at most $\Lambda$; this bounds the denominators of roots. Its
rates in each endpoint lie between $\kappa$ and $K$, and $B$ plays the
role of $m_+/m_-$ in~\eqref{five:density-bounds}.
For $0\le p<q\le N$ define the table entry, when it is first needed,
\[
 w_i(p,q)=\eta\left\lfloor
       \frac{v_i(p/N,q/N)}{\eta}\right\rfloor,
 \qquad w_i(p,q)=0\quad(p\ge q).
\]
Store each queried entry.  Subdivide every square of the table grid
by its southwest-to-northeast diagonal, and let $T_i$ be the continuous
affine interpolant on these triangles.  Thus, if
$p=\min(\lfloor Na\rfloor,N-1)$,
$q=\min(\lfloor Nb\rfloor,N-1)$,
$\alpha=Na-p$ and $\beta=Nb-q$, then for $\beta\ge\alpha$,
\[
 T_i(a,b)=(1-\beta)w_i(p,q)
       +(\beta-\alpha)w_i(p,q+1)+\alpha w_i(p+1,q+1),
\]
and for $\alpha\ge\beta$ it equals
\[
 (1-\alpha)w_i(p,q)+(\alpha-\beta)w_i(p+1,q)
                       +\beta w_i(p+1,q+1).
\]
Write $J(s)=\min(1,\max(0,s))$ for the clip to $[0,1]$. The
\emph{surrogate} valuation used by the protocol, including at reversed
intervals, is
\begin{equation}\label{extension:surrogate}
 W_i(a,b)=(1-\theta)T_i(J(a),J(b))+\theta(b-a)
 \qquad(a,b\in\R).
\end{equation}
The last term uses the original, unclipped endpoints. We call an
interval $[a,b]$ \emph{legal} if $0\le a\le b\le1$.

\begin{lemma}\label{extension:surrogate-properties}
For every $h>0$ and all real $a,b$,
\[
 \kappa h\le W_i(a,b+h)-W_i(a,b)\le Kh,
 \qquad
 \kappa h\le W_i(a-h,b)-W_i(a,b)\le Kh.
\]
Moreover $W_i(a,a)=0$, one evaluation of $W_i$ costs at most three
value queries, and on legal intervals
\begin{equation}\label{extension:surrogate-error}
 |W_i(a,b)-v_i(a,b)|\le
 \delta:=2L/N+\eta+\theta
 \le\eps/4+\eps/1024.
\end{equation}
On every affine region of $W_i$,
\begin{equation}\label{extension:integer-affine}
 E W_i(a,b)=\sigma a+\tau b+\rho,
 \quad \sigma,\tau,\rho\in\mathbb Z,
 \quad E\kappa\le-\sigma,\tau\le EK\le\Lambda.
\end{equation}
The coefficients $\sigma,\tau$ are multiples of $N$.
\end{lemma}

\begin{proof}
We prove the rate bounds, then~\eqref{extension:surrogate-error}, then
\eqref{extension:integer-affine}.
The derivative of $T_i$ in its second coordinate on a triangle is $N$
times a vertical edge difference of the table.  Inclusion monotonicity
and the Lipschitz assumption bound the corresponding true difference
between $0$ and $L/N$.  Rounding changes an edge difference by less than
$\eta$.  The derivative therefore belongs to
$[-\eta N,L+\eta N]$.  The negative of the first-coordinate derivative
has the same bound.  Clipping can replace these derivatives by zero.
Adding the linear term in~\eqref{extension:surrogate} gives rates between
$\theta-(1-\theta)\theta/8$ and
$(1-\theta)(L+\theta/8)+\theta$, hence between $\kappa$ and $K$.
Integrating along a coordinate line proves the asserted inequalities
across all region boundaries.  The interpolant is zero on and below
the diagonal: on a diagonal grid square its upper triangle equals
$N(b-a)w_i(p,p+1)$ and its lower triangle is zero.  Thus $W_i(a,a)=0$.

An interpolated value uses at most three table entries.  The vertices
of its triangle
are at endpoint distance at most $2/N$ from the queried point, so
$|T_i-v_i|\le2L/N+\eta$.  Since both $b-a$ and $v_i(a,b)$ belong to
$[0,1]$ on a legal interval, the tilt adds at most $\theta$ error.
Here $2L/N\le\eps/8$, $\theta\le\eps/8$ and
$\eta\le\eps/1024$.

Finally $T_i/\eta$ has integer affine coefficients, with its two
endpoint coefficients divisible by $N$.  Since
$E(1-\theta)T_i=(2^t-1)T_i/\eta$ and
$E\theta(b-a)=(b-a)/\eta$, the same assertion holds after the tilt;
$1/\eta=8N/\theta$ is divisible by $N$.  The coefficient bounds follow
from the rate bounds already proved.  This also covers clipped
regions, where the appropriate endpoint coefficient of $T_i$ is zero.
\end{proof}

\subsection{Calibration without additivity}

We now rebuild the calibrated partitions of
Section~\ref{five:chart-section}, valuing each piece directly by
$W_i$. Use the five pieces
\[
 L=[0,x],\quad P=[x,c],\quad U=[c,b],\quad
 M=[b,d],\quad R=[d,1].
\]
The piece $L$ is unrelated to the Lipschitz bound $L$.  For any
$(x,b,d)\in[0,1]^3$ put
\[
 f_i=\max\{W_i(0,x),W_i(b,d),W_i(d,1)\}.
\]
Define $\ell_i,u_i\in\R$ by
\[
 W_i(x,\ell_i)=f_i=W_i(u_i,b),
 \qquad \mathcal A(x,b,d)=\ell_{(2)}-u_{(4)},
\]
where subscripts in parentheses denote increasing order statistics:
$\ell_{(2)}$ is the second smallest $\ell_i$ and $u_{(4)}$ the second
largest $u_i$. We call $\ell_i,u_i$ \emph{thresholds}; they play the
role of the marks in~\eqref{five:calibration}, and $\mathcal A$ is the corresponding calibration function.
The signed extension in~\eqref{extension:surrogate} makes these roots
unique even if they lie outside the cake.  Moreover $f_i\ge0$,
$\ell_i\ge x$, and $u_i\le b$.

\begin{lemma}\label{extension:calibration}
Increasing $x$ by $h>0$ increases $\mathcal A$ by at least $h/B$.
Increasing $b$ by $h$ decreases $\mathcal A$ by at least $h/B$.
Increasing $d$ by $h$ in $\mathcal A(x,d,d)$ decreases it by at least $h/B$.
Every $\ell_i,u_i$ is $3B$-Lipschitz in $(x,b,d)$ for the maximum
norm, and $\mathcal A$ is $6B$-Lipschitz.  Also
\[
 \mathcal A(0,0,0)>0,\quad \mathcal A(0,1,1)=-1,\quad \mathcal A(x,x,d)>0
 \quad(0\le x\le d\le1).
\]
\end{lemma}

\begin{proof}
We follow the proof of Lemma~\ref{five:calibration-rates}, with the
rates $\kappa,K$ in place of $m_-,m_+$.
As $x$ increases, $f_i$ cannot decrease.  At the old $\ell_i$, moving
the first endpoint right by $h$ lowers the value by at least
$\kappa h$.  Recovering the new target requires increasing $\ell_i$
by at least $\kappa h/K=h/B$.  Meanwhile $u_i$ cannot increase.
As $b$ increases, $f_i$ cannot increase; $\ell_i$ cannot increase,
whereas $u_i$ increases by at least $h/B$.  If $b=d$, the middle
public value $W_i(b,d)$ is zero, and the same argument applies when
both $b,d$ increase.  Each assertion passes to order statistics.

A displacement of maximum norm $h$ changes $f_i$ by at most $2Kh$.
It changes the value at the old threshold, before adjusting the
threshold coordinate, by at most another $Kh$.  The rate in the
threshold coordinate is at least $\kappa$, giving $3Kh/\kappa=3Bh$.  At $(0,0,0)$, $f_i=W_i(0,1)>0$,
$\ell_i=1$ and $u_i<0$.  At $(0,1,1)$ all three public values are
zero, so $\ell_i=0,u_i=1$.  Finally, if $b=x$,
one of $[0,x],[x,d],[d,1]$ has length at least $1/3$, whence
$f_i\ge\kappa/3$ and $\ell_i>x>u_i$.
\end{proof}

There is consequently a unique $d_0\in(0,1)$ with $\mathcal A(0,d_0,d_0)=0$.
For $d\in[d_0,1]$ there is a unique $a(d)\in[0,d)$ with
$\mathcal A(a(d),d,d)=0$; $a(d_0)=0$ and $a(d)>0$ for $d>d_0$.
For $0\le x\le a(d)$ there is a unique $b=b(x,d)\in(x,d]$ with
$\mathcal A(x,b,d)=0$, and $b=d$ exactly when $x=a(d)$.  Set
$c(x,d)=\ell_{(2)}=u_{(4)}$ there.  Parametrise these divisions by
\begin{equation}\label{extension:monotone-chart}
 d=d_0+(1-d_0)t,\qquad x=s\,a(d),\qquad(s,t)\in[0,1]^2.
\end{equation}
As with~\eqref{five:chart}, we call this family the \emph{chart}; the search
uses its grid of mesh $1/H$ ($H$ chosen below), whose rows have
constant $d$.

We now verify properties \ref{five:h-assignment}--\ref{five:h-rows},
with surrogate piece values as scores.  For
property~\ref{five:h-assignment}, calibration gives a legal
assignment at every point of this square: distinct owners of $P,U$
receive global favourites (pieces of maximal score), and every other
agent has a public global favourite.  Except at $(x,b,d)=(0,1,1)$ all $f_i$ are
positive: if some $f_i=0$, positivity of $W_i$ on every interval of
positive length forces
$x=0,b=d=1$.  That exceptional point has $\mathcal A=-1$ and is not in the
chart.  It follows that $x<c<b$.  The private piece $P$ has at least
two weak claimants, meaning agents with value at least $f_i$, and at
most one strict claimant, with value more than $f_i$, since its
claimant conditions are $\ell_i\le c$ and $\ell_i<c$.  The
corresponding conditions for $U$ are $u_i\ge c$ and $u_i>c$, with the
same counts.  The case analysis in the proof of
Lemma~\ref{five:private-owners} uses only these counts, so it applies
verbatim.

For property~\ref{five:h-boundary}: on the left side $s=0$, $L$ is
empty; on the right side $s=1$, $M$ is empty; on the top $t=1$, $R$ is
empty; and on the bottom $t=0$, both $L,M$ are empty.  Such pieces
have value zero, while $f_i>0$, so they cannot be public favourites.
For property~\ref{five:h-rows}: on every row other than the bottom
one, increasing $s$ strictly increases $x$ and hence, by
Lemma~\ref{extension:calibration}, $b$.  Thus every agent's score for
$L$ strictly increases, her score for $M$ strictly decreases, and her
score for $R$ is constant.

The quantitative bounds follow from Lemma~\ref{extension:calibration}
as in the proof of Lemma~\ref{five:quantitative-chart}.  The root $a$
is $6B^2$-Lipschitz;
the map $(s,t)\mapsto x$ is $7B^2$-Lipschitz; and $b$ is
$6B^2$-Lipschitz as a function of $(x,d)$, hence $42B^4$-Lipschitz on
the square.  The cut $c$ is $126B^5$-Lipschitz.  Consequently every
score $W_i(\text{piece})$ is $W_*=252KB^5$-Lipschitz.  Choose
$H=2^\lambda$ to be the least power of two at least $8W_*/\eps$.
On a mesh cell the score oscillation is at most $W_*/H$, and
by~\eqref{five:corner-envy} the abstract search therefore gives
surrogate envy at most $2W_*/H\le\eps/4$.

\subsection{Exact roots with few value queries}\label{extension:roots-section}

An approximate calibration would not suffice for the assignment
argument at a vertex with ties, where the claimant counts need
$\ell_{(2)}=c=u_{(4)}$ exactly.  The rational interpolation allows us
to find the exact calibration instead, since the calibrated cuts have
bounded denominators (Lemma~\ref{extension:root-denominators}). Sign
bisection then finds $d_0$ and each $a(d)$, and a joint bisection finds
$b,c$ at each vertex.

\begin{lemma}\label{extension:root-denominators}
The denominator of $d_0$ is at most $3\Lambda^2$.
For rational $d$, the denominator of $a(d)$ divides
$A\operatorname{den}(d)$ for an integer $1\le A\le3\Lambda^2$.
For rational $x,d$ with $x<a(d)$, put
$D=\operatorname{lcm}(\operatorname{den}(x),\operatorname{den}(d))$.
Then the denominators of $b(x,d)$ and $c(x,d)$ divide $A'D$ and
$\tau_1 A'D$, respectively, for integers
$1\le A'\le3\Lambda^2$ and $1\le\tau_1\le\Lambda$.
At a mesh vertex it follows that
\[
 \operatorname{den}(b)\le Q_b:=27\Lambda^6H^2,
 \qquad
 \operatorname{den}(c)\le Q_c:=27\Lambda^7H^2.
\]
\end{lemma}

\begin{proof}
We eliminate $c$ between two threshold equations and bound the
integer coefficient of the remaining unknown.
At a calibrated point choose agents realising the two order statistics
and affine regions containing the two threshold evaluations.  With
integer coefficients as in~\eqref{extension:integer-affine}, their
equations have the form
\[
 \tau_1c+\sigma_1x+\rho_1=Ef_1,
 \qquad \sigma_2c+\tau_2b+\rho_2=Ef_2.
\]
Choose for each $f$ a maximising public branch.  Its coefficient of
$b$ is either zero or a negative integer, say $\phi,\phi'$ for the
two equations.  Eliminating $c$ leaves coefficient
\[
 A'=\tau_1\tau_2-\tau_1\phi'+\sigma_2\phi
\]
on $b$.  Every summand is nonnegative, the first is positive, and each
is at most $\Lambda^2$.  The remaining terms are integer combinations
of $x,d,1$.  This proves the assertion about $b$ and, by substitution,
the assertion about $c$.

At $b=d$, a public branch is $W_i(0,x)$, zero, or $W_i(d,1)$.
Let $\psi,\psi'\ge0$ be the coefficients of $x$ in the selected two
branches.  The same elimination gives coefficient
$A=\sigma_1\sigma_2-\sigma_2\psi+\tau_1\psi'$ on $x$.
Again $1\le A\le3\Lambda^2$, proving the assertion about $a(d)$.
At $x=0,b=d=d_0$, both public maxima are $W_i(d_0,1)$.
Writing their first-coordinate coefficients as $\sigma',\sigma''<0$
gives coefficient
$A_0=\sigma_2\sigma'-\tau_1\sigma''+\tau_1\tau_2$ on $d_0$,
with $1\le A_0\le3\Lambda^2$.
At row $j$, where $t=j/H$, $\operatorname{den}(d)$ divides $HA_0$;
at column $i$, where $s=i/H$, $\operatorname{den}(x)$ divides
$H^2AA_0$.
Thus $D\le9\Lambda^4H^2$, proving the interior bounds. On the right
boundary $b=d$, a threshold equation gives denominator at most
$\Lambda D$ for $c$, which also satisfies the displayed bounds.
The bottom case is included by $a(d_0)=0$.
\end{proof}

As in the proof of Lemma~\ref{five:rational-roots}, if a strictly
monotone function has a rational zero of denominator at most $Q$ in a
closed bracket of length at most one, exact sign bisection finds it in
$O(\log Q)$ steps, even if it is an endpoint: once the bracket is
shorter than $1/Q^2$, its rational of smallest denominator is the zero.
This rational can be found in polynomial bit time by continued fractions.

Next we evaluate $\mathcal A$ exactly.  One exact threshold costs $O(\log N)$
evaluations of $W_i$.
For example, for $\ell_i$ on the line $a=x$, $W_i(x,y)$ is affine
between the table grid lines and the diagonals of the triangles it
crosses, a list of at most $2N+1$ points known without queries.
Its entries are accessed by index in the interleaved progressions
$q/N$ and $x+q/N$ for integer $q$, restricted to $[0,1]$; the list
is never materialised.
If $f_i\ge W_i(x,1)$, the solution is
$1+(f_i-W_i(x,1))/\theta$.  Otherwise binary-search this list for
the affine interval containing the threshold and interpolate exactly.
The threshold $u_i$ in the left endpoint is symmetric, with an affine
tail below zero.
Each $W_i$ evaluation uses at most three original value queries.
Computing all five maxima and ten thresholds therefore evaluates $\mathcal A$
with $O(\log N)$ queries. This supplies exact signs for $d_0$ and
for $a(d)$ on every requested row.

Computing $b,c$ at every visited vertex by nested bisection would
introduce an unnecessary logarithm.  Instead, as in
Section~\ref{five:roots-section}, fix $x,d$ with
$d>d_0$ and $x<a(d)$, and write
$\ell(\beta)=\ell_{(2)}(x,\beta,d)$,
$u(\beta)=u_{(4)}(x,\beta,d)$.  The first is nonincreasing and the
second strictly increasing; they cross at $(b,c)$.
For a test pair $(\beta,\gamma)$ of trial values for $(b,c)$, the
signs of
$\gamma-\ell(\beta)$ and $\gamma-u(\beta)$ are available by direct
counts, without computing a threshold.  For example,
\[
 \gamma>\ell(\beta)
 \ \Longleftrightarrow\
 \#\{i:W_i(x,\gamma)>f_i(\beta)\}\ge2,
\]
with the weak inequality handled by a weak count.  Likewise
$\gamma<u(\beta)$ exactly when at least two agents have
$W_i(\gamma,\beta)>f_i(\beta)$.  The five middle public values
$W_i(\beta,d)$ and the ten trial private-piece values $W_i(x,\gamma)$,
$W_i(\gamma,\beta)$ cost
at most $45$ original queries per test; the ten fixed public values
$W_i(0,x)$, $W_i(d,1)$ cost at most $30$ queries once per vertex.

As in~\eqref{five:vertex-signs}, both signs positive imply
$\gamma>c$; both negative imply
$\gamma<c$.  Opposite weak signs, with the first nonpositive and the
second nonnegative and not both zero, imply $\beta<b$; the reverse
pair implies $\beta>b$.  Both zero identify the root.  To see these
implications, for $\beta\le b$ we have
$\ell(\beta)\ge c\ge u(\beta)$, with the inequalities reversed for
$\beta\ge b$.  The order of the two thresholds determines on which
side of the crossing $\beta$ lies.

Maintain open brackets for both $b,c$, initially $(x,d)$, and test
their midpoints.  Each verdict halves one bracket.  When a bracket
has length less than $1/Q_b^2$ or $1/Q_c^2$, freeze its coordinate at
the smallest-denominator rational in its closure.
Lemma~\ref{extension:root-denominators} makes that coordinate exact.  Once $b$ is fixed, all remaining
verdicts concern $c$; once $c$ is fixed, all remaining verdicts concern
$b$.  Thus no verdict contradicts a frozen coordinate, and there are
at most
\[
 2\log_2Q_b+2\log_2Q_c+3
 \le26\log_2\Lambda+8\lambda+23
\]
tests.  On the right side $b=d$ is already known, and on the bottom
row the single partition has $x=0,b=d_0$; in both cases the common
threshold $c=\ell_{(2)}=u_{(4)}$ costs just one evaluation of $\mathcal A$.

\begin{proof}[Completion of the proof of Theorem~\ref{extension:monotone-five}]
Run the abstract score search on the chart
\eqref{extension:monotone-chart}, computing only the rows and vertices
it requests.  At each requested vertex compute the exact chart point
as above and the $25$ scores.  Use the assignment rule of the proof of
Lemma~\ref{five:private-owners}, which applies as shown above, for
property~\ref{five:h-assignment}. The abstract search
returns a mesh cell of nonzero winding number, in the sense of
Section~\ref{five:abstract-section}. If this cell touches the bottom,
compute that row's one partition, its scores, and its legal
assignment once, costing $O(\log N)$ queries. The bottom row's
constant boundary vector $T$ alone does not supply the assignment. Use the same
assignment at both bottom
corners and find a matching in the union of the four assignments,
as in Lemma~\ref{five:abstract-search}. At any selected corner
the assigned surrogate values are envy-free to error $2W_*/H\le\eps/4$.
Transferring the two values in an envy comparison by
\eqref{extension:surrogate-error} gives total error
$\eps/4+2\delta<\eps$.

It remains to count queries. Write $\mathcal L=\max\{1,\log_2(L/\eps)\}$. From the
parameter choices,
$\log N$, $\log\Lambda$, and $\log H$ are all $O(\mathcal L)$:
$N=O(L/\eps)$, $E=8N/\theta^2$, and
$W_*=O(L^6/\eps^5)$. The initial root $d_0$ uses
$O(\log\Lambda)$ signs, each costing $O(\log N)$ queries, hence
$O(\mathcal L^2)$ queries. On each of $O(\log H)$ rows, computing $a(d)$
uses $O(\log\Lambda+\log H)$ signs at the same cost, totalling
$O(\mathcal L^3)$ queries. Without supplied row indices,
Lemma~\ref{five:abstract-search} evaluates $O(\log^2 H)$ vertices.
Each needs $O(\log\Lambda+\log H)$ constant-cost crossing tests
and $25$ scores, hence again $O(\mathcal L^3)$ queries in total.
The right-boundary and bottom routines cost $O(\log N)$ per use
and are covered by these bounds. Thus the complete protocol costs
$O(\mathcal L^3)$. Every original query is a cached table entry
$v_i(p/N,q/N)$, so its endpoints are dyadic. The exact rational
output cuts require no further queries.
\end{proof}

\section{Query models and valuation classes}\label{sec:other-models}
This appendix proves the transfers used in Section~\ref{sec:models}.
For additive approximate envy-freeness, restricted RW, unrestricted
RW and prefix cuts alone have the same polylogarithmic-versus-polynomial
classification. Removing cuts requires a known Lipschitz bound.
Section~\ref{sec:rw-conventions} proves these simulations;
Section~\ref{sec:monotone-to-additive} removes density bounds from an
additive input by changing coordinates. The main upper and lower bounds
are proved directly and do not depend on these transfers.

\subsection{Weaker access to additive valuations}\label{sec:rw-conventions}
The Robertson--Webb model of Section~\ref{sec:rw-model}, which we also
call the \emph{full model}, provides three resources: cut queries,
evaluation queries, and a free choice of coordinates, meaning that
queries may use, and output cuts may be placed at, arbitrary points of
the cake. Withdrawing one resource gives a weaker model, and we now
bound the cost of each withdrawal for the approximate problem.
\begin{itemize}
\item In \emph{restricted RW}, both kinds of query remain, but
evaluations may only be made at, and cut queries may only start at, the
endpoints $0,1$ and points returned by earlier cut queries to any agent,
and output cuts must be such points. The identities of
Section~\ref{sec:rw-model} remain legal at such points. This is the
model that Br\^anzei and Nisan call RW~\cite{BN} and that we call
restricted RW. Woeginger and Sgall also restrict evaluations and final cuts
to previously marked points~\cite[Section~2]{WS}. For example,
evaluating $F_i(1/2)$ is legal in restricted RW only after $1/2$ has
been returned as a cut; asking an agent for half her value generally
produces a different point.
\item With \emph{cut queries only}, there are no evaluation queries;
cuts may start at arbitrary points, and output cuts are arbitrary.
\item With \emph{value queries only}, an agent reports her value of any
specified interval, but there are no cut queries; queries and output
cuts may use arbitrary points. For additive
valuations this is the evaluation-only model of Br\^anzei and Nisan,
which they call $\mathrm{RW}^{-}$~\cite[Definition~5]{BN}. Hollender
and Rubinstein use the same access for monotone
valuations~\cite[Section~3]{HR}, and so does
Theorem~\ref{extension:monotone-five}.
\end{itemize}
Every protocol in a weaker model is also a Robertson--Webb protocol, at
no extra cost, so lower bounds in the full model hold in each of them.
The converse direction is harder: the identities of
Section~\ref{sec:rw-model} do not simulate
an arbitrary-coordinate
evaluation exactly in restricted RW, and a finite number of value
queries cannot in general locate an exact cut. For the
\emph{approximate allocation problem}, however, any two of the three
resources suffice to simulate the full model at a logarithmic cost per
query, with one
proviso: without cut queries, a Lipschitz bound on the valuations must
be known.
In fact cut queries suffice on their own (Corollary~\ref{cor:cut-only}). The simulations answer the
queries exactly for a nearby valuation profile, not approximately for
the original one; this is what allows an arbitrary exact-answer protocol
to be run on the simulated answers.
Proposition~\ref{prop:rw-transfer}, an implementation of
equal-value discretisation and lazy interpolation~\cite{BN,HR}, removes
the free choice of coordinates and makes the logarithmic overhead
explicit; it also turns a protocol for hungry additive valuations into
one for all additive valuations.

\begin{proposition}[From the Robertson--Webb model to restricted RW]
\label{prop:rw-transfer}
Suppose that for every $0<\eta\le1$ a deterministic
Robertson--Webb protocol finds a complete connected
$\eta$-envy-free allocation of every profile of $n$ hungry
additive valuations with a uniform bound $Q(n,\eta)$ on its
queries. For all additive valuations and
$0<\eps\le1$, a restricted-RW protocol finds such an allocation with
envy at most $\eps$ using
\[
 O\!\left((Q(n,\eps/2)+n^2)\log(1/\eps)\right)
\]
queries. Output boundaries are returned cuts or endpoints.
\end{proposition}
\begin{proof}
We run the given protocol on hungry piecewise-linear interpolants of the
valuations, built from lazily revealed quantile locations, and then round
its output cuts to returned points.
Put $K=\lceil16/\eps\rceil$. For each agent, store the first answer
$q_{ij}$ to $\operatorname{Cut}_i(0,j/K)$, querying it only when needed;
set $q_{i0}=0$ and $q_{iK}=1$. Distinct levels have strictly ordered
locations, even when the original valuation has flat intervals. Let
$\widehat F_i$ interpolate the points $(q_{ij},j/K)$ linearly.
It is a hungry cumulative function and
$\|F_i-\widehat F_i\|_\infty\le1/K$.

At any $x\in[0,1]$, binary search among the $K+1$ quantile indices finds
two consecutive locations bracketing $x$. Their stored values determine
$\widehat F_i(x)$ exactly, using $O(\log K)$ original Cut queries.
Given a value target, at most two quantile locations determine
$\widehat F_i^{-1}$ exactly by interpolation. Together these operations
also implement arbitrary-start cuts for the interpolant. They use no
original evaluation at an unreturned coordinate.

Run the given protocol at tolerance $\eps/2$ on these interpolants.
Lazy revelation does not change a previously supplied answer, because
that answer was determined by consecutive quantile levels. Completing
all unqueried levels therefore gives a fixed hungry profile consistent
with the entire transcript. The uniform bound $Q(n,\eps/2)$ applies:
a transcript with more queries could be completed to such a profile,
contradicting that bound. Interval values differ from their original
values by at most $2/K$, so the output has original envy at most
$\eps/2+4/K$.

Finally, for each output cut $y$ and each agent, binary search finds
that agent's largest quantile location at most $y$. Replace $y$ by the largest of
these $n$ locations, denoted $r(y)$. This is the predecessor of $y$ in
the union of the grids, so $r$ is nondecreasing and
$0\le F_i(y)-F_i(r(y))\le1/K$ for every $i$.
The rounded allocation remains complete and ordered. Each piece value
changes by at most $2/K$, adding at most $4/K$ to envy. The total is
at most $\eps/2+8/K\le\eps$. Rounding the $n-1$ interior cuts costs
$O(n^2\log K)$ queries and gives legal output boundaries.
\end{proof}

\begin{corollary}\label{cor:cut-only}
The protocol of Proposition~\ref{prop:rw-transfer} makes only prefix cut
queries $\operatorname{Cut}_i(0,j/K)$, where $K=\lceil16/\eps\rceil$.
Hence the bound of that
proposition is also achieved by a protocol that makes no evaluation queries,
with output cuts at returned points or endpoints.
\end{corollary}
\begin{proof}
The interpolant values, its inverse values, the simulated cuts from
arbitrary starting points and the output rounding are all computed from
the locations $q_{ij}$, each obtained by one such query.
\end{proof}

Cut queries cannot be withdrawn in the same way for all additive
valuations: without a bound on their Lipschitz constants, value queries
alone cannot give any bound that is independent of the
valuations~\cite[Section~8]{BN}, even for two agents~\cite[Section~6]{HR}.
When the densities are bounded above by a constant,
Br\^anzei and Nisan use $O(\log(1/\eps))$ prefix evaluations to find
a cut whose prefix value differs from the target by at most
$\eps$~\cite[Theorem~8.2]{BN}. Such
approximate answers need not be consistent with any single profile, so
inserting them into an arbitrary exact-answer protocol is not justified
in general. With a known Lipschitz bound, interpolation on a fixed grid
instead answers every query exactly for a nearby hungry profile, as the
next proposition shows.

\begin{proposition}[Approximate allocation with value queries]
\label{prop:value-transfer}
Assume the hypothesis of Proposition~\ref{prop:rw-transfer}. Let $L$ be
a known common Lipschitz bound for the cumulative functions $F_i$ of
$n$ additive valuations. For every $0<\eps\le1$, a
deterministic protocol that makes only value queries finds a complete
connected $\eps$-envy-free allocation using
\[
 O\!\left(Q(n,\eps/2)\log(L/\eps)\right)
\]
queries.
\end{proposition}
\begin{proof}
Necessarily $L\ge F_i(1)-F_i(0)=1$. Let $G$ be the least power of two
with $G\ge16L/\eps$, and put $\theta=\eps/16$. For each agent, let
$\widetilde F_i$ interpolate the points $(k/G,F_i(k/G))$, $0\le k\le G$,
linearly, and set
\[
 \widehat F_i(x)=(1-\theta)\widetilde F_i(x)+\theta x .
\]
Each $\widehat F_i$ is continuous, piecewise affine with breakpoints in
the grid, strictly increasing, and satisfies $\widehat F_i(0)=0$ and
$\widehat F_i(1)=1$; it is the cumulative function of a hungry additive
valuation. On a grid cell $[k/G,(k+1)/G]$, both $F_i$ and
$\widetilde F_i$ take values between $F_i(k/G)$ and $F_i((k+1)/G)$,
which differ by at most $L/G\le\eps/16$. Moreover
$|\widehat F_i-\widetilde F_i|\le\theta$. Hence
$\|F_i-\widehat F_i\|_\infty\le\eps/8$, interval values differ by at
most $\eps/4$, and each envy comparison changes by at most $\eps/2$.

Each grid value $F_i(k/G)$ costs one value query and is stored. An
evaluation of $\widehat F_i$ at an arbitrary point, and hence an interval
evaluation, is computed exactly from at most four grid values, after
locating the relevant cells by free computation. A cut
$\operatorname{Cut}_i(a,t)$ for $\widehat F_i$ asks for
$\widehat F_i^{-1}(\widehat F_i(a)+t)$, which is unique because
$\widehat F_i$ is strictly increasing; whether the target is feasible is
decided exactly from $\widehat F_i(a)$. Binary search over the grid
indices finds the cell whose endpoint values bracket the target, using
$O(\log G)$ grid values, and linear interpolation in that cell gives the
exact answer. Every query to the profile $(\widehat F_i)$ therefore costs
$O(\log G)=O(\log(L/\eps))$ value queries.

The profile $(\widehat F_i)$ is a fixed profile of hungry additive
valuations, and the simulated answers are its exact answers.
Running the given protocol at tolerance $\eps/2$ thus makes at most
$Q(n,\eps/2)$ simulated queries and returns an allocation that is
$\eps/2$-envy-free for this profile. Its cuts may be arbitrary points,
which the value-query model permits as outputs. For the original
valuations its envy is at most $\eps/2+\eps/2=\eps$.
\end{proof}

\begin{table}[ht]
\centering
\begin{tabular}{@{}p{.25\linewidth}p{.27\linewidth}p{.38\linewidth}@{}}
\toprule
Resources kept & Model & Simulation of the full model\\
\midrule
Cut, Eval, coordinates & Robertson--Webb (full) & ---\\
Cut, Eval & Restricted RW & Proposition~\ref{prop:rw-transfer}\\
Cut, coordinates & Cut queries only & Corollary~\ref{cor:cut-only}\\
Eval, coordinates & Value queries only & Proposition~\ref{prop:value-transfer}, given a Lipschitz bound; impossible without one~\cite{BN,HR}\\
Cut & Prefix cuts, outputs at returned points & Corollary~\ref{cor:cut-only}\\
\bottomrule
\end{tabular}
\caption{Simulating the Robertson--Webb model for additive valuations, in the approximate allocation problem, at a logarithmic cost per query. The resources are cut queries (Cut), evaluation queries (Eval) and the free choice of coordinates.}
\label{tab:resources}
\end{table}
\Cref{tab:resources} summarises the simulations.

\paragraph{Consequences for our results.}
The five-agent protocol of Theorem~\ref{five:main} needs no simulation,
because it already works in restricted RW.

\begin{proposition}\label{prop:five-restricted}
The protocol of Theorem~\ref{five:main} is a restricted-RW protocol: all
its cut queries start at $0$, it evaluates only at endpoints of the cake
and at points returned by earlier cut queries, and every output cut is an endpoint of the cake or
a point returned by a cut query. Hence five agents with additive
valuations admit a restricted-RW protocol with $O(\log^2(1/\eps))$
queries.
\end{proposition}
\begin{proof}
The protocol queries the valuations only through probes
(Lemma~\ref{five:rank}); all other steps of Section~\ref{five:section}
compute with their results. A probe asks for prefix cuts
$\operatorname{Cut}_i(0,j/K)$, which start at the endpoint $0$, and for
prefix values $F_h(z)$ at a point $z=y_{i,j}$ returned by an earlier
cut query. The output cuts are the images $\rho(t)=p_{\lfloor tJ\rfloor}$
of auxiliary cuts whose ranks are found by Locate
(Section~\ref{five:protocol-section}), so each output cut is an
endpoint or a quantile mark returned by a cut query.
\end{proof}

The four-agent protocol of Theorem~\ref{four:main} evaluates at arbitrary
points. Applied to it, Proposition~\ref{prop:rw-transfer} gives a
restricted-RW protocol with $O(\log^2(1/\eps))$ queries; we do not obtain
a logarithmic bound in restricted RW. The lower bound of
Theorem~\ref{seven:main} is proved in the full model, so it holds in
every weaker one. By the remark after that theorem, it holds even for
protocols that need to succeed only on valuations with densities in
$[\ell_n,u_n]$, which have the Lipschitz bound $u_n$. Since a value
query is an evaluation query of the full model, this gives the
$\Omega(\eps^{-1/3})$ deterministic lower bound for value queries,
even when a Lipschitz bound $L\ge u_n$ is supplied.

For each fixed population, the existence of a uniform deterministic
polylogarithmic query bound for additive approximate envy-freeness is
therefore the same in the Robertson--Webb model, in restricted RW and
with cut queries only. The proofs of
Propositions~\ref{prop:rw-transfer} and~\ref{prop:value-transfer} run
the given protocol only on profiles with positive piecewise-constant
densities, so they also apply to the protocols cited from~\cite{BN} in
\Cref{tab:oracle}, which are proved for valuations with densities. The
existence of such a bound is also unchanged if the valuations are
required to be hungry: Theorem~\ref{seven:main} already concerns hungry
valuations, and Proposition~\ref{prop:rw-transfer} turns a protocol for
hungry additive valuations into one for all additive valuations. For
additive valuations with a known Lipschitz bound $L$,
Proposition~\ref{prop:value-transfer} turns the polylogarithmic
protocols of \Cref{tab:oracle} (cut and choose, the three-agent protocol
of~\cite{BN}, and Theorems~\ref{four:main} and~\ref{five:main}) into
value-query protocols with an extra factor $O(\log(L/\eps))$, and, when
$L\ge u_n$, the lower bound of Theorem~\ref{seven:main} holds for value
queries. So, with value queries only, two to five agents admit
polylogarithmic protocols for every known Lipschitz bound, and every
fixed $n\ge7$ needs $\Omega(\eps^{-1/3})$ queries for every known bound
$L\ge u_n$, and $\Omega(\eps^{-1/3}/\log(1/\eps))$ queries for every
known bound $L\ge n$ (Section~\ref{sec:monotone-to-additive}). Some lower limit on $L$ is needed here: an additive $1$-Lipschitz valuation
satisfies $F_i(t)\le t$ and $1-F_i(t)\le1-t$, so it is uniform, and the
cuts $k/n$ are envy-free without any query.

\subsection{From monotone to additive valuations}\label{sec:monotone-to-additive}
A protocol for monotone valuations with a known Lipschitz bound~$L$ is,
in particular, a protocol for additive valuations whose densities are at
most~$L$. The following proposition turns a value-query protocol for
monotone $n$-Lipschitz valuations into a restricted-RW protocol for all
additive valuations, which also makes cut queries, at a logarithmic cost
per query. It measures positions on the cake by the merged order of the agents' quantile marks instead of by
length, as in Section~\ref{five:quantiles-section}. In these
coordinates the additive valuations of $n$ agents lie within $O(\eps)$
of auxiliary valuations that are additive and $n$-Lipschitz, and whose
values can be computed exactly with few queries.

\begin{proposition}[From monotone to additive valuations]
\label{prop:monotone-to-additive}
Suppose that for every $0<\eta<1$ a deterministic value-query protocol
finds a complete connected $\eta$-envy-free allocation for every profile
of $n$ monotone $n$-Lipschitz valuations, using at most $Q(n,\eta)$
queries. Then for additive valuations and every $0<\eps\le1$, a
restricted-RW protocol finds a complete connected $\eps$-envy-free
allocation using
\[
 O\bigl(n^2(Q(n,\eps/2)+n)\log(1/\eps)\bigr)
\]
queries.
\end{proposition}
\begin{proof}
We run the given protocol on an auxiliary profile that is additive and
$n$-Lipschitz, answer each of its value queries with
$O(n^2\log(1/\eps))$ queries to the original valuations, and map its
output back to the cake.

Let $K=2^\kappa$ be the least power of two with $K\ge24/\eps$. Define
the quantile marks, entries, merged order, ranks, positions $p_k$,
counts $c_h(k)$ and auxiliary cumulative functions $G_h$ as in
Section~\ref{five:quantiles-section}, for $n$ agents instead of five, so
that $J=n(K-1)+1$ and $G_h(k/J)=Q_{h,k}$ is given
by~\eqref{five:virtual-values}. The proof of
Lemma~\ref{five:virtual-transfer} applies verbatim except for the upper
slope bound: on each knot cell $K^2G_h$ has slope $K$ or $(K-1)J+K$,
and
\[
 \frac{(K-1)J+K}{K^2}=n-\frac{2n-2}{K}+\frac{n-1}{K^2}\le n .
\]
Hence each $G_h$ is strictly increasing with slopes between $1/K$ and
$n$, the auxiliary valuations $G_h(b)-G_h(a)$, $a\le b$, are additive
and $n$-Lipschitz, and for the projection
$\rho(t)=p_{\lfloor tJ\rfloor}$ we have $|G_h(t)-F_h(\rho(t))|\le3/K$,
where $F_h$ is the original cumulative function of agent~$h$.

The probes and Locate of Section~\ref{five:access-section} work in the
same way for $n$ agents. A probe needs at most $2n-1$ queries: one cut,
$n-1$ evaluations at the returned point and at most $n-1$ further cuts
(Lemma~\ref{five:rank}). Locate needs at most $n(\kappa+1)$ probes
(Lemma~\ref{five:flattened-access}); it uses $t$ only through
$\lfloor tJ\rfloor$, so it applies to every $t\in[0,1)$, and at $t=1$
everything is known. Answer each value query of the given protocol, for
agent $h$ at endpoints $a\le b$, by $G_h(b)-G_h(a)$, computed from the
affine forms that Locate gives at $a$ and $b$. This costs at most
$2n(\kappa+1)(2n-1)$ queries, all of them prefix cuts or evaluations at
returned points.

Cache every answer. Once the run, including the final Locates below, is
over, complete all unasked marks by arbitrary admissible answers.
Distinct quantile levels of one agent have strictly ordered marks, so
this fixes all $n(K-1)$ quantile marks of the original valuations, and
with them one auxiliary profile, consistent with every cached answer; as
in Section~\ref{five:protocol-section}, this holds also for adaptive cut
answers on flat intervals, so every answer given to the protocol is the
exact answer for this profile. The auxiliary profile consists of $n$
monotone $n$-Lipschitz valuations, so the given protocol, run with
$\eta=\eps/2$, makes at most $Q(n,\eps/2)$ value queries; a run with
more could be completed in the same way after $Q(n,\eps/2)+1$ queries,
contradicting that bound. It returns cuts $t_1\le\cdots\le t_{n-1}$ and
an assignment with envy at most $\eps/2$ for the auxiliary profile.
Locate each $t_k$ and output the cuts $\rho(t_k)$ with the same
assignment. They are endpoints or
returned quantile marks, and they are ordered because $\rho$ is
nondecreasing. Each piece value changes by at most $6/K$, and each envy
comparison by at most $12/K\le\eps/2$, so the output is
$\eps$-envy-free. Since $\kappa=O(\log(1/\eps))$, the total is
$O(n^2(Q(n,\eps/2)+n)\log(1/\eps))$ queries.
\end{proof}

For five agents, Theorem~\ref{extension:monotone-five} with $L=5$ gives
$Q(5,\eta)=O(\log^3(1/\eta))$, so the proposition gives a restricted-RW
protocol with $O(\log^4(1/\eps))$ queries for arbitrary additive
valuations: polylogarithmic, though weaker than Theorem~\ref{five:main}.

The proof runs the given protocol only on additive valuations with
densities at most $n$, so the proposition also holds for a value-query
protocol that is only required to succeed on such valuations. With
Proposition~\ref{prop:value-transfer} this gives an equivalence for
additive valuations: for each $n$ and each fixed $L\ge n$, a
polylogarithmic Robertson--Webb protocol for $n$ agents exists if and
only if a polylogarithmic value-query protocol exists for $n$ agents
whose valuations have densities at most $L$. Indeed,
Proposition~\ref{prop:value-transfer} gives the second from the first,
and a value-query protocol for densities at most $L$ works in particular
for densities at most $n$, so the strengthened proposition gives the
first from the second.

The proposition also carries lower bounds over to value queries. Let
$n\ge7$, and suppose that a deterministic value-query protocol finds,
for every $0<\eta<1$, a complete connected $\eta$-envy-free allocation
for every profile of $n$ additive valuations with densities at most
$n$, using at most $Q(n,\eta)$ queries; every value-query protocol for
additive valuations with a known Lipschitz bound $L\ge n$ does so. The
restricted-RW protocol given by the strengthened proposition succeeds on
all additive valuations, so by Theorem~\ref{seven:main} it makes at least
$c_n\eps^{-1/3}$ queries on some profile when $\eps<\eps_n$. Hence
$O(n^2(Q(n,\eps/2)+n)\log(1/\eps))\ge c_n\eps^{-1/3}$, that is,
$Q(n,\eta)=\Omega(\eta^{-1/3}/\log(1/\eta))$. So for additive
valuations with a known Lipschitz bound $L\ge n$, value-query protocols
need $\Omega(\eps^{-1/3}/\log(1/\eps))$ queries, with constants
depending only on $n$. This is weaker by a logarithmic factor than the
bound for $L\ge u_n$ in Section~\ref{sec:rw-conventions}, but it holds
for every $L\ge n$.

\section{The randomized lower bound}\label{app:randomized-lower}

Chen and Teng established randomized query lower bounds for discrete
Brouwer fixed points~\cite{CT}. For our cake-cutting lower bound, we use
the construction of Section~\ref{seven:section} and prove a distributional
bound for its particular hidden-path oracle.

\begin{lemma}\label{seven:random-path}
In the hidden-path problem of Section~\ref{seven:source-section}, let
$L_2,\ldots,L_N$ be independent and uniform on $\{1,\ldots,N\}$, with
$L_1=1$. Every deterministic strategy that makes at most $q\le N-2$
probes and then outputs a label satisfies
\[
 \Pr[\text{output}=L_N]\le\frac{q+1}{N-q}.
\]
\end{lemma}
\begin{proof}
Initially only $L_1$ is \emph{exposed}. A probe at an active vertex
also exposes its own label and its neighbours' labels.
Conditional on a transcript of $m$ probes, every
unexposed label remains uniform among the candidates not excluded by
inactive replies in its layer. These sets have size at least $N-m$,
and the unexposed labels remain independent: the transcript fixes
exposed labels and separately excludes candidates in the other layers.

A \emph{lucky hit} probes the active vertex of a layer whose label is
not yet exposed. At probe $m+1$, conditional on no earlier lucky hit,
its probability is at most $1/(N-m)$. Thus the probability of any
lucky hit in $q$ probes is at most $q/(N-q+1)$. Without a lucky hit,
the exposed layers form a prefix: a probe at an exposed active vertex
can extend it by at most one layer. Its length is at most $q+1<N$,
so $L_N$ remains unexposed. Conditional on any such transcript, a final
guess succeeds with probability at most $1/(N-q)$. Adding the two
possibilities gives
\[
 \Pr[\text{output}=L_N]
 \le\frac{q}{N-q+1}+\frac{1}{N-q}
 \le\frac{q+1}{N-q}.\qedhere
\]
\end{proof}

\begin{proof}[Proof of Proposition~\ref{seven:randomized}]
Fix $n$ and choose $N=\Theta_n(\eps^{-1/3})$ as in the proof of
Theorem~\ref{seven:main}. For each hidden path, use the same encoded
valuation profile. Every successful allocation reveals $L_N$, and
simulating $Q$ cake queries and decoding the output costs at most
$a_nQ+b_n$ path probes. This bound holds for each fixed input and
random seed; it does not require the cake protocol to be deterministic.

Nonterminating runs have infinite cost. If an encoded profile has
positive probability of nontermination, the conclusion already holds.
Otherwise, since the path family is finite, outside one null set of
seeds the protocol terminates on every encoded profile.

For sufficiently large $N$, put
\[
 q_N=\lfloor N/4\rfloor-1,\qquad
 t_N=\left\lfloor\frac{q_N-b_n}{a_n}\right\rfloor=\Omega_n(N).
\]
Truncate the simulation when it requests a $(t_N+1)$-st cake query,
and return an arbitrary label if it has not produced a decodable
allocation. The source strategy uses at most $q_N$ probes. Applying
Lemma~\ref{seven:random-path} for each remaining seed and averaging gives
\[
 \Pr[\text{successful cake output within }t_N\text{ queries}]
 \le\frac{q_N+1}{N-q_N}\le\frac13,
\]
where the probability includes both the random path and the seed.
Consequently some fixed path has early-success probability at most
$1/3$ over the seed alone. On its encoded profile, the protocol succeeds
with probability at least $2/3$, so its query count $Q$ satisfies
\[
 \Pr[Q>t_N]\ge\frac13,\qquad
 \mathbb E[Q]\ge\frac{t_N}{3}=\Omega_n(\eps^{-1/3}).
\]
The density bounds are those of the unchanged construction and depend
only on $n$.
\end{proof}

\end{document}